\documentclass[usegraphicx,usenatbib]{mn2e}
\usepackage{amsmath}
\usepackage{times}
\usepackage{subfigure}
\usepackage{epsfig}

\renewcommand{\descriptionlabel}[1]%
 {\hspace{\labelsep}\textbf{#1}}

\title[A Survey for Planetary Transits in the Field of NGC~7789]
  {A Survey for Planetary Transits in the Field of NGC~7789}

\author[D.M. Bramich et al.]
  {D.M.~Bramich,$^{1,2}$\thanks{E-mail: dmb7@st-and.ac.uk}
   Keith~Horne,$^1$
   I.A.~Bond,$^{3,4}$
   R.A.~Street,$^5$
   A.~Collier~Cameron,$^1$
   \newauthor
   B.~Hood,$^1$
   J.~Cooke,$^6$
   D.~James,$^{7,8}$
   T.A.~Lister,$^1$
   D.~Mitchell,$^6$
   \newauthor
   K.~Pearson,$^9$
   A.~Penny,$^{10}$
   A.~Quirrenbach,$^{11}$
   N.~Safizadeh$^6$ and
   Y.~Tsapras$^{12}$   
 \medskip  
 \\$^1$School of Physics \& Astronomy,
   University of St.~Andrews, North Haugh,
   St.~Andrews, Fife,
   KY16 9SS, UK
 \\$^2$Instituto de Astrofisica de Canarias,
       C/ Via Lactea s/n, E-38200, La Laguna, Tenerife, 
       Spain
 \\$^3$Institute for Astronomy,
       University of Edinburgh, Royal Observatory,
       Blackford Hill, Edinburgh,
       EH9 3HJ, UK
 \\$^4$Institute for Information and Mathematical Sciences,
       Massey University, Auckland, New Zealand
 \\$^5$APS Division, Dept. of Pure \& Applied Physics,
       Queen's University, University Road,
       Belfast, BT7 8NQ, UK
 \\$^6$Center for Astrophysics \& Space Sciences (CASS),
       Uni. of California, San Diego,
       9500 Gilman Drive, La Jolla,
       CA 92093-0424, USA
 \\$^7$Laboratoire d'Astrophysique, Observatoire de Grenoble,
       BP53, F-38041 Grenoble,
       Cedex 9, France
 \\$^8$Dept. of Physics \& Astronomy, Vanderbilt University,
       1807 Station B, Nashville,
       TN 37235, USA
 \\$^9$Louisiana State University,
       Dept. of Physics \& Astronomy,
       202 Nicholson Hall, Tower Drive, Baton Rouge,
       Louisiana 70803-4001, USA
 \\$^{10}$Rutherford Appleton Laboratory, Chilton, Didcot,
       Oxon, OX11 0QX, UK
 \\$^{11}$Leiden Observatory, P.O. Box 9513, NL-2300 RA Leiden, 
          The Netherlands
 \\$^{12}$School of Mathematical Sciences, Queen Mary University of London,
          Mile End Road, London, E1 4NS, UK}

\begin{document}

\date{Accepted 2005 ???. Received 2004 October 18; in original form 2004
      October 1}

\pagerange{\pageref{firstpage}--\pageref{lastpage}} \pubyear{2004}

\maketitle

\label{firstpage}

\begin{abstract} 
We present results from 30 nights of observations of the open cluster 
NGC~7789 with the WFC camera on the INT telescope in La Palma. From
$\sim$900 epochs, we obtained lightcurves and Sloan $r^{\prime} - i^{\prime}$ colours for $\sim$33000 
stars, with $\sim$2400 stars with better than 1\% precision. We expected to detect
$\sim$2 transiting hot Jupiter planets if 1\% of stars host such a companion and that a typical hot Jupiter 
radius is $\sim1.2R_{\mbox{\small J}}$. We find 24 transit candidates, 14 of which we can 
assign a period. We rule out the transiting planet model for 21 of these candidates using various robust arguments.
For 2 candidates we are unable to decide on their nature, although it seems most likely that they are eclipsing binaries 
as well. We have one candidate exhibiting a single eclipse for which we 
derive a radius of 
$1.81^{+0.09}_{-0.00} R_{\mbox{\small J}}$. Three 
candidates remain that require follow-up observations in order to determine their nature.
\end{abstract} 


\section{Introduction}

The surprising existence of short period ($\sim$4 days) Jupiter mass
extra-solar planets (termed ``hot Jupiters'') confirmed by
radial velocity measurements in the last decade has shown
us that planetary systems exist in patterns unlike that of our own Solar System. The  
class of hot Jupiter planets ($P \la 10$ days and $M \sin{i} \la 10 M_{\mbox{\small J}}$)
makes up $\sim$19\% (27 out of 140 as of 01/02/05) of the planets 
discovered to date \citep{sch1996} and $\sim$1\% of nearby solar type stars host 
such a companion \citep{but2000}. Recently we are starting to see the fruits of current transit surveys. OGLE have produced over
100 transit candidates during two seasons (\citealt{uda2002a}; \citealt{uda2002b}; \citealt{uda2003}), by far
the most prolific transit survey. EXPLORE/OC have produced a handful of transit candidates that are currently being
followed up spectroscopically \citep{mal2003} and a search of the MACHO photometry database has revealed nine
transit candidates \citep{dra2004}.
To date there are seven confirmed transiting extra-solar planets,
HD 209458b (\citealt{cha2000}; \citealt{bro2001}) discovered first by 
the radial velocity method, 
and OGLE-TR-56b \citep{kon2003}, OGLE-TR-113b (\citealt{bou2004}; \citealt{kon2004a}), OGLE-TR-132b \citep{bou2004}, 
OGLE-TR-111b \citep{pon2004}, TrES-1 \citep{alo2004} and OGLE-TR-10b \citep{kon2004b} 
discovered first by the transit method. The spectroscopic follow-up of the OGLE transit candidates has revealed
a new class of short period planets called the ``very hot Jupiters''. Such planets have periods less than the 
3 day cut-off identified in the sample of radial velocity planets. 

In order to discover transiting hot Jupiters through a photometric survey, one requires high cadence, high 
accuracy observations ($\la1$\% accuracy per data point with a duty cycle $\ga8$ data points per hour) of many 
stars ($>10^{4}$) simultaneously over long observing runs ($>10$ nights). Any transit candidates (stars that show at 
least one eclipse event) may be subsequently 
followed up by radial velocity (RV) measurements in order to determine the companion mass, or at least an upper limit to the 
mass. However, due to the long integration times on large telescopes 
required for RV follow-up and the high frequency 
of planetary transit mimics, it is prudent to try and rule out the transiting planet model for as many transit 
candidates as possible via simple supplementary observations and/or further analysis of the lightcurve 
(\citealt{cha2003}; \citealt{sir2003}; \citealt{dra2003}; \citealt{cha2004}).

The study of open clusters for transiting planets has a number of advantages over fields
in other parts of the sky or Galactic plane. While providing a relatively large
concentration of stars on the sky (but not so large as to cause blending problems
as in the case of globular clusters observed from the ground), they also provide
a set of common stellar parameters for the cluster members. These are metallicity, age, stellar crowding and radiation 
density. Also, the fainter cluster members are smaller stars and therefore they are likely to show deeper transit 
signatures, helping to offset sky noise contributions. The identification of the cluster main sequence
in the colour magnitude diagram allows the assignment of a model-dependent mass and radius to each photometric cluster 
member, and assuming a law relating extinction to distance for the field allows the assignment of a 
model-dependent mass, radius and distance to 
all stars in the field under the assumption that they are main sequence stars. Transit candidates with well defined 
phased lightcurves may therefore be analysed in detail as to whether they are consistent with a transiting planet 
model. An estimate of the fraction of stars
hosting a hot Jupiter (referred to as the hot Jupiter fraction) may be obtained by comparing the number 
of hot Jupiters that are actually detected to how many one would expect to detect using the knowledge of the star 
properties and the lightcurves themselves. The dependence of the hot Jupiter fraction on the cluster parameters 
may then be investigated by extending the experiment to other open clusters.

Currently there are a number of groups searching for transiting planets in open clusters \citep{bra2004a}. These include UStAPS 
(\citealt{str2003}; this paper; \citealt{hoo2005}), 
EXPLORE/OC (\citealt{mal2003}; \citealt{lee2004}; \citealt{bra2004b}), STEPSS \citep{bur2004} 
and PISCES (\citealt{moc2002}; \citealt{moc2004}). A number of transit candidates have been put forward by these groups but none have 
been confirmed as transiting planets so far. In this survey, as with open cluster transit surveys in general, we are photometrically
observing faint stars (16 to 21 mag). This makes follow-up observations difficult but not impossible. However, many of the stars 
observed are of later spectral types than those probed by the RV surveys since RV surveys are limited to bright solar neighbourhood FGK 
stars. In particular, our survey of NGC~7789 probes the spectral types F5 to M5 (see Fig. 2) including a large proportion of K and M 
stars. Furthermore, we are searching for planets out to distances well beyond the solar neighbourhood.

The observations of open cluster NGC~7789 is the subject of this paper. The main parameters of the 
cluster are shown in Table 1. For a good review of previous relevant work on this cluster see \citet{gim1998}. 
In Section 2 we report on the observations made, in Section 3 we present in detail the data 
reduction/photometry, in Section 4 we present the astrometry and colour data, in Section 5 we describe the 
transit detection algorithm and in Section 6 we present a detailed analysis of the transit candidates.
Finally, in Section 7 we outline our conclusions and future work.

\begin{table}
\begin{center}
\caption{Properties of the open cluster NGC~7789. Data taken from 
         http://obswww.unige.ch/webda by Mermilliod, J.C. and the SIMBAD database.}
\begin{tabular}{cc}
\hline
RA (J2000.0)        & 23$^{\mbox{\small h}}$ 57$^{\mbox{\small m}}$    \\
Dec (J2000.0)       & +56\degr 43\arcmin   \\
$l$                 & 115\fdg48            \\
$b$                 & $-$5\fdg37             \\
Distance (pc)       & 2337                 \\
Radius              & $\sim$16\arcmin      \\
Age (Gyr)           & 1.7                  \\
$[$Fe/H$]$          & $-$0.08                \\
$E(B-V)$            & 0.217                \\
\hline
\end{tabular}
\end{center}
\end{table}

\section{Observations}

We observed the open cluster NGC~7789 (see Table 1) using the 2.5m Isaac Newton Telescope (INT) of the Observatorio 
del Roque de los Muchachos, La Palma, in the Canary Islands during three bright runs with dates 1999 June 22-30, 
1999 July 22-31 and 2000 September 10-20. 
For brevity, these runs shall be refered to from now on as 1999-06, 1999-07 and 2000-09 respectively.
We used the Wide Field Camera (WFC) which consists of a 4 EEV CCD mosaic where each CCD is 2048x4096 
pixels \citep{wal2001}. The pixel scale is 0.33\arcsec/pix and field of view $\sim$0.5\degr x 0.5\degr. The gain and 
readout noise values for each chip were calculated automatically during the preprocessing stage of 
the data reduction (see Section 3.1). The mosaic field was centred on NGC~7789 at 
$\alpha=23^{\mbox{\small h}}57^{\mbox{\small m}}30^{\mbox{\small s}}$ 
and $\delta=+56\degr43\arcmin41\arcsec$. 

The usual procedure for each night was to obtain $\sim$5 bias frames and $\sim$8 sky flat frames at both 
the beginning and end of the night. Observations on NGC~7789 in the runs 1999-06 and 1999-07 consisted 
of $\sim$6 pairs of 300s exposures taken every $\sim$50 minutes during the later part of each 
night. Observations in the 
2000-09 run consisted of sequences of ten consecutive 300s exposures 
followed by a bias frame, repeated continuously throughout the 
whole of each night. With a readout time of 100s and various losses due to bad weather/seeing and 
telescope jumps, this resulted in a total of $880\times300$~s exposures in Sloan $r^{\prime}$ over the three runs, 
with 691 of 
these exposures from the 2000-09 run alone. During the 2000-09 run, we also took 5 images of NGC~7789 with varying 
exposure times in Sloan $i^{\prime}$, along with 5 sky flat frames, in order to provide us with the necessary colour 
information.  

\section{CCD Reductions}

\subsection{Preprocessing: CCD Calibrations}

\begin{figure*}
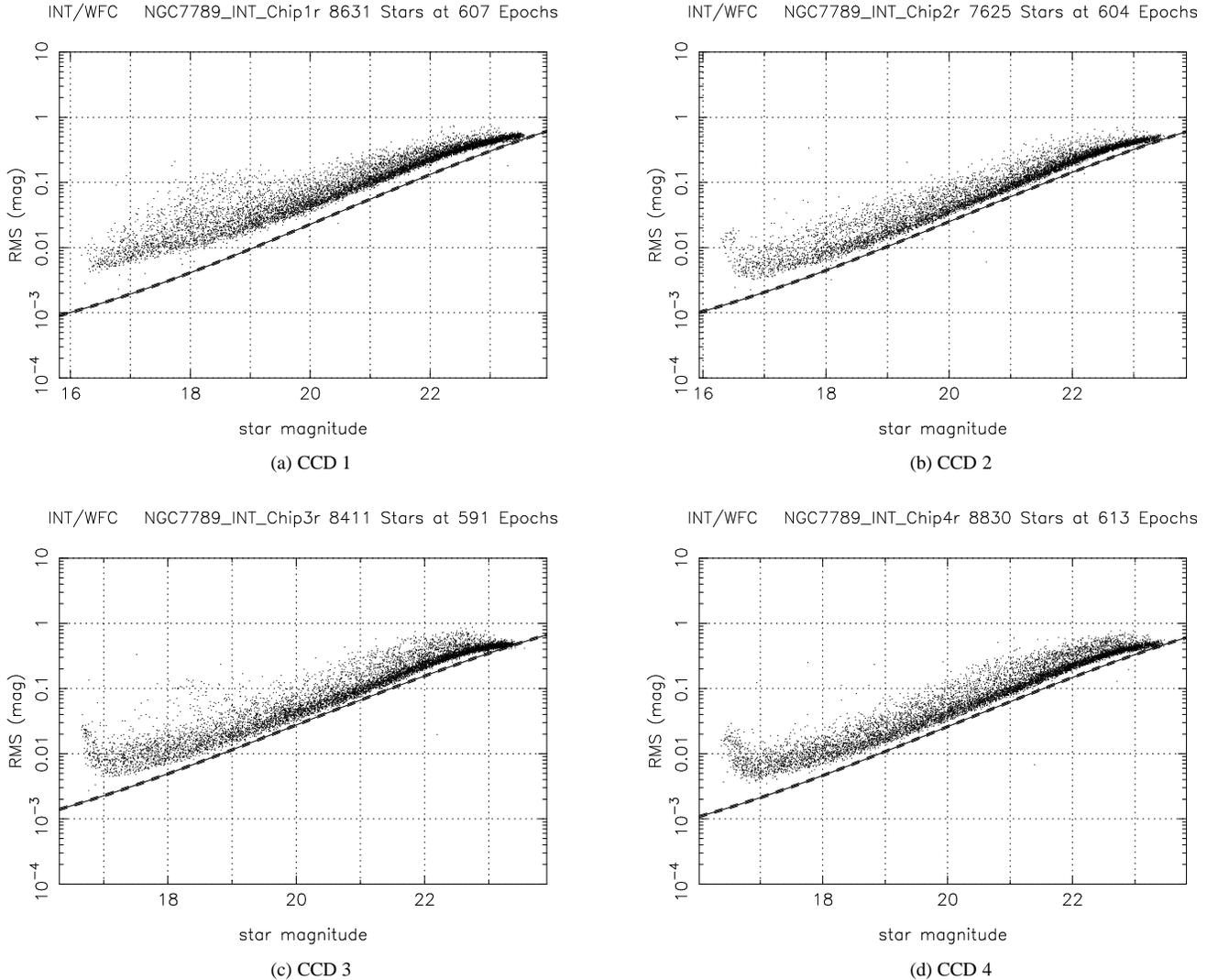

\def\subfigtopskip{4pt}
\def\subfigbottomskip{8pt}
\def\subfigcapskip{4pt}
\centering
\begin{tabular}{cc}
\subfigure[CCD~1]{\epsfig{file=rms_chip1.ps,angle=270.0,width=0.5\linewidth}} &
\subfigure[CCD~2]{\epsfig{file=rms_chip2.ps,angle=270.0,width=0.5\linewidth}} \\
\subfigure[CCD~3]{\epsfig{file=rms_chip3.ps,angle=270.0,width=0.5\linewidth}} &
\subfigure[CCD~4]{\epsfig{file=rms_chip4.ps,angle=270.0,width=0.5\linewidth}} \\
\end{tabular}
\caption{Plots of standard deviation (RMS) of the lightcurves against mean instrumental
         Sloan $r^{\prime}$ magnitude for all stars from each chip for the 2000-09 run.
         The lower curve in each diagram represents the theoretical noise limit for photon and readout
         noise.
\label{fig:acc}}
\end{figure*}

Each run and each chip was treated independently for the purpose of the reductions. 
The reduction process was carried out by a single C-shell/IRAF script that runs
according to a user-defined parameter file. Bad pixels were flagged in a user-defined detector bad
pixel mask, and ignored where relevant. The reductions consisted of the standard bias subtraction and flat 
fielding of the science frames using appropriate master calibration frames. The gain and readout noise of each chip 
were also determined during the reductions.

\subsection{Photometry: Difference Image Analysis}

Differential photometry on the reduced science frames in the Sloan $r^{\prime}$ filter was accomplished using the 
method of difference image analysis (\citealt{ala1998}; \citealt{ala2000}). Our implementation of this procedure was
adapted from the code written for the MOA project \citep{bon2001}, and it consists of three automated
scripts. The process is described below in relative detail since the same procedure has subsequently been used on many
other data sets (for example \citealt{fer2004}). Bad pixels are ignored 
in the operations that the scripts perform.

\begin{enumerate}
  \item The first script constructs a reference frame from selected frames with good seeing, and a star 
        list from the reference frame. First, stars are detected and matched between the best 
        seeing frames in order to derive a set of linear transformations and geometrically align the 
        frames. The frames are then combined into a mean reference frame using the exposure times of 
        the individual images as weights. The reference frame is analysed using IRAF's DAOPhot 
        package \citep{ste1987}. The package identifies stars on the reference frame and chooses a set 
        of 175 point-spread function (PSF) stars. A ``penny2'' PSF function that varies quadratically 
        with position, along 
        with a lookup table of residuals, is solved for. The neighbours of the PSF stars are then 
        subtracted using this solution, and a new PSF function is solved for. This new solution is 
        used to measure the instrumental fluxes and positions of all stars on the reference frame. The result 
        is a reference frame with a corresponding star list. We used 13 consecutive best seeing 
        images ($\sim$1\arcsec) to construct the reference frame. 
  \item The reference frame is used to produce a set of difference images. The main idea behind 
        difference image analysis is that an image frame $I(x,y)$ is related to the reference frame 
        $R(x,y)$ via the following equation:
        \begin{equation}
	\label{eqn:convolve}
        I(x,y) = R \otimes K(x,y) + B(x,y) 
        \end{equation}
        where
        \begin{equation}
        R \otimes K(x,y) \equiv \int\int R(x-u,y-v)\ K(u,v,x,y)\ {\rm d}u\ {\rm d}v 
        \end{equation}
	Here $B(x,y)$ represents the change in the sky background and $K(u,v,x,y)$ is a spatially-varying convolution 
        kernel relating the point-spread function on the reference frame to the point-spread function on the image
        frame at spatial position $x,y$. We model the convolution kernel:
	\begin{equation}
	K(u,v,x,y) = \sum_i a_i(x,y) b_i(u,v)
	\end{equation}
	as the sum of a set of basis functions $b_i(u,v)$ each formed as a product of
        a two-dimensional Gaussian function of $u$ and $v$ with a polynomial of degree 2 in $u$ and $v$. 
        For the basis functions we use 3 Gaussian components with sigmas of 2.1 pix, 1.3 pix and 0.7 pix and
	associated polynomial degrees of 2, 4 and 6 respectively.
        To allow for the kernel's spatial dependence, the coefficients $a_i(x,y)$
	are polynomials of degree 2 in $x$ and $y$. The kernel is also normalised
	to a constant integral over $u$ and $v$ for each $x$ and $y$, thus ensuring a constant
        photometric scale factor between the reference frame and image frame. We model the 
        differential sky background $B(x,y)$ as a polynomial of degree 2 in $x$ and $y$.
        We solve for $K(u,v,x,y)$ and $B(x,y)$ in the least-squares sense for each science frame by fitting
	to pixel boxes around selected bright stars distributed uniformly across the reference frame.
	The kernel is assumed to be independent of $x$ and $y$ within each box. 
        A difference image is then constructed for each science frame 
        by rearranging Eqn.~\ref{eqn:convolve} to the following form and using the solutions for $K(u,v,x,y)$ 
	and $B(x,y)$: 
        \begin{equation}
        D(x,y) = I(x,y) - R \otimes K(x,y) - B(x,y)
        \end{equation}
        The difference image $D(x,y)$ should simply be an image representative of 
        the Poisson noise in $I(x,y)$. However, any objects that have varied in 
        brightness in comparison to the reference frame should show up as positive or negative pixel 
        areas on the difference image which may be measured to obtain the differential flux. In our 
        analysis, each chip was split up into 8 square sections and the difference image constructed 
        from solving for the kernel and differential sky background in each section. Also, a high 
        signal-to-noise empirical PSF for the reference frame is constructed in each section by 
        stacking up a set of stamps centred on suitable bright stars.
  \item The third script measures the differential flux on each difference image via optimal PSF 
        scaling at the position of each star. The normalised and sky-subtracted empirical PSF 
        constructed for each square section of the reference frame in step~(ii) is convolved with the 
        kernel corresponding to the current difference image section. The convolved PSF is optimally scaled, 
        at the position of the current star, to the difference image. A $3 \sigma$ clip on the 
        residuals of the scaling is performed, and one pixel rejected. The scaling and rejection is 
        repeated until no more pixels are rejected. The differential flux is measured as the integral 
        of this scaled PSF.
\end{enumerate}

\begin{table}
\begin{center}
\caption{Magnitude offsets $\overline{\Delta m_{1}}$ and $\overline{\Delta m_{2}}$ added to the
         lightcurve data points from the 1999-06 and 1999-07 runs respectively.}
\begin{tabular}{ccccc}
\hline
Chip No. & $\overline{\Delta m_{1}}$ & $\sigma_{1}$ &
           $\overline{\Delta m_{2}}$ & $\sigma_{2}$ \\
\hline
1 & -0.832 & 0.037 & -0.753 & 0.041 \\
2 & -0.698 & 0.036 & -0.843 & 0.043 \\
3 & -0.596 & 0.031 & -0.404 & 0.054 \\
4 & -0.527 & 0.036 & -0.470 & 0.037 \\
\hline
Run: & 1999-06 & & 1999-07 & \\
\hline
\end{tabular}
\end{center}
\end{table}

A lightcurve for each star was constructed by the addition of the differential fluxes to the star fluxes 
as measured on the reference frame. The following equations were used:
\begin{equation}
f_{\mbox{\small tot}}(t) = f_{\mbox{\small ref}} + \frac{f_{\mbox{\small diff}}(t)}{p(t)}
\end{equation}
\begin{equation}
m(t) = 25.0 - 2.5 \log (f_{\mbox{\small tot}}(t))
\end{equation} 
where $f_{\mbox{\small tot}}(t)$ is the star flux (ADU/s) at time $t$, $f_{\mbox{\small ref}}$ is the star flux (ADU/s) 
as measured on the reference frame, $f_{\mbox{\small diff}}(t)$ is the differential flux (ADU/s) at time $t$ as 
measured on the difference image, $p(t)$ is the photometric scale factor (the integral of the kernel 
solution over $u$ and $v$) at time $t$ and $m(t)$ is the magnitude of the star at time $t$.
Uncertainties are propagated in the correct analytical fashion. 

Flux measurements were rejected for a $\chi^{2} \mbox{pix}^{-1} \geq 5.0$ for the PSF scaling, and for PSFs with a 
FWHM$\geq$7.0 pix, in order to remove bad measurements. Hence, all the stars have differing numbers of photometric 
measurements. In each run, lightcurves with less than half of the total possible epochs were 
rejected. For the 2000-09 run this analysis produced 8631 lightcurves on Chip 1, 7625 
lightcurves on Chip 2, 8411 lightcurves on Chip 3 and 8830 lightcurves on Chip 4 (centred on the 
cluster). Fig.~\ref{fig:acc} shows a diagram of the RMS scatter in the lightcurves against instrumental magnitude for 
the 
2000-09 run for each chip. Similar diagrams were produced for the 1999-06 and 1999-07 runs but 
are not shown here for brevity. 

Since each run was treated independently for the reductions, each chip has three different reference frames and hence each 
star has three different reference magnitudes. For a particular star, let us denote the reference magnitude from the 
2000-09 run minus the reference magnitude from the 1999-06 run by $\Delta m_{1}$ and the reference magnitude from the 
2000-09 run minus the reference magnitude from the 1999-07 run by $\Delta m_{2}$. For each chip, we have calculated the 
unweighted mean of $\Delta m_{1}$ and $\Delta m_{2}$ over all stars on that chip. We then added the resulting 
$\overline{\Delta m_{1}}$ and $\overline{\Delta m_{2}}$ to the lightcurve data points in the 1999-06 and 1999-07 runs 
respectively. The values of the means $\overline{\Delta m_{1}}$ and $\overline{\Delta m_{2}}$ for each chip along with 
the standard deviations about the means $\sigma_{1}$ and $\sigma_{2}$ respectively are presented in Table 2.
 
As can be seen from Fig.~\ref{fig:acc}, we have obtained high precision photometry with an RMS accuracy of 
$\sim$3-5mmag at the bright end. Most stars are limited by sky noise because all three runs were during 
bright time. However, the ``backbone'' of points on each diagram lies above the 
theoretical limit by a factor of $\sim$1.5-2.0 depending on the chip being considered. 
We put this down to systematic errors in the data due to a 
subset of low quality difference images and/or sections of difference images that were produced from science frames 
taken on nights of poor quality seeing/atmospheric conditions. 

\section{Astrometry And Colour Data}

\subsection{Astrometry}

Astrometry was undertaken by matching 358 stars from the four reference frames (one for each chip) with 
the USNO-B1.0 star catalogue \citep{mon2003} using a field overlay in the image display tool 
GAIA \citep{dra2000}. The WFC suffers from pincushion distortion, hence it was necessary to fit a 
9 parameter astrometric solution to the reference frames in order to obtain sufficiently
accurate celestial coordinates for all the stars. The 9 parameters are made up of 6 parameters to 
define the linear transformation between pixel coordinates and celestial coordinates, 2 parameters to 
define the plate centre and 1 parameter to define the radial distortion coefficient. The 
starlink package ASTROM \citep{wal1998} was used to do the fit and the achieved accuracy was $\sim$0.4 
arcsec RMS radially for the 358 matching stars. The astrometric fit was then used to calculate 
the J2000.0 celestial coordinates for all stars with a lightcurve.

\subsection{Colour Indices}

The best image in the Sloan $i^{\prime}$ filter was aligned with the Sloan $r^{\prime}$ reference 
frame for each chip and the magnitudes of the stars were measured using DAOPhot PSF fitting
in the same way as they were measured on the reference frame in Section 3.2. Table 3 shows the number 
of stars with lightcurves that have Sloan $r^{\prime} - i^{\prime}$ colour indices as a result.

\subsection{Colour Magnitude Diagrams}

\begin{table}
\begin{center}
\caption{Number of stars with a Sloan $r^{\prime}$ lightcurve, and the number of such stars with a 
         Sloan $r^{\prime} - i^{\prime}$ colour index.}
\begin{tabular}{cccc}
\hline
     &                & No. Stars With A                       &            \\
Chip & No. Stars With & Lightcurve And An                      &            \\
No.  & A Lightcurve   & $r^{\prime} - i^{\prime}$ Colour Index & Percentage \\
\hline
1 & 8631 & 8497 & 98.4\% \\
2 & 7625 & 7576 & 99.4\% \\
3 & 8411 & 8290 & 98.6\% \\
4 & 8830 & 8672 & 98.2\% \\
\hline
Total: & 33497 & 33035 & 98.6\% \\
\hline
\end{tabular}
\end{center}
\end{table}

Figs.~2a-d show an instrumental colour magnitude diagram (CMD) for each chip. The cluster main sequence 
is 
clearly visible. Chip 4 is centred on the cluster and as expected shows the strongest cluster main sequence.
A theoretical cluster main sequence is plotted on each diagram over the cluster main sequence. We have used the 
theoretical models of \citet{bar1998} for the stellar mass range $0.60M_{\sun} \leq M_{*} \leq 1.40M_{\sun}$, the age 
of the cluster (1.7Gyr) and solar-type metallicity [M/H] $=$ 0 in order to predict the main sequence absolute 
magnitudes, colours and radii. 
Below a mass of 0.60$M_{\sun}$ the Baraffe model predicts $R-I$ colours 
substantially bluer than the observed cluster main sequence, a limitation noted in \citet{bar1998}. As a result we used 
data from \citet{lan1992} for the stellar mass range $0.08M_{\sun} \leq M_{*} \leq 0.60M_{\sun}$. The combined 
model for the cluster 
main sequence supplies an absolute magnitude $M_{R}$, an absolute magnitude $M_{I}$ and a stellar radius $R_{*}$ 
for the stellar mass range $0.08M_{\sun} \leq M_{*} \leq 1.40M_{\sun}$. We interpolated this combined model with 
cubic splines.

The interstellar medium (ISM) in the Milky Way is mostly concentrated in the Galactic plane and the density law 
governing its mean distribution (ignoring small scale variations) can be modelled by an Einasto law:
\begin{equation}
\rho(R,z) = \rho_{0} \exp \left( - \left( \frac{R - R_{\sun}}{h_{R}} \right) \right) \exp \left( - \frac{|z|}{h_{z}} \right)
\end{equation}
where $R$ is the Galactocentric distance, $z$ is the height above the Galactic plane, $\rho_{0}$ is the local density
of the ISM, $R_{\sun}$ is the distance of the Sun from the Galactic centre, $h_{R}$ is the ISM density scale height in 
the $R$ direction and $h_{z}$ is the ISM density scale height in the $z$ direction. One may derive the density $\rho$ 
of the ISM as a function of distance $d$ from the Sun in the direction of the open cluster NGC~7789 by using 
trigonometrical arguments to rewrite $R$ and $z$ as functions of $d$. In this derivation, we have assumed that the Sun 
has Galactic coordinates $(R,z) = (8.5$kpc$,0.015$kpc$)$ and that $\rho_{0} = 0.021 M_{\sun}$pc$^{-3}$, 
$h_{R} = 4.5$kpc, $h_{z} = 0.14$kpc as given in \citet{rob2003}. 
In any wave band, the total extinction $A$ as a function of $d$ is proportional to the integral of $\rho(d)$ over 
$d$. Hence, absorbing the constant $\rho_{0}$ into a new constant $K$ we have:
\begin{equation}
\label{eqn:extinction}
A(d) = K \int_{0}^{d} \rho(u) \,\text{d}u
\end{equation}
Adopting $E(B-V) = 0.217$ for the cluster (Table 1), we calculate the corresponding extinction to be
$A_{R} \approx 0.547$ and $A_{I} \approx 0.429$ in the $R$ and $I$ bands respectively, evaluated with a synthetic
photometry code (XCAL) using a Galactic extinction curve from \citet{sea1979}. This extinction applies to stars at
the cluster distance $d_{\mbox{\small c}} = 2337$pc, and hence, by numerically evaluating the integral in 
Eqn.~\ref{eqn:extinction}, 
we may calculate values for $K$ that apply to the $R$ and $I$ bands as $K_{R} = 
2.20\times10^{-2}$mag$M_{\sun}^{-1}$pc$^{2}$ and $K_{I} = 1.73\times10^{-2}$mag$M_{\sun}^{-1}$pc$^{2}$ respectively. 

We have used the law relating extinction to distance as given in Eqn.~\ref{eqn:extinction} to correct the absolute 
magnitudes $M_{R}$ and $M_{I}$ of the theoretical main sequence to the observed magnitudes $R(d)$ and $I(d)$ 
respectively. In the following equations, the distance $d$ has units of parsecs (pc):
\begin{equation}
R(d) = M_{R} + 5\log(d) - 5 + A_{R}(d)
\end{equation}
\begin{equation}
I(d) = M_{I} + 5\log(d) - 5 + A_{I}(d)
\end{equation}
where $A_{R}(d)$ and $A_{I}(d)$ are versions of Eqn.~\ref{eqn:extinction}
with $K = K_{R}$ and $K = K_{I}$ respectively.

Conversions between the Johnson-Cousins $R$ and $I$ magnitudes and the Sloan $r^{\prime}$ and $i^{\prime}$ 
magnitudes were done using the following predetermined relations presented on the Cambridge Astronomical Survey 
Unit (CASU) webpage\footnote{http://www.ast.cam.ac.uk/$\sim$wfcsur/index.php}:
\begin{equation}
r^{\prime} = R + 0.275(R-I) + 0.008
\end{equation}
\begin{equation}
r^{\prime} - i^{\prime} = 1.052(R-I) + 0.004
\end{equation}

Due to the lack of observations of standard stars, it was necessary to fit the interpolated theoretical main sequence to
the cluster main sequence on the CMD for each chip by eye, after correcting for the cluster distance and extinction, 
via simple $r^{\prime}$ and $r^{\prime}-i^{\prime}$ offsets. These offsets are displayed in Table 4.
Note that the required horizontal and vertical shifts are correlated, since shifts parallel to the main sequence
would have no effect if the main sequence were a straight line. Fortunately, the kink (change of slope)
in the main sequence near the spectral type K0 (0.8$M_{\sun}$) allows us to estimate both vertical and horizontal 
shifts. This feature is clearly visible on all 4 chips.

\begin{figure*}
\def\subfigtopskip{4pt}
\def\subfigbottomskip{8pt}
\def\subfigcapskip{4pt}
\centering
\begin{tabular}{cc}
\subfigure[CCD~1]{\epsfig{file=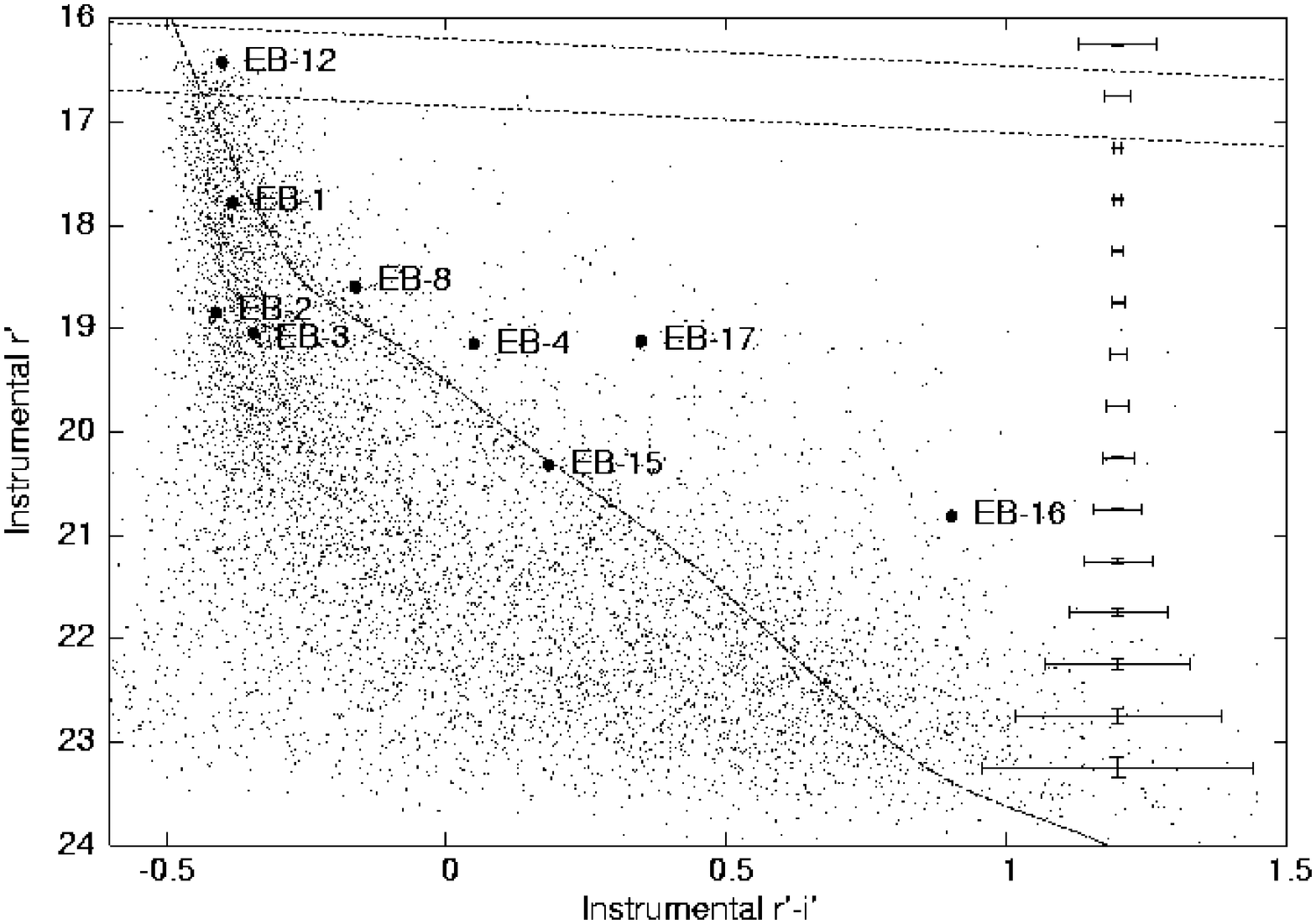,angle=270.0,width=0.45\linewidth}} &
\subfigure[CCD~2]{\epsfig{file=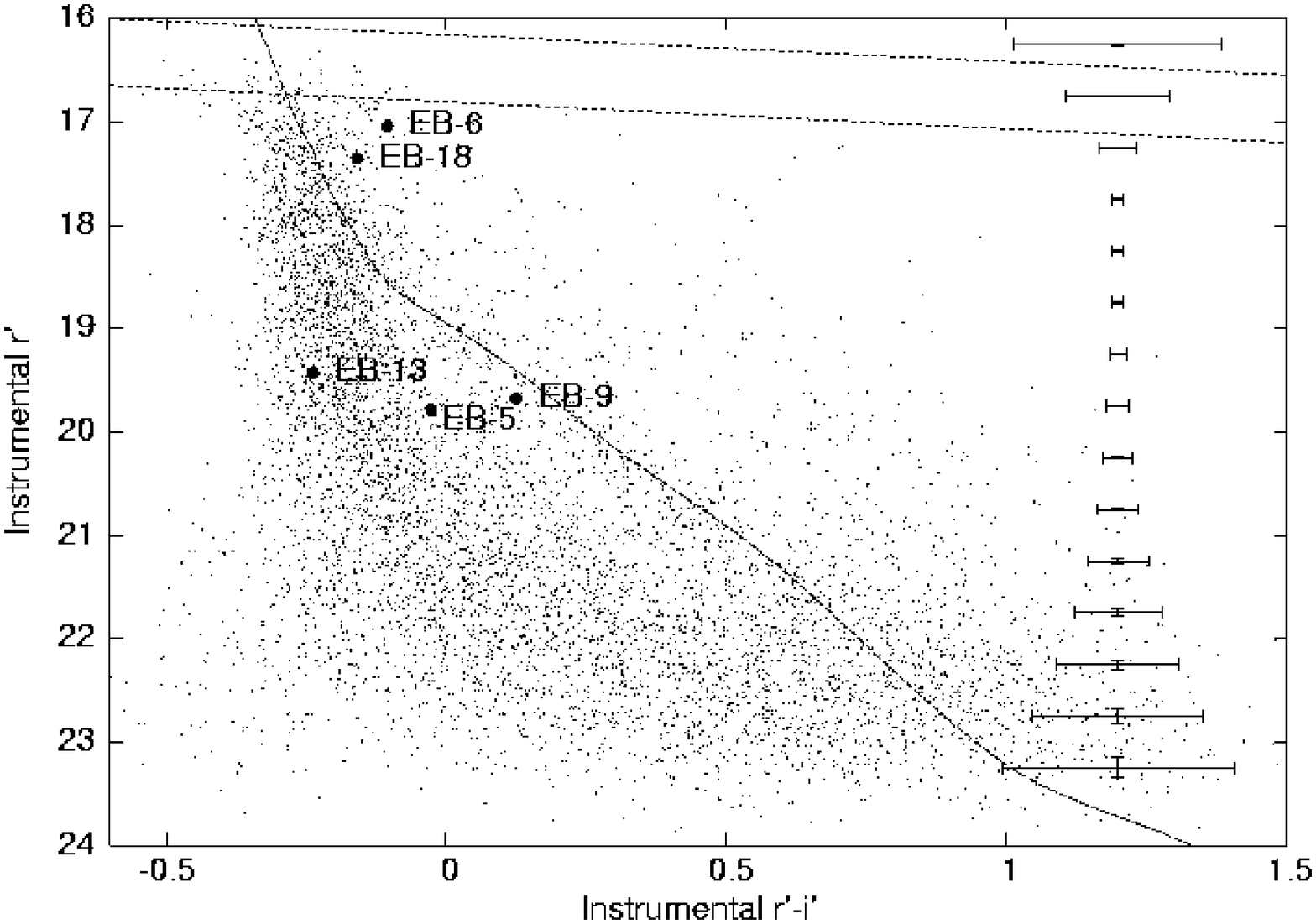,angle=270.0,width=0.45\linewidth}} \\
\subfigure[CCD~3]{\epsfig{file=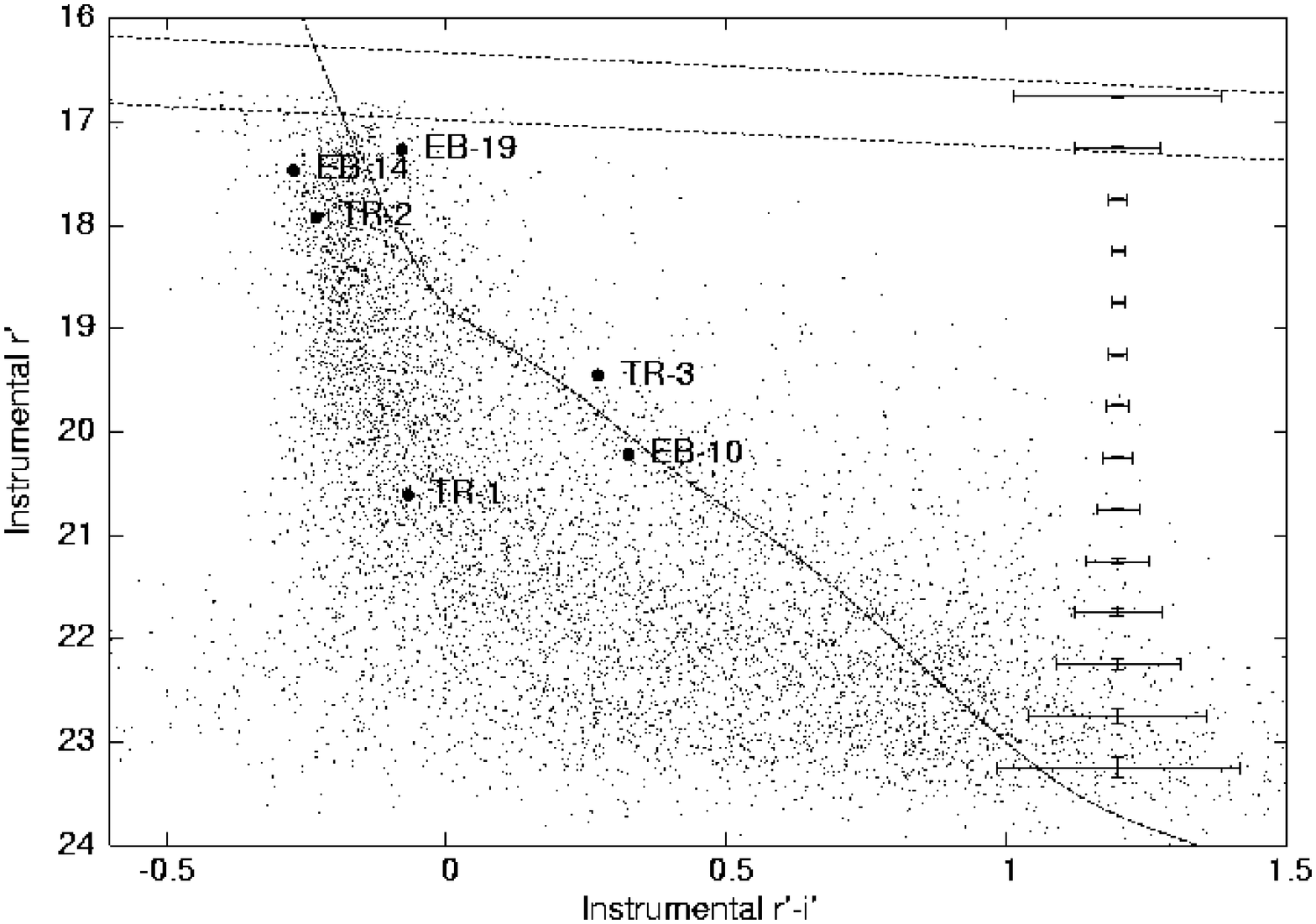,angle=270.0,width=0.45\linewidth}} &
\subfigure[CCD~4]{\epsfig{file=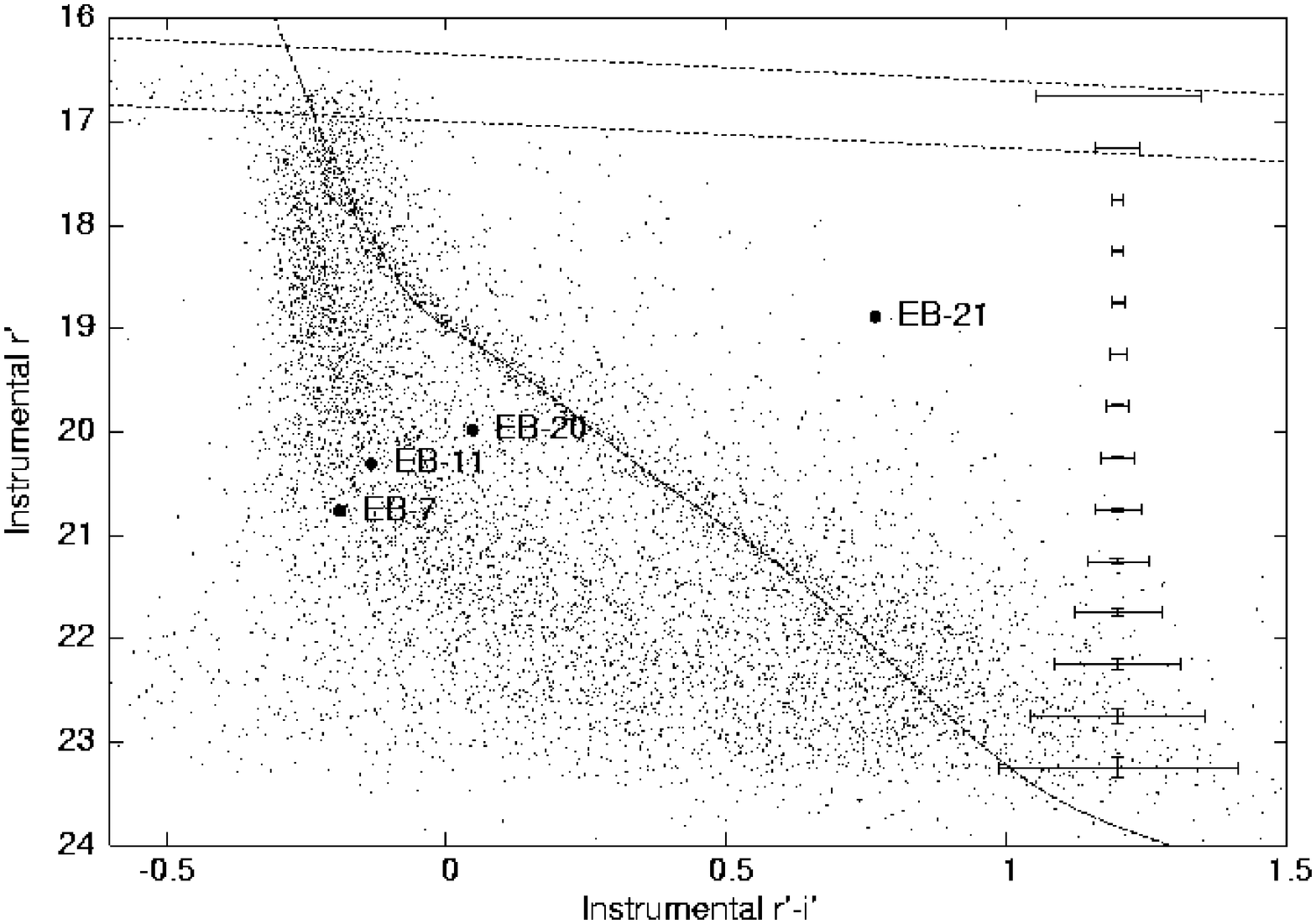,angle=270.0,width=0.45\linewidth}} \\
\end{tabular}
\begin{tabular}{c}
\subfigure[CCD~4]{\epsfig{file=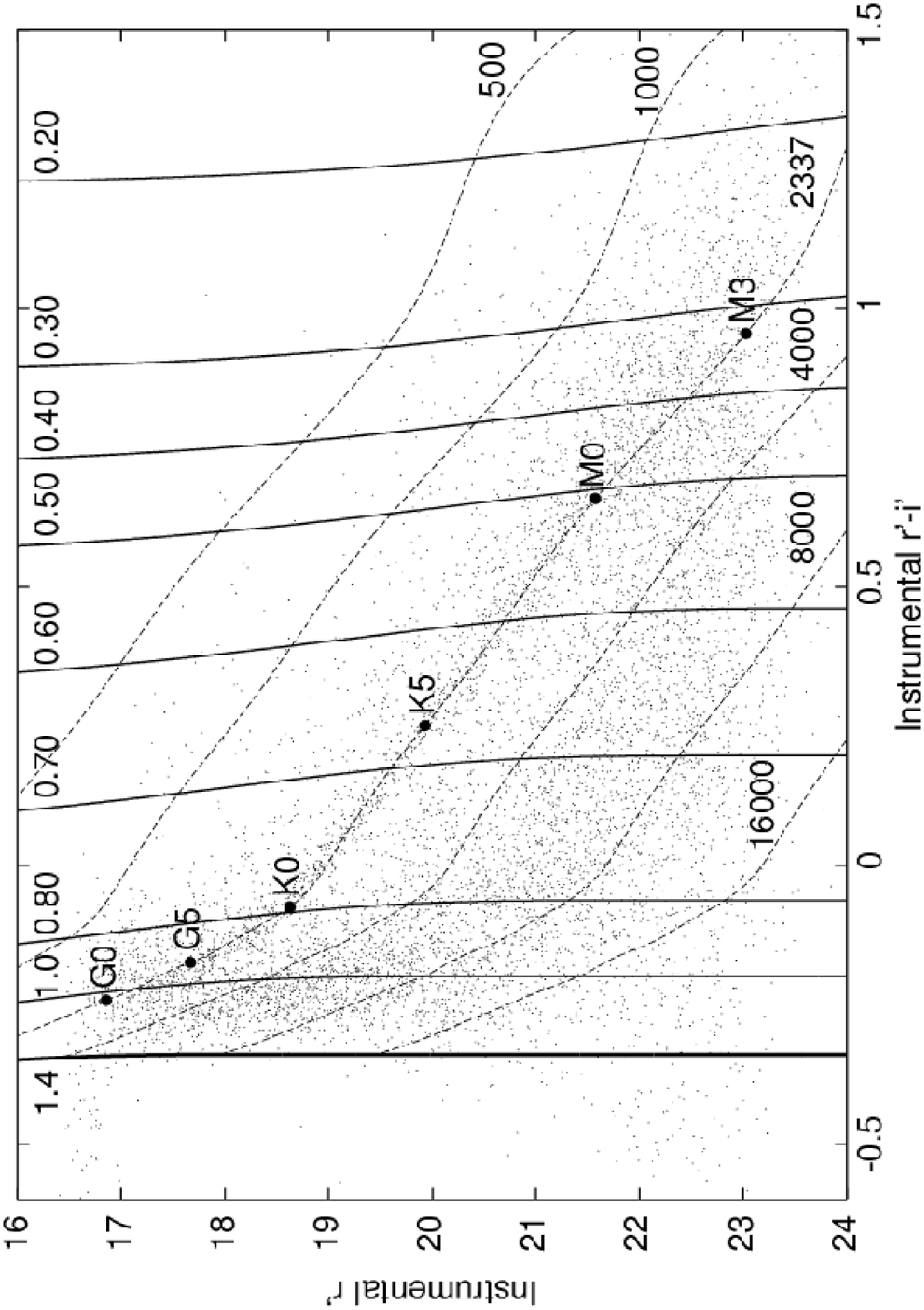,angle=270.0,width=0.65\linewidth}} \\
\end{tabular}
\caption{(a)-(d): Instrumental CMDs for all stars from each chip for the 2000-09 run.
         The main sequence is visible on each chip, and the theoretical cluster main sequence is overlayed as the 
         dashed line. The straight dotted lines are the faint limits for giant stars assuming no extinction and 
         the law relating extinction to distance in Eqn.~\ref{eqn:extinction}. The transit candidates of Section 6 
	are marked on
         as solid circles. The errorbars on the right hand side of each diagram represent the mean
         error bar on each measurement for 0.5 magnitude bins. (e): Instrumental CMD
         for Chip 4 (see the text in Section 4.4). The numbers along the top are masses in units of $M_{\sun}$ and
         the numbers along the bottom and right are distances in parsecs.
\label{fig:cmd}}
\end{figure*}

\begin{table}
\begin{center}
\caption{Offsets determined by eye between observed $r_{\mbox{\scriptsize obs}}^{\prime}$ and 
         $r_{\mbox{\scriptsize obs}}^{\prime}~-~i_{\mbox{\scriptsize obs}}^{\prime}$
         magnitudes, and theoretical $r^{\prime}(d_{\mbox{\scriptsize c}})$ and 
         $r^{\prime}(d_{\mbox{\scriptsize c}})~-~i^{\prime}(d_{\mbox{\scriptsize c}})$ 
         main sequence magnitudes for the cluster distance $d_{\mbox{\scriptsize c}}$.}
\begin{tabular}{ccc}
\hline
Chip No. & $r_{\mbox{\scriptsize obs}}^{\prime}-r^{\prime}(d_{\mbox{\scriptsize c}})$ & 
           $(r_{\mbox{\scriptsize obs}}^{\prime}-i_{\mbox{\scriptsize obs}}^{\prime})-(r^{\prime}(d_{\mbox{\scriptsize 
           c}})-i^{\prime}(d_{\mbox{\scriptsize c}}))$ \\
\hline
1 & -0.3 & -0.95 \\
2 & -0.3 & -0.80 \\
3 & -0.1 & -0.70 \\
4 & -0.1 & -0.75 \\
\hline
Estimated & & \\
Error:    & $\pm$0.1 & $\pm$0.05 \\
\hline
\end{tabular}
\end{center}
\end{table}

\subsection{Stellar Masses, Radii And Distances}

The identification of the cluster main sequence on each CMD allows a model-dependent mass, radius and distance for each 
star to be determined using the theoretical main sequence, assuming that each star is a main sequence star.
Giant stars (MK Luminosity Class III) have absolute magnitudes in the range $1.7 \le M_{V} \sim M_{R} \le -6.5$ 
\citep{lan1992}. Assuming that the Sun lies in the Galactic plane at a distance of 8.5kpc from the 
Galactic centre (the IAU value) and assuming that the Galactic disk has a radius of 14.0kpc \citep{rob1992}, 
then the distance to the edge of the Galaxy in 
the direction of NGC~7789 may be calculated as $\sim$8.1kpc using elementary trigonometry. The magnitude of the 
dimmest 
giant at 8.1kpc assuming no extinction is $R =$ 16.2 and, assuming the law relating extinction to distance in 
Eqn.~\ref{eqn:extinction}, the dimmest giant has a magnitude of $R =$ 16.9. These faint limits are marked on the CMDs in 
Figs.~2a-d as dotted lines. From this simple argument it can be seen that only the brightest stars in our sample 
will be contaminated with giant stars.

In principle, for each star on the CMD, it is possible to choose a value for the distance parameter $d$ in Equations 9 
and 10 such that the theoretical main sequence passes through the star's position on the CMD. The solution $d = d_{*}$ 
is then the distance to the star. The star mass $M_{*}$ and radius $R_{*}$ may subsequently be determined from where 
the star lies on the theoretical main sequence at a distance of $d_{*}$. Fig.~\ref{fig:cmd}e shows the grid of star 
masses and distances 
used for Chip 4. The solid vertical lines represent lines of constant stellar mass (and radius), and are labelled at the 
top of the diagram in units of $M_{\sun}$. The ``diagonal'' dashed lines represent theoretical main sequence models 
at different distances, and the distances are labelled to the right and bottom of the diagram in units of parsecs. 
Fiducial spectral types are marked on the cluster theoretical main sequence for clarity. 

Due to the steepness of the theoretical main sequence in the CMD for star masses greater than 0.80$M_{\sun}$, 
the determined star properties become more uncertain above 0.80$M_{\sun}$. 
Also, the theoretical main sequence that we have used terminates at a mass 
of 1.40$M_{\sun}$, which leads to a small region where there are no solutions for $d_{*}$. In Fig.~\ref{fig:cmd}e, 
this region
is blueward of the thick continuous line (corresponding to a mass of 1.40$M_{\sun}$). Stars with no solution for
$d_{*}$ have masses greater than 1.40$M_{\sun}$ (and radii greater than 1.70$R_{\sun}$) and large distances. It is 
around these stars that it is hardest to detect a transiting planet and hence a lack of solution for $d_{*}$, $M_{*}$ 
and $R_{*}$ will hardly affect the completeness of our survey. Table 6 shows the star masses, radii and distances 
obtained by the above procedure for the transit candidates discussed in Section 6. The star 
$r^{\prime}-i^{\prime}$ colours have been corrected where necessary for any lightcurve variations (since 
the reference frame from which the $r^{\prime}$ magnitude was determined has a different epoch to the $i^{\prime}$ 
frame from which the $i^{\prime}$ magnitude was determined).

\begin{figure*}
\def\subfigtopskip{4pt}
\def\subfigbottomskip{8pt}
\def\subfigcapskip{4pt}
\centering
\begin{tabular}{cc}
\subfigure[]{\epsfig{file=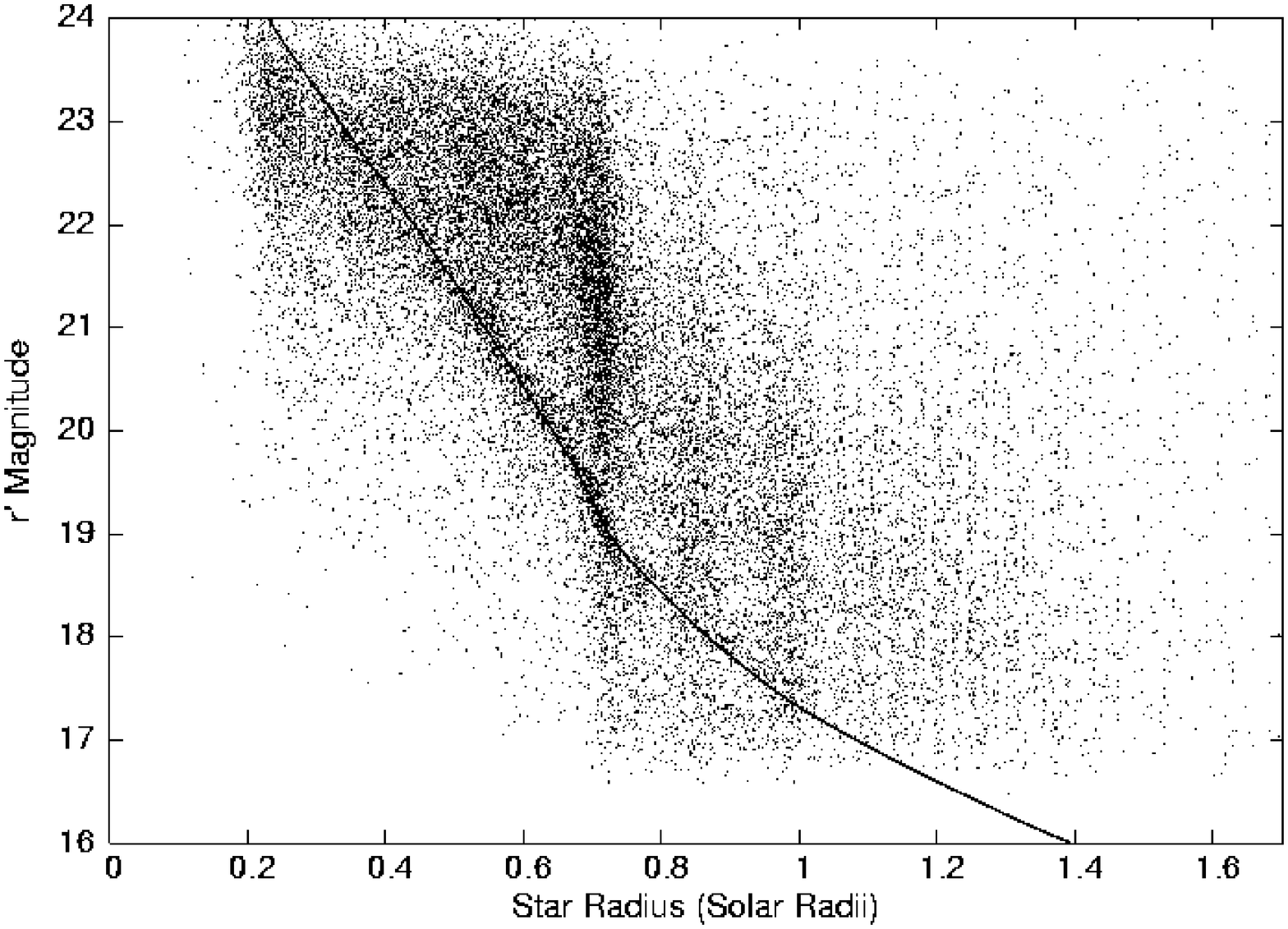,angle=270.0,width=0.5\linewidth}} &
\subfigure[]{\epsfig{file=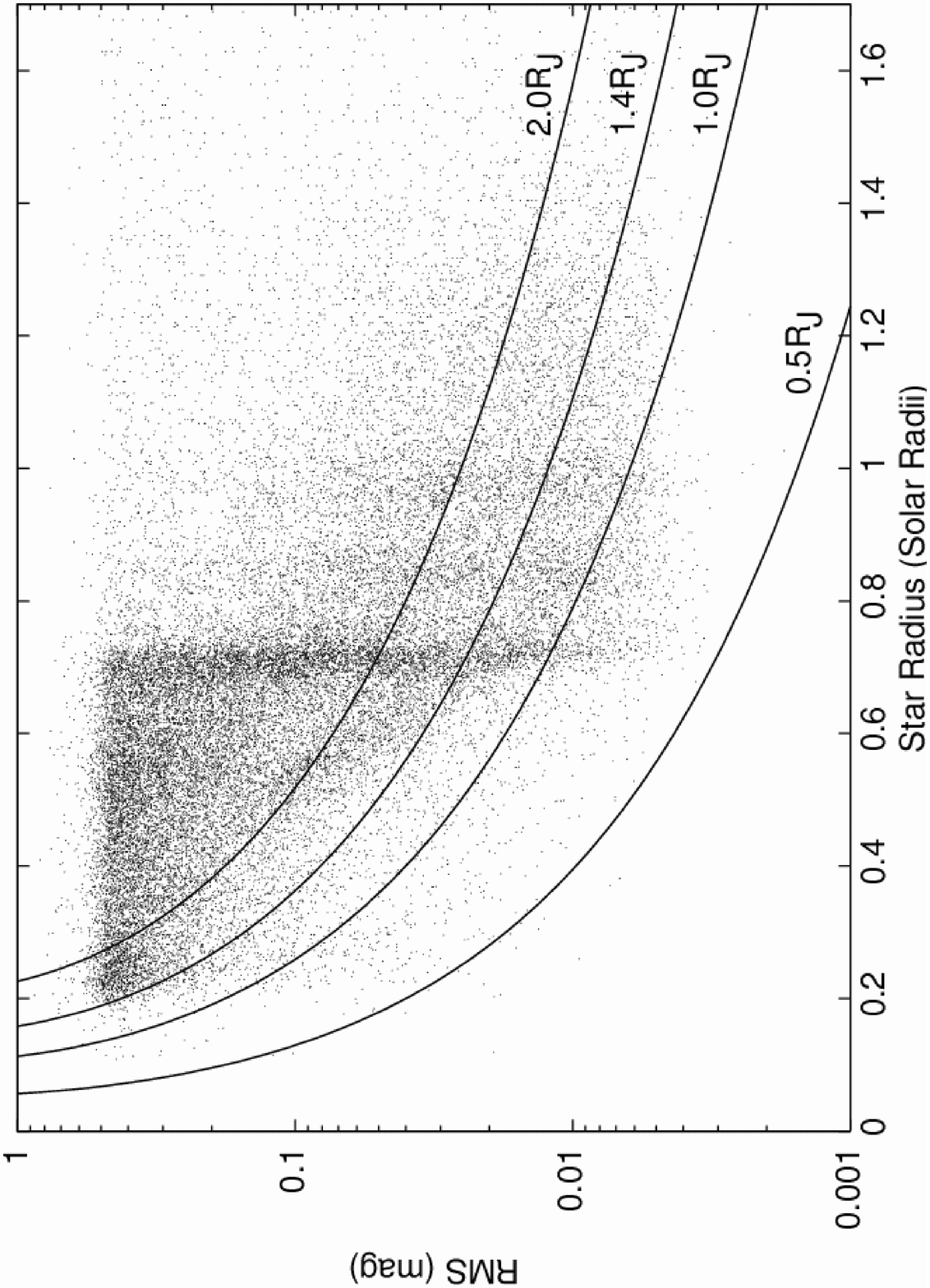,angle=270.0,width=0.5\linewidth}} \\
\subfigure[]{\epsfig{file=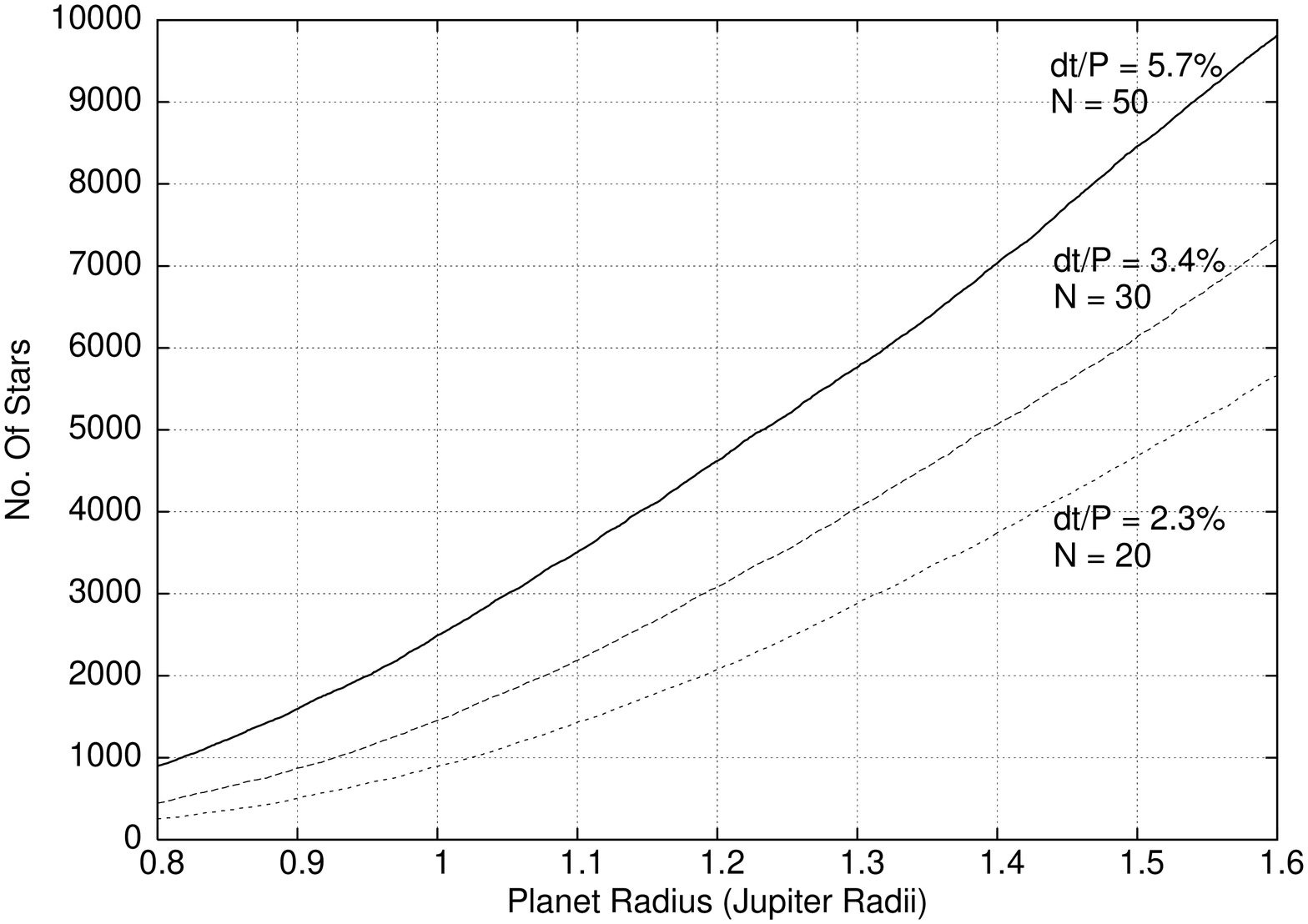,angle=270.0,width=0.5\linewidth}} &
\subfigure[]{\epsfig{file=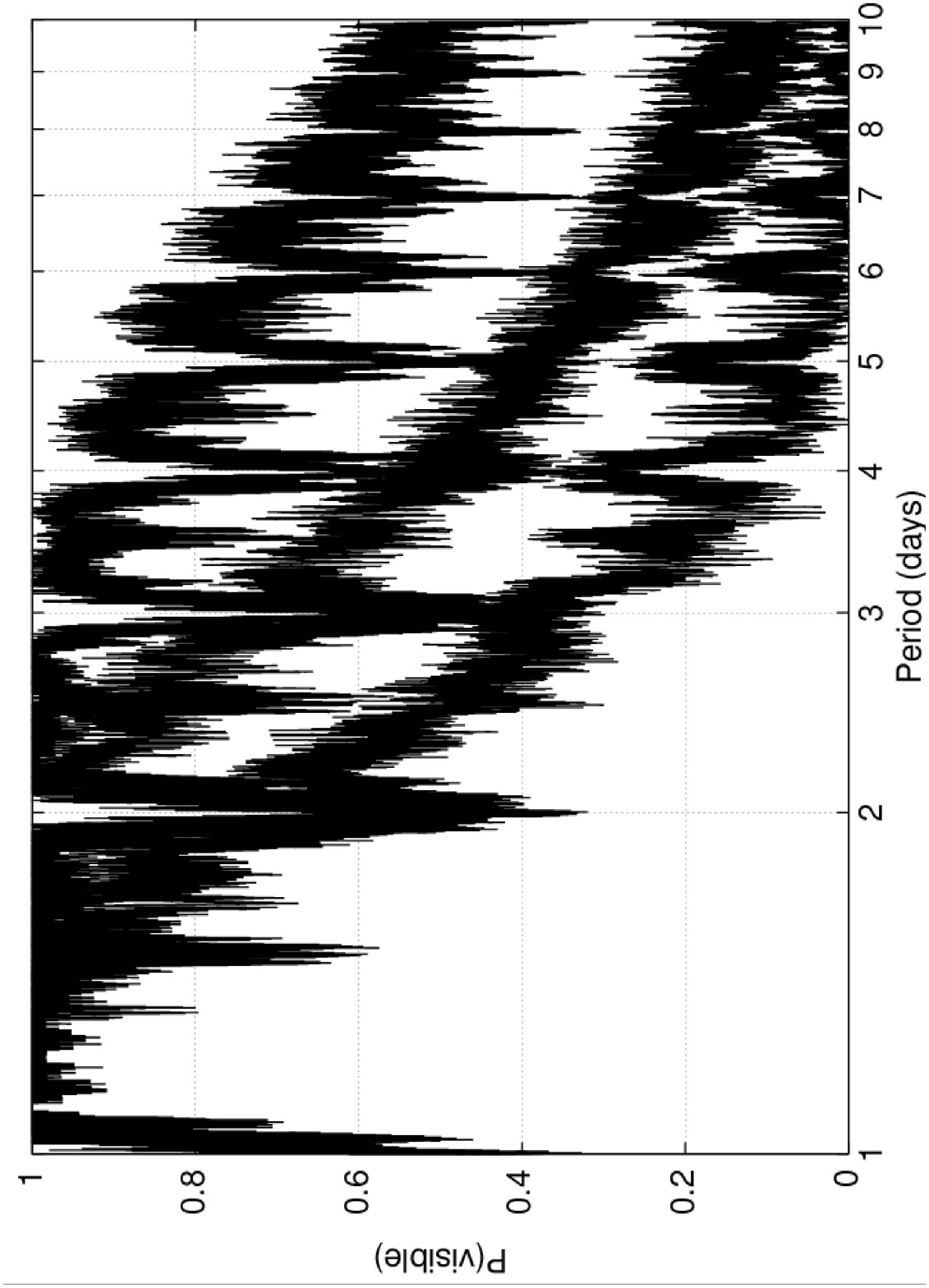,angle=270.0,width=0.5\linewidth}} \\
\end{tabular}
\caption{(a): A plot of $r^{\prime}$ magnitude against star radius for all four chips. The continuous line
         is the theoretical cluster main sequence.
         (b): Standard deviation (RMS) of the lightcurves against star radius for all four chips.
         The continuous curves are the detection limits for a HD209458-like planet of the quoted radius as 
         a function of star radius (see Section 4.5).
         (c): Number of stars in our survey that have an RMS accuracy better than that required for a
         S/N$\approx$10 detection with $N$ data points during transit as a function of planetary radius.
         (d): A plot of P(visible) against orbital period for $M = 1$ (upper curve), $M = 2$ (middle curve)
         and $M = 3$ (lower curve). P(visible) is the probability that at least M transits have a time
         of mid-transit that occurs during our observations.
\label{fig:rad}}
\end{figure*}

In Figs.~\ref{fig:rad}a-b we plot the $r^{\prime}$ magnitude and standard deviation (RMS) of the lightcurve of each star
versus the stellar radius derived from the main-sequence model and the observed $r^\prime-i^\prime$ colour index.
A vertical stripe of stars is evident with $R_{*}\sim0.75~R_{\sun}$.  
This arises because of a relatively rapid change in the colour index
with mass for the theoretical main sequence in this mass range. This
effect is also evident in Fig.~\ref{fig:cmd}e, where the vertical iso-mass lines are
more widely spaced for $0.5M_{\sun} < M_{*} < 0.8M_{\sun}$. If the mass function
and mass-radius relationship for main sequence stars are both smooth, then
this effect represents a deficiency in the $R-I$ colour index of the
stellar models.

\subsection{The Expected Number of Transiting Planets}

In Fig.~\ref{fig:rad}b each point below one of the curves represents a star
whose lightcurve has sufficient accuracy in our dataset
to reveal transits by HD 209458b-like planets of the indicated radius.
To obtain these detection threshold curves, we note that
the signal-to-noise ratio for detecting a transit is roughly:
\begin{equation} \label{eqn:snr}
	\text{S/N} \approx \frac{ \Delta m \sqrt{N}}{ \sigma }
\end{equation}
where $\Delta m$ is the transit depth,
$\sigma$ is the standard deviation of the photometric measurements,
and $N$ is the number of data points acquired during the transit.
The transit depth in magnitudes is approximately:
\begin{equation} \label{eqn:dm}
	\Delta m \approx -2.5 \log{ \left( 1 - \left( \frac{ R_{\mbox{\small p}} }{ R_{*} } \right)^{2} \right)}
\end{equation}
where $R_{\mbox{\small p}}$ is the radius of the planet and $R_{*}$ is the radius of the host star.
For random sampling of the orbital period $P$,
the probability that a given data point catches a transit is $\Delta t / P$,
where $\Delta t$ is the transit duration.
For a HD~209458b-like system ($\Delta m \approx 1.5$\%, $\Delta t \approx 3$~h, $P \approx 3.5$~d),
this fraction is $\Delta t/P=3.5$\% and thus $N\sim$30 of our 880 lightcurve data points would catch a transit.
For an OGLE-TR-56b-like system ($\Delta m \approx 1.3$\%, $\Delta t \approx 1.7$~h, $P~\approx~1.2$~d),
$\Delta~t/P~=~5.9$\% and thus $N\sim$52. In Section 5 we adopt a conservative transit detection threshold of 
S/N$\approx$10.
By rearranging Eqn.~\ref{eqn:snr} and using Eqn.~\ref{eqn:dm}, we have plotted in Fig.~\ref{fig:rad}b 
the required RMS accuracy $\sigma$ to detect
at S/N$=$10 a HD~209458b-like system ($N = 30$) as a function
of stellar radius $R_{*}$ for planetary radii of 0.5$R_{\mbox{\small J}}$, 1.0$R_{\mbox{\small J}}$,
1.4$R_{\mbox{\small J}}$ and 2.0$R_{\mbox{\small J}}$.

In fact, we may count the number of stars in our survey that have an RMS accuracy better than that required for a
S/N$\approx$10 detection with $N$ data points during transit. In Fig.~\ref{fig:rad}c we plot this number against 
$R_{\mbox{\small p}}$ for $N = 20$, $N = 30$ and $N = 50$. The curves $N = 30$ and $N = 50$ are representative
of a typical hot Jupiter (HD~209458b) and a typical very hot Jupiter (OGLE-TR-56b) respectively. 
In Fig.~\ref{fig:rad}d we plot P(visible) against orbital period where 
P(visible) is the probability that at least $M$ transits
have a time of mid-transit that occurs during our observations. The values $M = 1$, $M = 2$ and $M = 3$
are represented by the upper, middle and lower curves respectively.

We may use Figs.~\ref{fig:rad}c-d to estimate the expected number of transiting planets. 
Assuming a typical hot Jupiter radius of 1.2$R_{\mbox{\small J}}$, then we estimate that there are 3083 and 4625 stars
(Fig.~\ref{fig:rad}c) whose lightcurves have sufficient accuracy in order to detect HD~209458b-like and OGLE-TR-56b-like planets
respectively. Fig.~\ref{fig:rad}d shows that 1 to 3 day planets are almost guaranteed to transit during our observations and
that 3 to 5 day planets have $\sim$80\% probability. Assuming that $\sim$1\% of stars host a hot Jupiter companion, and that
$\sim$10\% of such systems exhibit transits, then we may expect $\sim$2 stars in our sample to reveal HD~209458b-like planetary transits
or $\sim$4 stars in our sample to reveal OGLE-TR-56b-like planetary transits.
In forthcoming work (Bramich \& Horne 2005) we report results
of more detailed modelling of the planet detection capabilities of our survey
based on Monte Carlo simulations that are consistent with these results.

\section{Transit Detection}

\begin{figure}
\epsfig{file=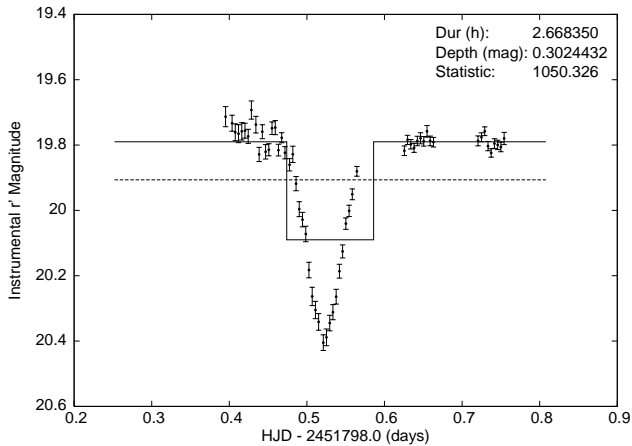,angle=270.0,width=\linewidth}
\caption{An example boxcar transit fit showing the in-transit and out-of-transit zones (continuous line) 
         and the constant fit (dashed line). The 
         horizontal axis is time (days) and the vertical axis instrumental Sloan $r^{\prime}$ magnitude.
         ``Statistic'' is the value of the transit statistic $S_{\mbox{\scriptsize tra}}^{2}$ for this fit.
\label{fig:box}}
\end{figure}

We used a matched filter algorithm to search for transits in the lightcurves.
Adopting a square ``boxcar'' shape for the transit lightcurve,
the transit model has 4 parameters:
the out-of-eclipse magnitude $m_0$, the time of mid-transit $t_0$, duration $\Delta t$ and depth $\Delta m$.
We search for transits with durations ranging from 0.5~h to 5~h,
spanning this range with 12 values of $\Delta t$ spaced by factors of 1.23.
We move the transit centroid $t_0$ through the data in steps of $\Delta t/4$.
As illustrated in Fig.~\ref{fig:box},
we fit both a constant and a boxcar transit lightcurve
to the data points in a window of width $5\Delta t$ centred on each value of $t_0$.
Our transit detection statistic is:
\begin{equation}
S_{\mbox{\small tra}}^{2} \equiv \frac{ \chi^{2}_{\mbox{\small const}} - \chi^{2}_{\mbox{\small tra}} }
            { \left(\frac{ \displaystyle \chi^{2}_{\mbox{\small out}} }
		{ \displaystyle N_{\mbox{\small out}} - 1 }\right) }
\end{equation}
where $\chi^{2}_{\mbox{\small tra}}$ is the chi squared of the boxcar transit fit, $\chi^{2}_{\mbox{\small const}}$ is 
the chi squared of the constant fit, $\chi^{2}_{\mbox{\small out}}$ is the chi squared of
the boxcar transit fit for the $N_{\mbox{\small out}}$ out-of-transit data points.
The statistic $S_{\mbox{\small tra}}^{2}$ is effectively the squared signal-to-noise ratio 
of the fitted transit signal renormalised to the reduced chi squared of the out-of-transit data points.
This modified matched filter algorithm was designed to help downweight systematic errors with
$\chi^{2}_{\mbox{\small out}}/\left( N_{\mbox{\small out}} - 1 \right)~>~1$ (and serendipitously, variables), since 
transit signals should have $\chi^{2}_{\mbox{\small out}}/\left( N_{\mbox{\small out}} - 1 \right)~\sim~1$.

The transit detection algorithm outlined above was applied to the 1999-07 and 2000-09 runs.
Initial tests with $S_{\mbox{\small tra}}^{2}$ generated many spurious transit candidates in which the transit fit
matched low data points at the beginning or end of a night. To suppress these we
introduced additional requirements on the number of in-transit and out-of-transit lightcurve data points.
For the densely sampled 2000-09 run we required at least 3 in-transit and 8 out-of-transit lightcurve data points
for a transit detection. For the more sparsely sampled 1999-07 run we required at least 2 in-transit and 6 
out-of-transit lightcurve data points for a transit detection. The time sampling in the 1999-06 run was too sparse to 
support transit hunting via the above technique.

In each lightcurve the highest value of $S_{\mbox{\small tra}}^{2}$ on each night
was identified, and those with $S_{\mbox{\small tra}}^{2} \geq 100$
(equivalent to S/N $\geq$ 10) were retained for closer examination.
Table~\ref{tab:raw} lists for each chip the number of raw candidate transits thereby selected
over the two runs.
Despite the high signal-to-noise threshold for detection, 2182 raw transit candidates were found.
A careful visual inspection of the corresponding lightcurves lead us to reject the majority of these
based on a number of criteria.
The majority of the raw transit candidates (61.8\%) were rejected because
they appeared to represent a single much fainter data point resulting from a ``bad'' section in 
one of the difference images.  Such cases were readily identifiable because the 
lightcurves of many stars triggered a transit detection at the same epoch.
A large number of variable stars were picked up 
($\sim 100$ lightcurves $\equiv$ 19.5\% of the raw transit candidates), which we plan to 
present in a forthcoming paper.
Lightcurves showing eclipses with clearly different depths 
were also assigned as variable stars since a stellar binary 
is indicated in this case. 

For the remaining raw transit candidates we examined the star on the reference frame.
This revealed that many of the remaining transit signatures were caused 
by the following (in order of most common occurence):
\begin{enumerate}
  \item Image defects detected as ``stars'' (4.5\%).
  \item Stars lying on or close to image defects, bad columns and/or saturation spikes (4.0\%).
  \item Stars close to saturation (3.7\%).
  \item Very closely blended stars (2.1\%).
  \item Stars close to the edge of the CCD (0.8\%).
\end{enumerate}
The reference image for each chip contained a large number of saturated stars along with large 
saturation spikes which unfortunately increased the incidence of such false alarms. 

For the transit candidate lightcurves that survived to this point, we checked the difference images for the night(s) 
of the suspected transit(s) by constructing a difference image movie. This revealed that a handful of the candidates 
(0.6\%) were the result of a consecutive set of poor subtractions at the star position. The other candidates clearly 
showed a flat difference image followed by a growing and then diminishing ``dimple'', indicating a drop and then 
recovery in the brightness of the star. 

\begin{figure}
\epsfig{file=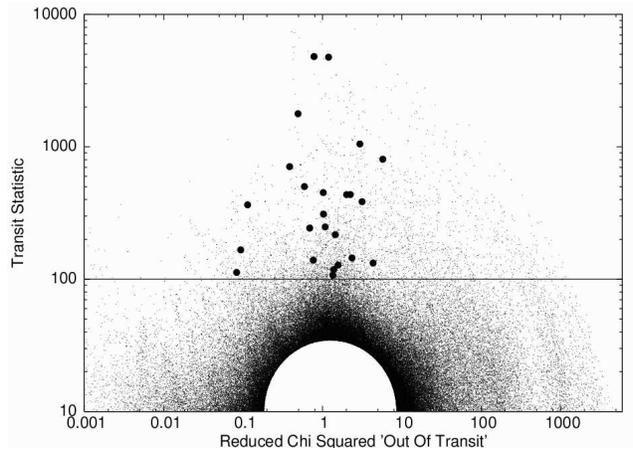,angle=270.0,width=\linewidth}
\caption{The transit detection statistic $S_{\mbox{\scriptsize tra}}^{2}$ against the out-of-transit reduced chi squared 
         $\chi^{2}_{\mbox{\scriptsize out}} / (N_{\mbox{\scriptsize out}} - 1)$.
	 The initial transit detection threshold shown by the horizontal line
         is set at $S_{\mbox{\scriptsize tra}}^{2} = 100$.
         The strongest transits detected in the lightcurves
	 of the 24 stars that survived subsequent data quality tests
	 are plotted as solid circles.
         The blank semicircular region is saturated with test points.
\label{fig:det}}
\end{figure}

We discuss in Section~6 below the 24 transit candidate lightcurves that passed all of the 
data quality tests outlined above.
Reference to a transit candidate from now on refers only to one of these transit candidate
lightcurves. 
Fig.~\ref{fig:det} shows all tests for which $S_{\mbox{\small tra}}^{2} \geq$ 10 and highlights the 
eclipse with the greatest value of $S_{\mbox{\small tra}}^{2}$ for each transit candidate.
Table 6 details the number of fully and partially observed eclipses that are present for each transit candidate 
and how these eclipses are distributed between the three runs. Table 6 also lists the J2000.0 celestial 
coordinates for each transit candidate.

\begin{table}
\begin{center}
\caption{The number of raw transit candidates and remaining transit candidate lightcurves after the weeding process 
         for each chip over the two runs.
\label{tab:raw}}    
\begin{tabular}{ccc}
\hline
         & No. Raw Transit & No. Remaining Transit   \\
Chip No. & Candidates      & Candidate Lightcurves \\
\hline
1 & 617 & 9  \\
2 & 311 & 5  \\
3 & 721 & 6  \\
4 & 533 & 4  \\
\hline
Total: & 2182 & 24 \\
\hline
\end{tabular}
\end{center}
\end{table}

\section{Transit Candidates}

\subsection{Theoretical Models}

\begin{table*}
\centering
\caption{Star and lightcurve properties for the transit candidates. The number in brackets indicates the uncertainty on 
         the last decimal place. Columns 2 and 3 are calibrated $r^{\prime}$ and 
         $r^{\prime}-i^{\prime}$ using the shifts in Table 4. Columns 7-9 indicate the number of
         fully and partially observed eclipses (f,p) that are present in the lightcurve of the corresponding run.
         The eclipsing stellar binaries are classified into the following categories: E = Eclipsing binary,
         EA = Algol type eclipsing binary, RS = RS Canum Venaticorum type eclipsing binary, CV = 
         Cataclysmic variable.}
\begin{tabular}{@{}lllllrc@{}c@{}c@{}ccc}
\hline
Star & $r^{\prime}$ & $r^{\prime} - i^{\prime}$ & Mass & Radius & Distance & $\;$1999-06$\;$ & $\;$1999-07$\;$ & 
$\;$2000-09$\;$ & RA & Dec & Variable \\
 & (mag) & (mag) & ($M_{\sun}$) & ($R_{\sun}$) & (pc)     & f,p     & f,p     & f,p     & (J2000.0) & (J2000.0) & Class \\
\hline
EB-1 & 18.077 & 0.569(17) & 0.949(33) & 0.947(42) & 2908(13)  & 0,0 & 0,1 & 0,0 & 23 58 36.75 & $+$56 26 56.1 & E
\\ EB-2 & 19.140 & 0.538(9) & 1.028(22) & 1.062(35) & 5762(11)  & 0,0 & 0,1 & 0,2 & 23 58 33.88 & $+$56 37 04.9 & E     
\\ EB-3 & 19.346 & 0.606(12)$^a$ & 0.897(23) & 0.882(29) & 4346(31)  & 0,0 & 0,1 & 0,1 & 23 57 11.92 & $+$56 31 24.9 & RS
\\ EB-4 & 19.447 & 0.999(10) & 0.666(4)  & 0.635(4)  & 1799(4)   & 0,0 & 0,0 & 0,1 & 23 56 46.79 & $+$56 36 13.8 & E
\\ EB-5 & 20.088 & 0.773(18)$^a$ & 0.751(5)  & 0.716(5)  & 3564(20)  & 0,0 & 1,1 & 1,1 & 23 55 58.93 & $+$56 40 29.6 & E      
\\ EB-6 & 17.338 & 0.695(45) & 0.759(23) & 0.723(24) & 1168(2)   & 0,1 & 0,1 & 0,1 & 23 55 42.99 & $+$56 39 14.9 & E          
\\ EB-7 & 20.865 & 0.560(37) & 0.98(8) & 1.00(13)  & 11312(52) & 0,0 & 1,0 & 1,0 & 23 58 27.34 & $+$56 46 36.4 & E        
\\ EB-8 & 18.900 & 0.788(9)$^a$ & 0.740(3) & 0.707(3)  & 2037(39)  & 0,5 & 2,3 & 5,2 & 23 58 29.19 & $+$56 32 42.7 & RS    
\\ EB-9 & 19.981 & 0.925(12)$^a$ & 0.701(4) & 0.671(4) & 2647(51) & 1,2 & 3,1 & 5,2 & 23 55 18.41 & $+$56 43 14.3 & RS     
\\ EB-10 & 20.323 & 1.025(19)$^b$ & 0.662(8) & 0.630(8)  & 2545(79)  & 0,0 & 1,1 & 3,2 & 23 57 29.99 & $+$56 57 34.3 & RS   
\\ EB-11 & 20.402 & 0.616(24) & 0.89(5) & 0.87(6)   & 6784(22)  & 0,0 & 0,0 & 3,2 & 23 58 13.37 & $+$56 45 36.1 & E        
\\ EB-12 & 16.724 & 0.55(8)$^b$ & 0.94(17) & 0.93(25)  & 1589(17)  & 0,1 & 1,2 & 0,0 & 23 57 12.86 & $+$56 31 26.5 & EA       
\\ EB-13 & 19.728 & 0.562(13)$^a$ & 0.980(23) & 0.988(34) & 6619(59)  & 1,1 & 3,0 & 2,1 & 23 55 59.17 & $+$56 45 14.1 & EA    
\\ EB-14 & 17.571 & 0.427(10)$^a$ & 1.33(4) & 1.57(7) & 5599(85)  & 0,3 & 1,2 & 3,1 & 23 57 18.04 & $+$56 51 12.0 & EA        
\\ EB-15 & 20.619 & 1.133(18)$^a$ & 0.620(9) & 0.586(10) & 2378(133) & 1,1 & 2,2 & 5,1 & 23 58 39.38 & $+$56 36 44.7 & RS or CV
\\ EB-16 & 21.120 & 1.853(19) & 0.239(6) & 0.245(4) & 752(4) & 0,1 & 1,0 & 1,1 & 23 56 57.23 & $+$56 34 03.4 & E         
\\ EB-17 & 19.420 & 1.298(9)$^a$ & 0.544(5) & 0.507(4)  & 1043(14)  & 1,0 & 2,0 & 2,0 & 23 56 47.64 & $+$56 36 28.7 & RS   
\\ EB-18 & 17.650 & 0.64(9)$^a$ & 0.81(12) & 0.78(14) & 1578(10)  & 0,0 & 0,1 & 0,2 & 23 56 01.68 & $+$56 43 08.3 & RS      
\\ EB-19 & 17.368 & 0.62(7) & 0.82(10)  & 0.79(12)  & 1477(2)   & 0,1 & 0,1 & 0,2 & 23 56 36.67 & $+$56 52 43.4 & RS    
\\ EB-20 & 20.080 & 0.797(18)$^a$ & 0.743(6)  & 0.710(5)  & 3423(84)  & 0,0 & 1,0 & 2,1 & 23 56 11.77 & $+$56 45 55.6 & RS    
\\ EB-21 & 18.984 & 1.516(9) & 0.399(6)  & 0.379(5)  & 526(2)    & 0,1 & 0,0 & 1,0 & 23 57 51.16 & $+$56 42 03.2 & E
\\ \hline
   TR-1 & 20.703 & 0.632(32) & 0.87(5) & 0.84(6) & 7291(30) & 0,0 & 0,1 & 1,0 & 23 57 45.06 & $+$56 55 36.6 & 
\\ TR-2 & 18.024 & 0.467(8) & 1.195(22) & 1.348(37) & 5348(7) & 0,0 & 0,0 & 1,0 & 23 57 24.52 & $+$56 55 17.6 & 
\\ TR-3 & 19.553 & 0.971(9) & 0.679(4) & 0.649(4) & 1995(5) & 1,0 & 0,0 & 2,1 & 23 56 47.11 & $+$56 51 10.2 &
\\ \hline
\end{tabular}
\raggedright
$^a$Colour corrected using the lightcurve model. $^b$Colour corrected using interpolation of the lightcurve.
\end{table*}

The lightcurves of the 24 transit candidates selected in Section 5 were modelled as a star and planet system
in the following way. We assume spherical stars, a luminous primary of radius $R_{*}$ and a dark massless companion
of radius $R_{\mbox{\small c}}$ in a circular orbit with radius $a$ and period $P$ inclined by the
inclination $i$ relative to our line of sight. The time $t_0$ is the time of mid-eclipse of the primary by the
companion. Since we already know $R_{*}$, the parameters that need to be constrained for such a system are $P$,
$t_{0}$, $i$, $R_{\mbox{\small c}}$ and a constant magnitude $m_{0}$.
Periodic variations in the apparent brightness of the
star were also accounted for in three different ways, leading to three competing planetary transit models:
\begin{gather}
f_{1}(t) = f_{0}(1 - f_{\mbox{\small c}}(t))
\tag*{Model 1} \\
f_{2}(t) = f_{1}(t) \Biggl( 1 +
           A \sin \left( \frac{2\pi(t-t_{0})}{P_{\mbox{\small var}}} + \phi \right) \Biggr)
\tag*{Model 2} \\
\begin{split}
f_{3}(t) = f_{1}(t) \Biggl( 1 &- C_{\mbox{\small e}} \cos \left(
                                   \frac{4\pi(t-t_{0})}{P} \right)  \\
                                &- C_{\mbox{\small h}} \cos \left(
                                   \frac{2\pi(t-t_{0})}{P} \right) \Biggr) \\
\end{split}
\tag*{Model 3} \\
\end{gather}
The function $f_{n}(t)$ is the predicted stellar flux at time $t$ for Model $n$, $f_{0}$ is a constant flux value and
$f_{\mbox{\small c}}(t)$ is the fraction of the total stellar flux obscured by the companion at time $t$. The function
$f_{1}(t)$ is calculated in a numerical fashion by creating a grid for the observed stellar disk and calculating
the flux from each grid element taking into account the apparent
position of the companion at time $t$ and the effect of linear limb darkening with $u = 0.5$.

Model 1 is therefore appropriate for a star with a constant brightness. Model 2 incorporates sinusoidal stellar flux 
variations of semi-amplitude $A$ and phase $\phi$ which do not necessarily have the same period as the orbital period 
of the companion. Such variations may be present for stars with a lot of star spot activity. Model 3 incorporates 
stellar flux variations due to two effects. The first effect, modelled by the $C_{\mbox{\small e}}$ cosine term, is 
due to the star being tidally distorted into an ellipsoidal shape by the companion and rotationally synchronised. The 
value of $C_{\mbox{\small e}}$ quantifies the semi-amplitude of such ellipsoidal flux variations. The second effect, 
modelled by the $C_{\mbox{\small h}}$ cosine term, is due to heating on one side of the companion 
caused by irradiation by the star. The value of $C_{\mbox{\small h}}$ quantifies the semi-amplitude of the heating 
term.

If stellar flux variations exist, and they are best modelled by ellipsoidal and/or heating terms, then this 
favours a stellar rather than a planetary companion. A planet does not have enough mass to distort the shape of the 
star, and neither does a planet 
emit enough radiation to be detectable in the lightcurve. Eclipses with different depths
also imply a stellar companion since the secondary eclipse is caused by occultation of the companion, 
indicating that the companion is emitting enough radiation to contribute to the observed brightness of the system.

\begin{figure*}
\def\subfigtopskip{4pt}
\def\subfigbottomskip{8pt}
\def\subfigcapskip{4pt}
\centering
\begin{tabular}{cc}
\subfigure[{\bf EB-1} - L \& CTFIT - 1999-07 Night 3]
{\epsfig{file=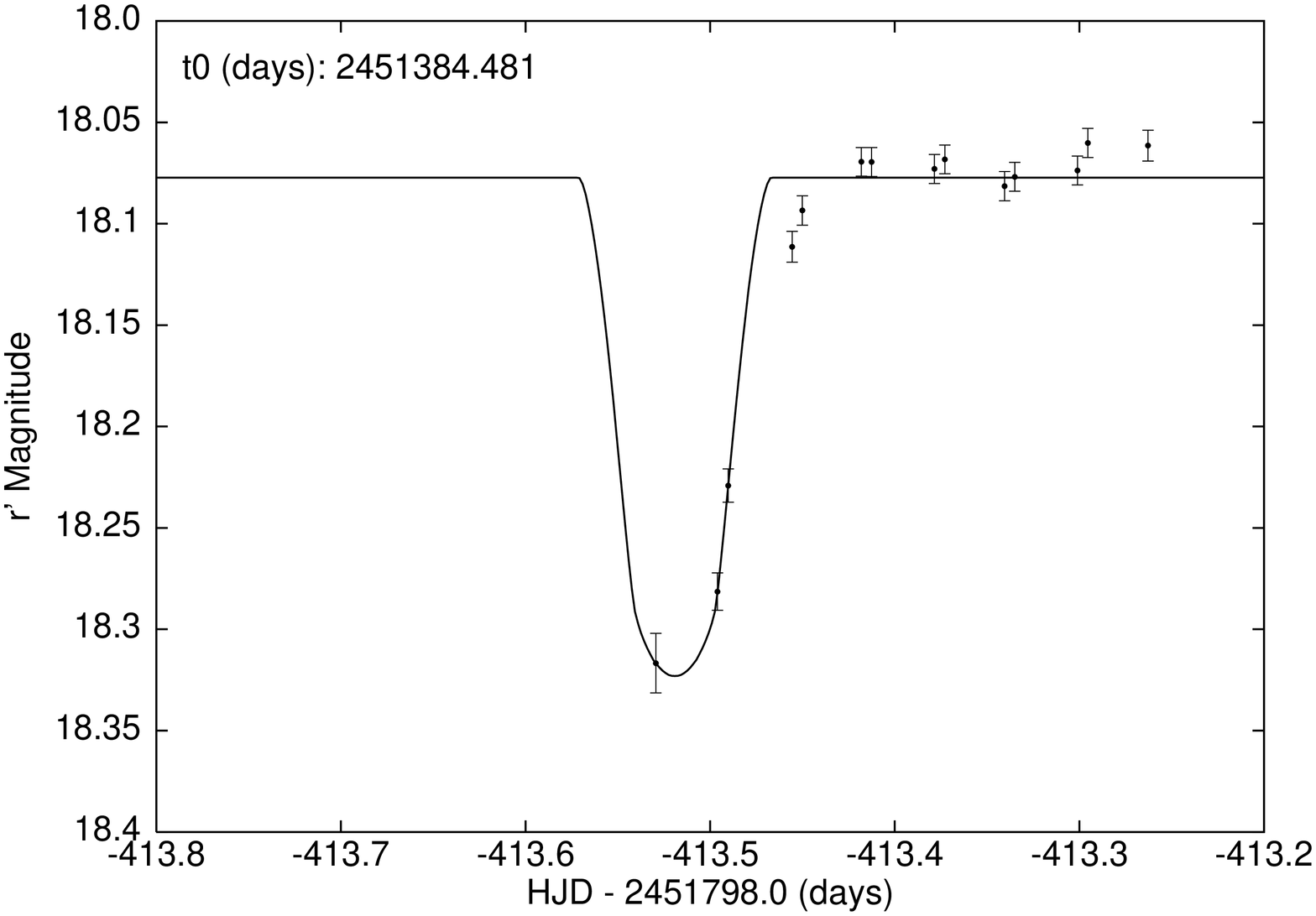,angle=270.0,width=0.4\linewidth}} &
\subfigure[{\bf EB-2} - CPL \& CTFIT - 2000-09]
{\epsfig{file=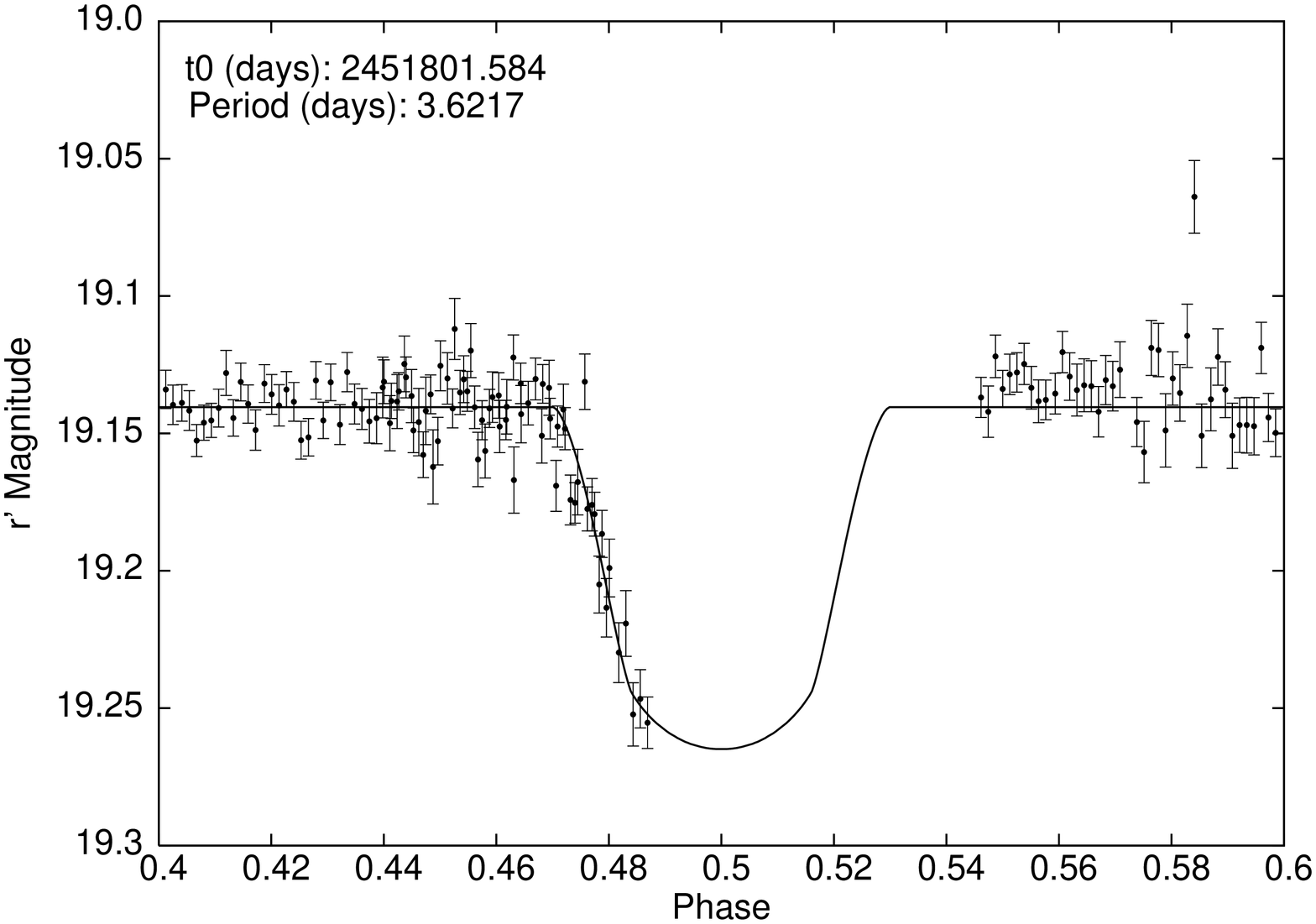,angle=270.0,width=0.4\linewidth}} \\
\subfigure[{\bf EB-3} - L \& CTFIT - 2000-09 Night 1]
{\epsfig{file=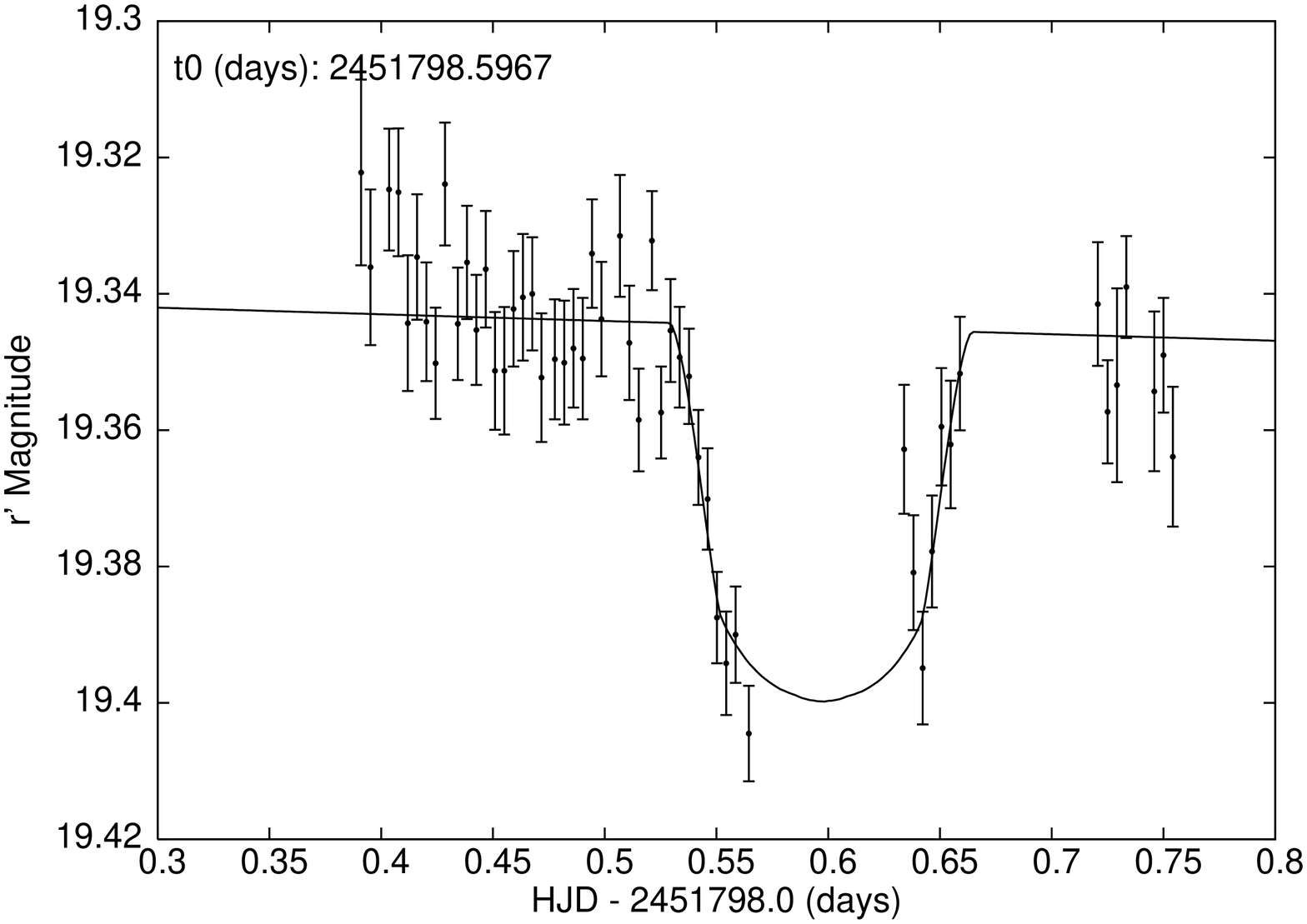,angle=270.0,width=0.4\linewidth}} &
\subfigure[{\bf EB-4} - L \& CTFIT - 2000-09 Night 8]
{\epsfig{file=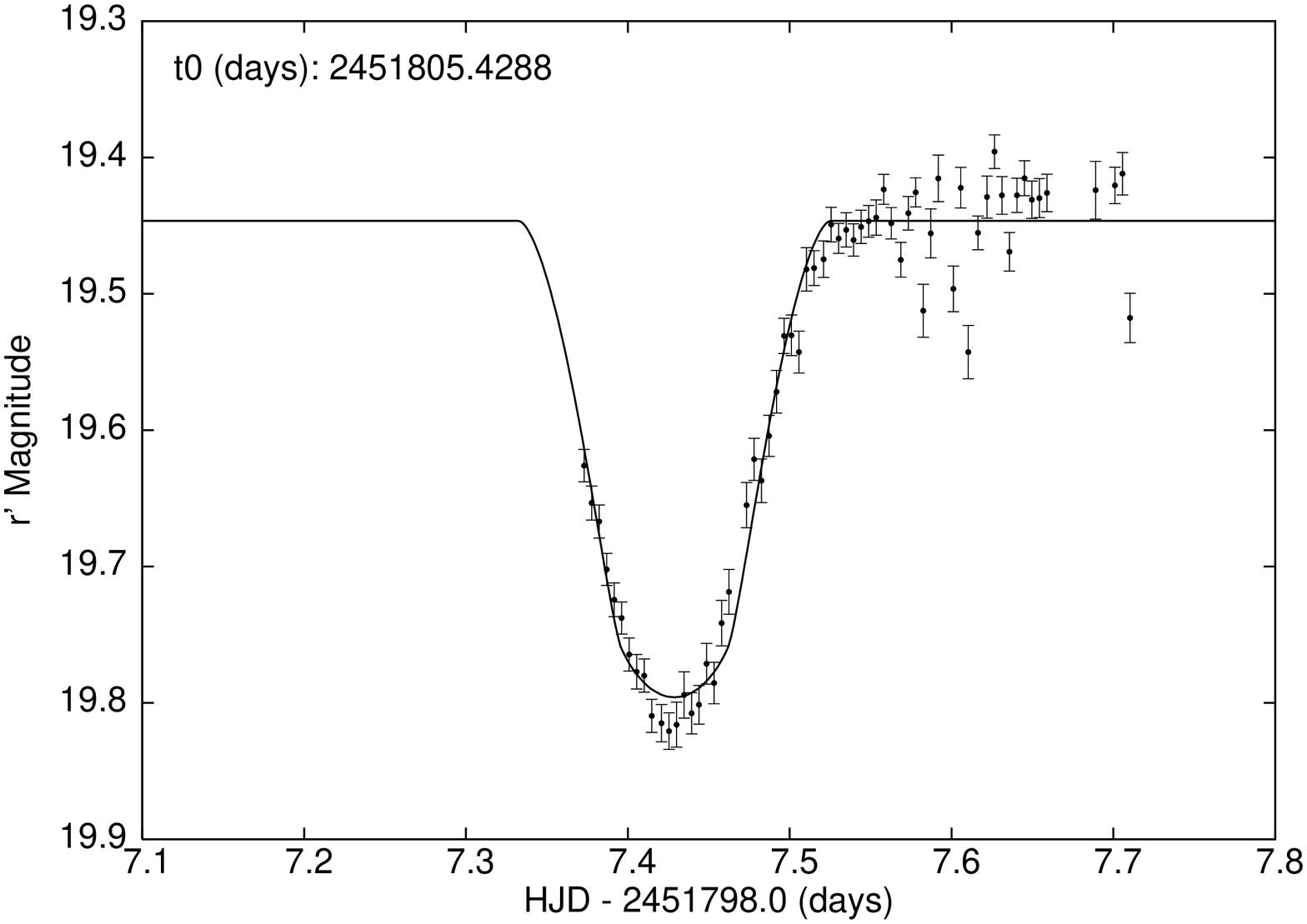,angle=270.0,width=0.4\linewidth}} \\
\subfigure[{\bf EB-5} - L \& CTFIT - 2000-09 Night 1]
{\epsfig{file=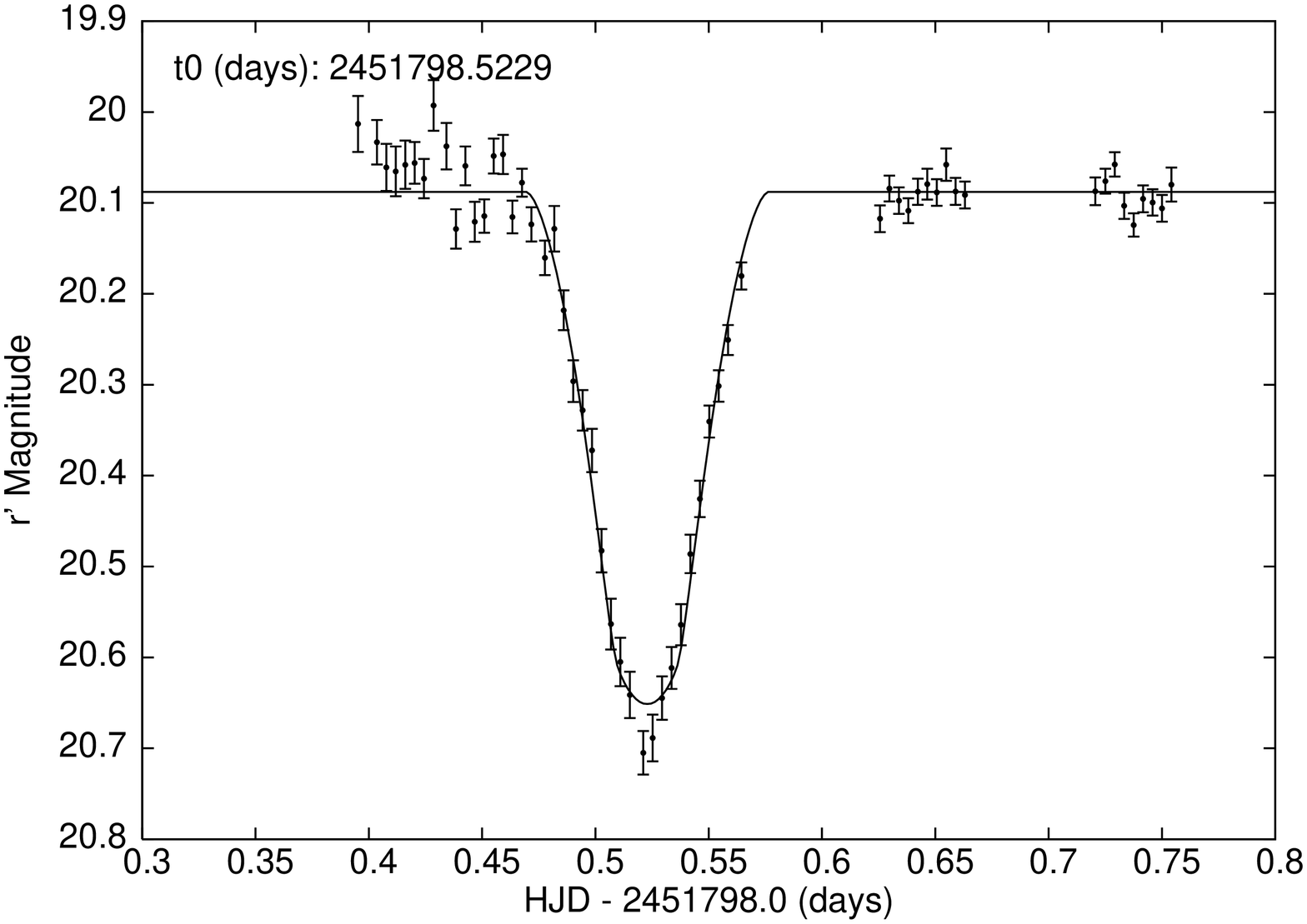,angle=270.0,width=0.4\linewidth}} &
\subfigure[{\bf EB-6} - L \& CTFIT - 1999-07 Night 6]
{\epsfig{file=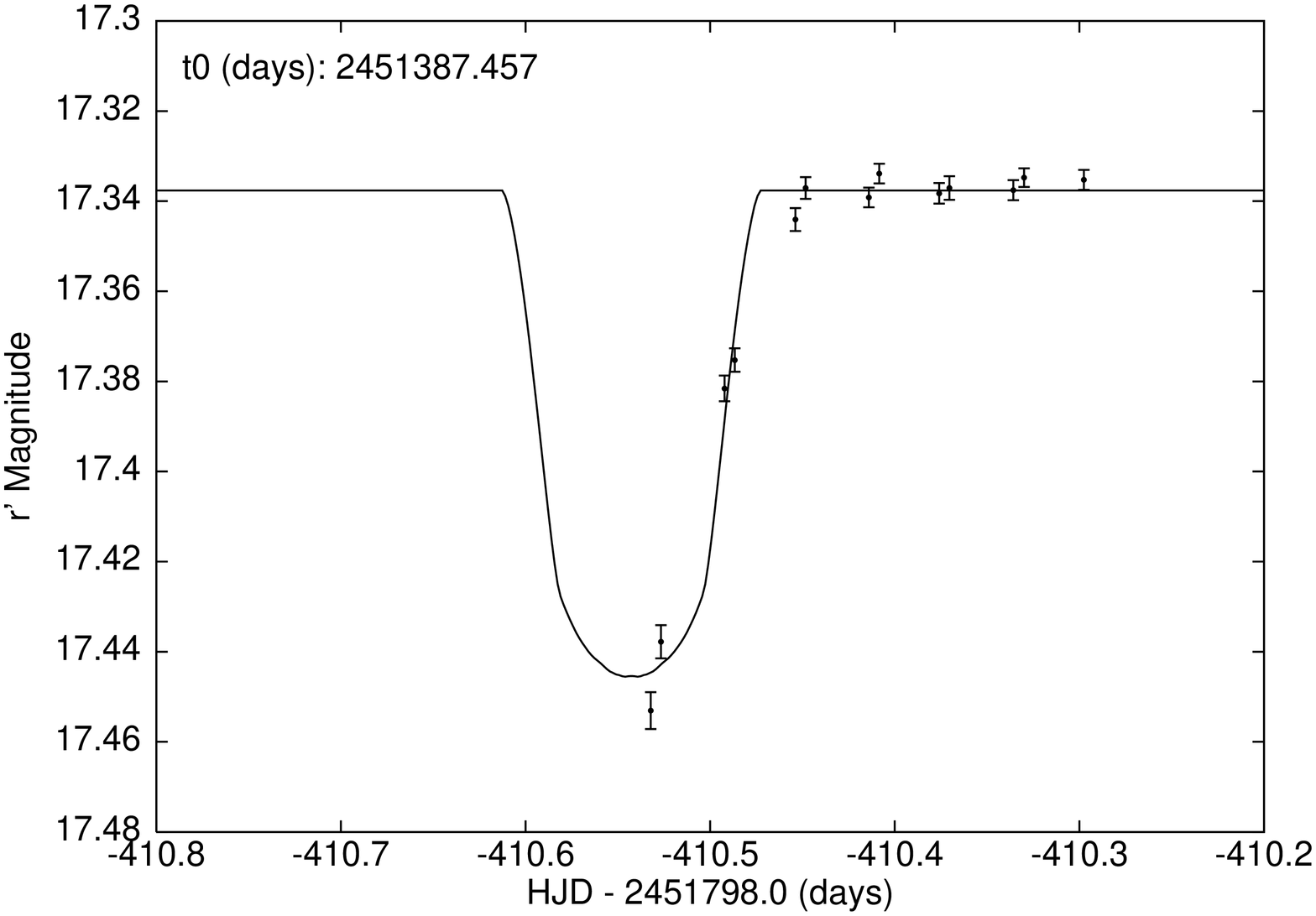,angle=270.0,width=0.4\linewidth}} \\
\end{tabular}
\begin{tabular}{cc}
\subfigure[{\bf EB-7} - L \& CTFIT - 2000-09 Night 10]
{\epsfig{file=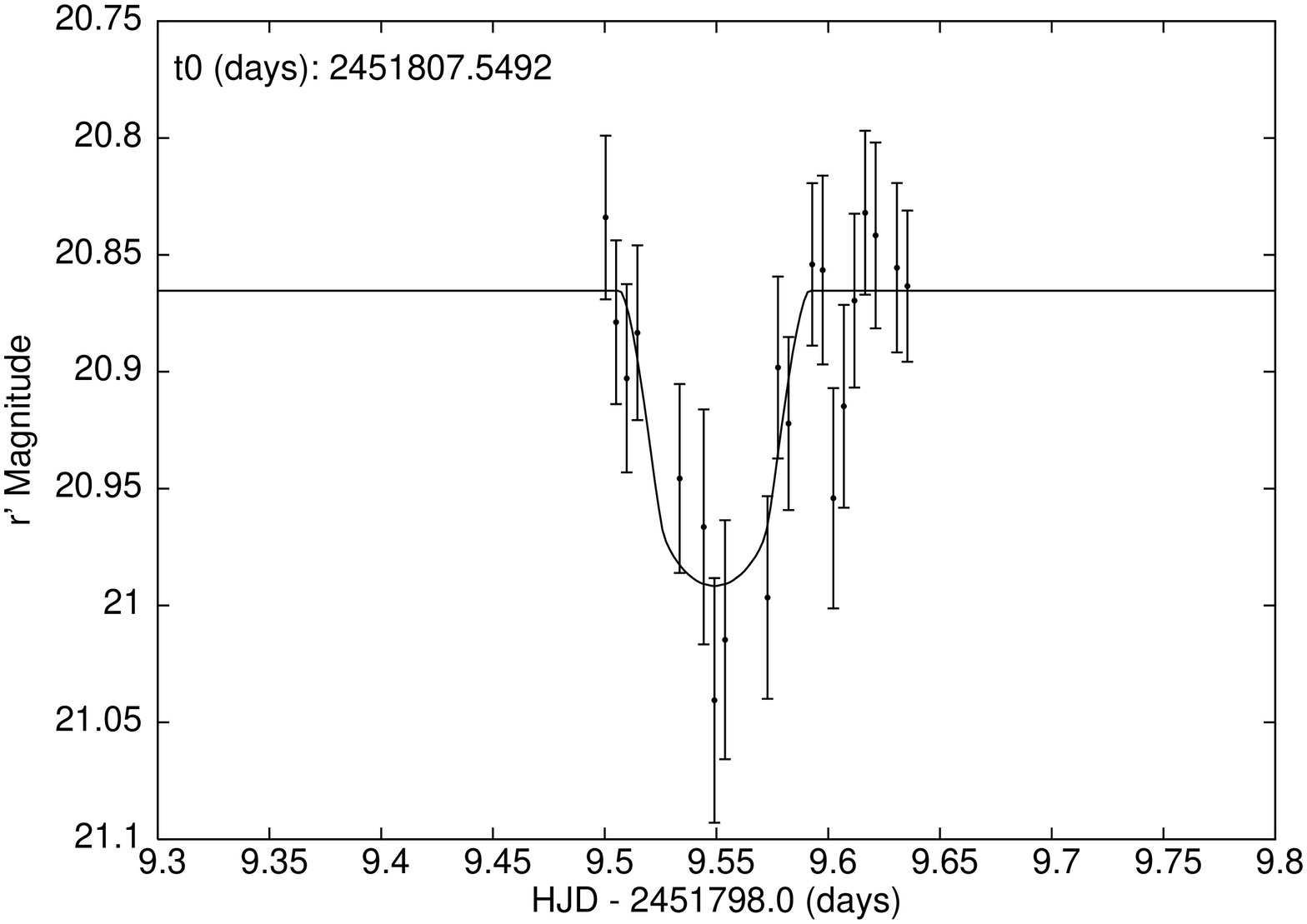,angle=270.0,width=0.4\linewidth}} \\
\end{tabular}
\caption{Eclipsing binaries with undetermined periods.}
\end{figure*}

In the following sections, the transit candidates have been organised into groups depending on the lightcurve
properties, and analysed accordingly. Relevant parameters are shown on the lightcurve plots. Periods were initially
determined using a periodogram, and then refined during the fitting of the appropriate transit model. We refer to 
any transit fit for which all 5 parameters have been optimised as a full transit fit. 
For brevity, the following labelling format has been adopted for the plots in these 
sections:

\small $<$Star No.$>$ - $<$Plot Code$>$ - $<$Run(s) To Which The Plot Applies$>$

\normalsize The plot codes are as follows:
\begin{enumerate}
  \item L - Lightcurve
  \item PL - Phased lightcurve
  \item CPL - Close up of phased lightcurve
  \item BL - Binned lightcurve
  \item CTFIT - Central transit fit
  \item FTFIT - Full transit fit
  \item CM - Chi squared contour map showing the best fit solution with a cross and the 1, 2 and 3$\sigma$
             confidence regions with solid, dashed and shorter dashed lines respectively. Annular, grazing and no
             eclipse regions are separated by thick solid lines.
\end{enumerate}

\subsection{Eclipsing Binaries With Undetermined Periods}

In Fig. 6 we present 7 transit candidates for which we were unable to determine a period, although we were able
to classify them as eclipsing binaries. The reason for not being able to determine the period was due to either the
presence of only one fully/partially observed eclipse in the lightcurve and/or cycle ambiguity between eclipses.
We fitted the best-defined eclipse for each candidate with Model 1 keeping the inclination fixed at 90.0$\degr$ (we call this
a central transit fit). This fit determines a minimum radius of the companion given the eclipse profile, since
at lower inclination values the same size companion obscures a smaller fraction of the total stellar flux
due to limb darkening effects and the possibility that the eclipse is grazing instead of annular. Hence, at lower 
inclinations a larger companion radius is required to account for the observed eclipse depth.

\begin{figure*}
\def\subfigtopskip{4pt}
\def\subfigbottomskip{8pt}
\def\subfigcapskip{4pt}
\centering
\begin{tabular}{cc}
\subfigure[{\bf EB-8} - PL, BL \& FTFIT - 2000-09 (Top), 1999-07 (Middle with +0.1~mag offset) \&
           1999-06 (Bottom with +0.2~mag offset)]
{\epsfig{file=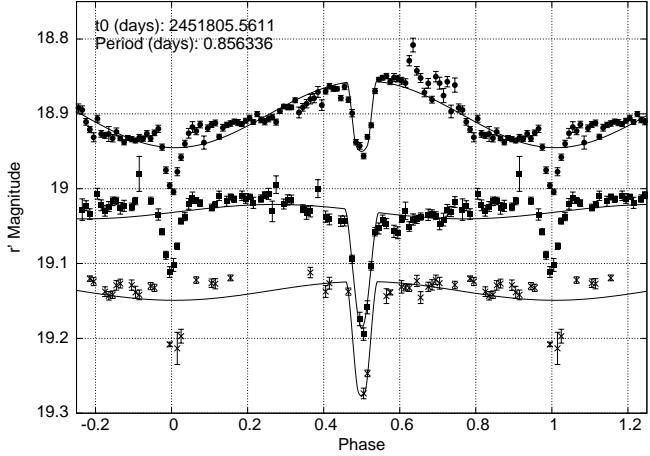,angle=270.0,width=0.5\linewidth}} &
\subfigure[{\bf EB-9} - PL, BL \& FTFIT - 2000-09 (Top), 1999-07 (Middle with +0.2~mag offset) \&
           1999-06 (Bottom with +0.4~mag offset) ]
{\epsfig{file=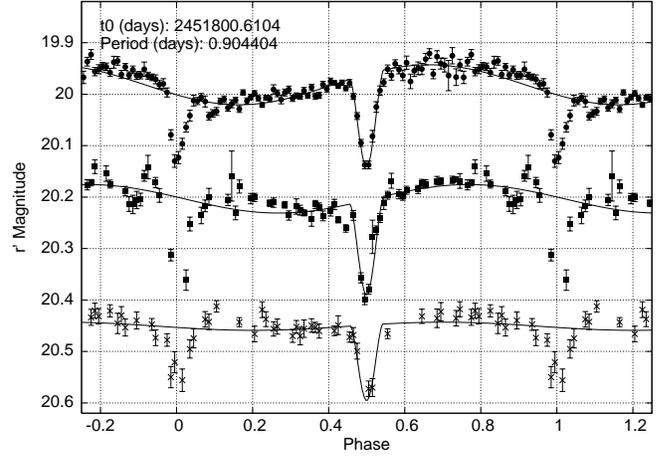,angle=270.0,width=0.5\linewidth}} \\
\subfigure[{\bf EB-10} - PL, BL \& FTFIT - 2000-09 (Top) \& 1999-07 (Bottom with +0.4~mag offset)]
{\epsfig{file=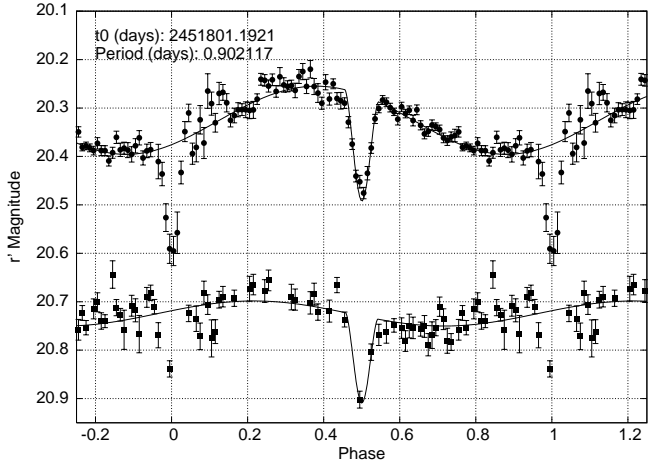,angle=270.0,width=0.5\linewidth}} &
\subfigure[{\bf EB-11} - PL, BL \& FTFIT - 2000-09]
{\epsfig{file=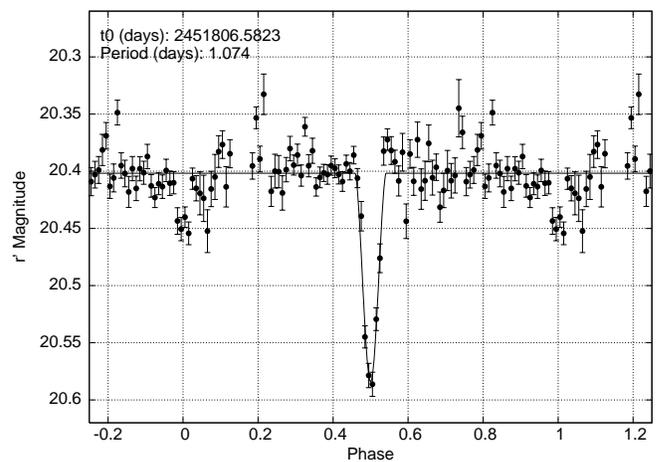,angle=270.0,width=0.5\linewidth}} \\
\end{tabular}
\caption{Eclipsing binaries exhibiting secondary eclipses.}
\end{figure*}

For each of the 7 transit candidates except EB-3, the central transit fit yields a minimum companion radius 
that is greater than 0.2$R_{\sun}$. This favours a stellar rather than a planetary companion. The lack of out-of-eclipse 
lightcurve variations leads us to conclude that these are eclipsing binaries. For {\bf EB-2}, a periodogram 
reveals that there are only two possible periods (3.6216$\pm$0.0053 d and 7.233$\pm$0.010 d). We plot the 2000-09 lightcurve folded
on the shorter period in Fig. 6b.

In the case of {\bf EB-3}, 
the derived minimum companion radius of 0.18$R_{\sun}$ is most likely an under estimate since the eclipse is possibly 
deeper than the fit shown in Fig. 6f. Also, 
the lightcurve shows sinusoidal out-of-eclipse variations. Therefore we class this system as a RS CVn type  
eclipsing binary.

\begin{figure*}
\def\subfigtopskip{4pt}
\def\subfigbottomskip{8pt}
\def\subfigcapskip{4pt}
\centering
\begin{tabular}{cc}
\subfigure[{\bf EB-12} - PL, BL \& FTFIT - 1999-07 (Top) \& 1999-06 (Bottom with +0.1~mag offset)]
{\epsfig{file=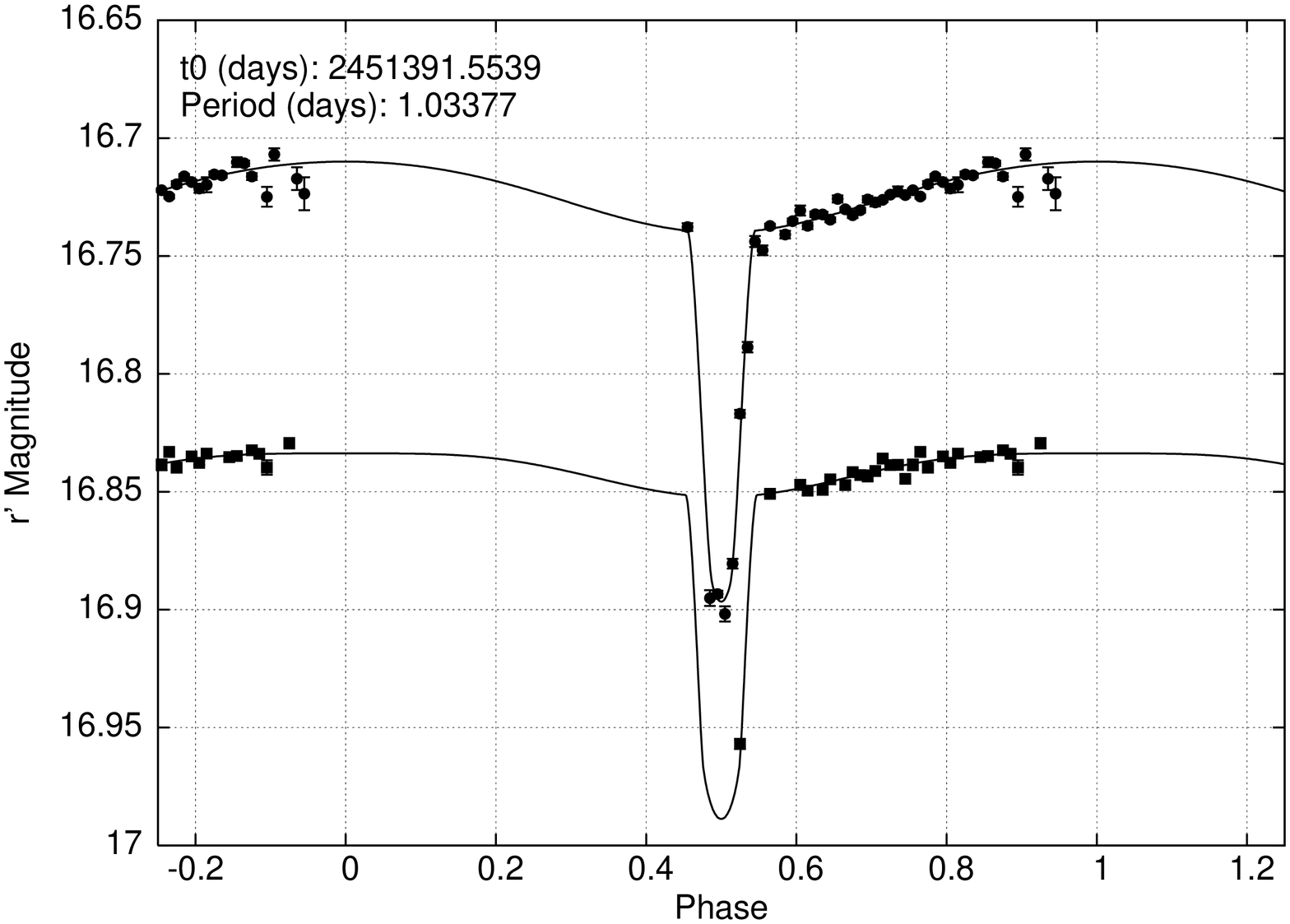,angle=270.0,width=0.5\linewidth}} &
\subfigure[{\bf EB-13} - PL, BL \& FTFIT - 1999-06, 1999-07 \& 2000-09]
{\epsfig{file=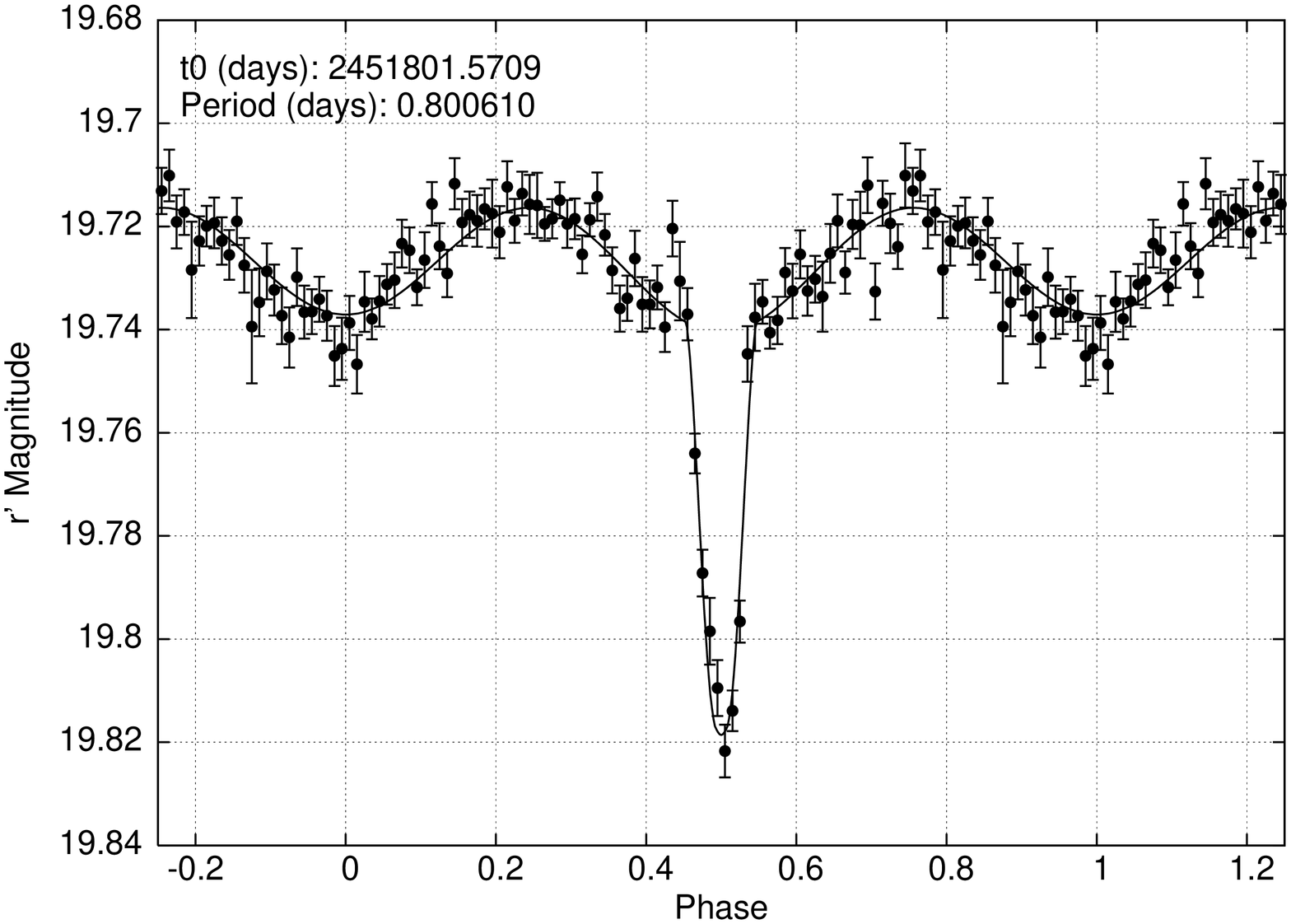,angle=270.0,width=0.5\linewidth}} \\
\end{tabular}
\begin{tabular}{c}
\subfigure[{\bf EB-14} - PL, BL \& FTFIT - 2000-09 (Top), 1999-07 (Middle with +0.1~mag offset) \&
           1999-06 (Bottom with +0.2~mag offset)]
{\epsfig{file=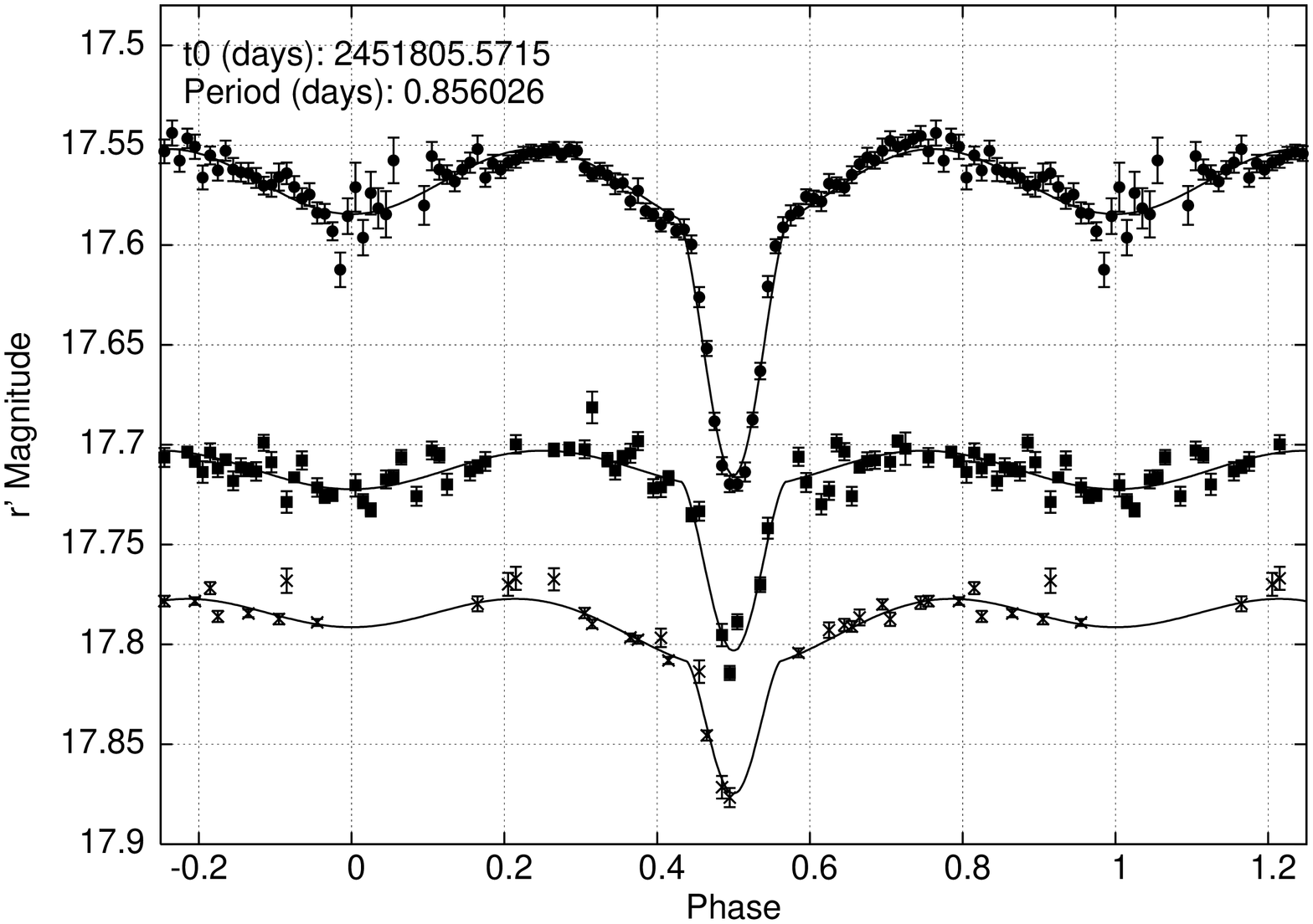,angle=270.0,width=0.5\linewidth}} \\
\end{tabular}
\caption{Eclipsing binaries exhibiting ellipsoidal variations and heating effects.}
\end{figure*}

\subsection{Eclipsing Binaries Exhibiting Secondary Eclipses}

In Fig.~7 we present 4 transit candidates that exhibit secondary eclipses in their lightcurves implying
that the companion is luminous. Fig.~7 shows the folded lightcurve for each transit candidate 
along with the best fit transiting planet model. The lightcurves from the different runs are offset vertically (in magnitude) from each 
other in order to highlight any changes in the out-of-eclipse variations. 

\begin{figure*}
\def\subfigtopskip{4pt}
\def\subfigbottomskip{8pt}
\def\subfigcapskip{4pt}
\centering
\begin{tabular}{cc}
\subfigure[{\bf EB-15} - PL, BL \& FTFIT - 2000-09 (Top), 1999-07 (Middle with +0.3~mag offset) \& 1999-06
           (Bottom with +0.6~mag offset)]
{\epsfig{file=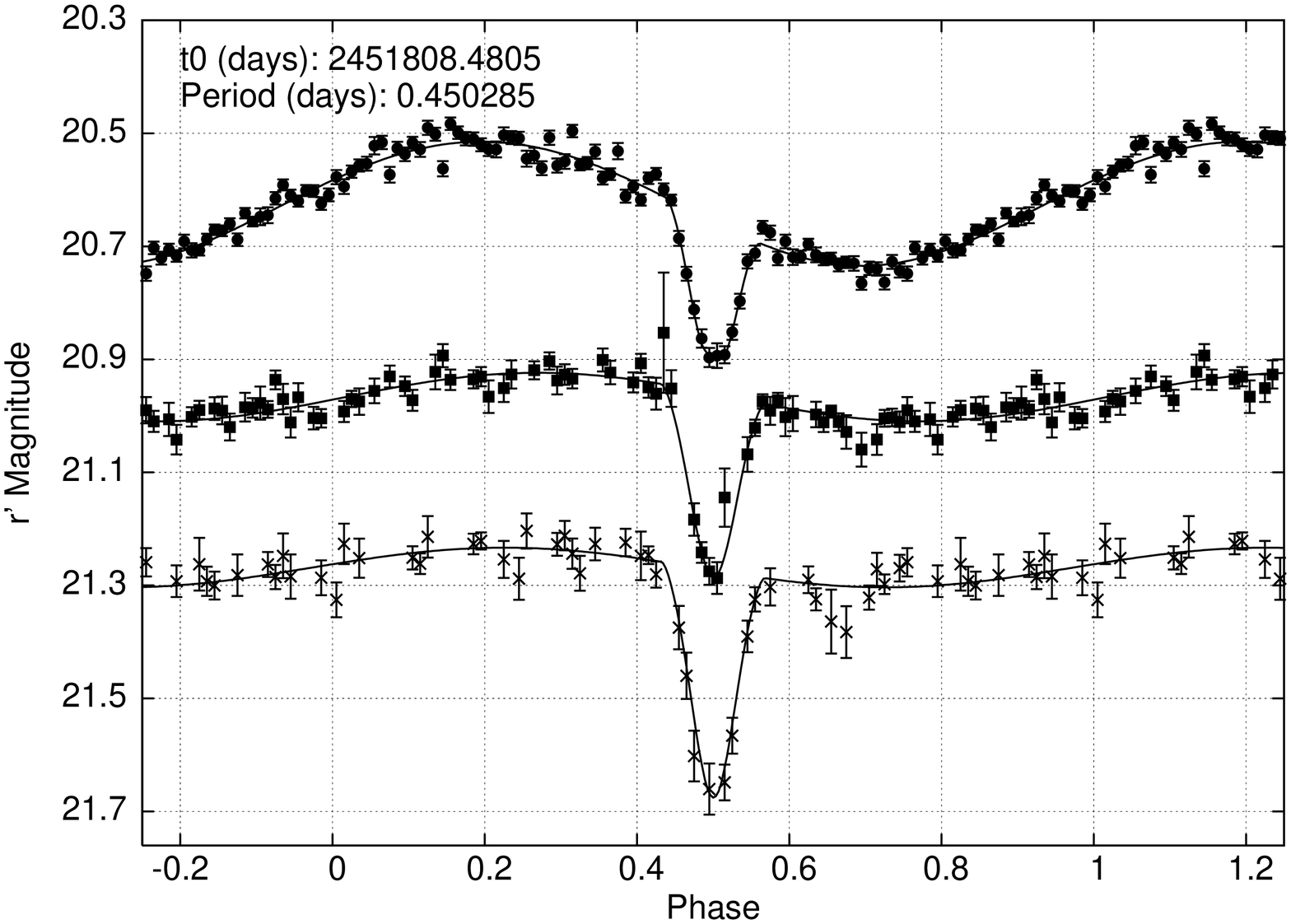,angle=270.0,width=0.5\linewidth}} &
\subfigure[{\bf EB-15} - CPL \& FTFIT - 2000-09 (Top), 1999-07 (Middle with +0.3~mag offset) \& 1999-06
           (Bottom with +0.6~mag offset)]
{\epsfig{file=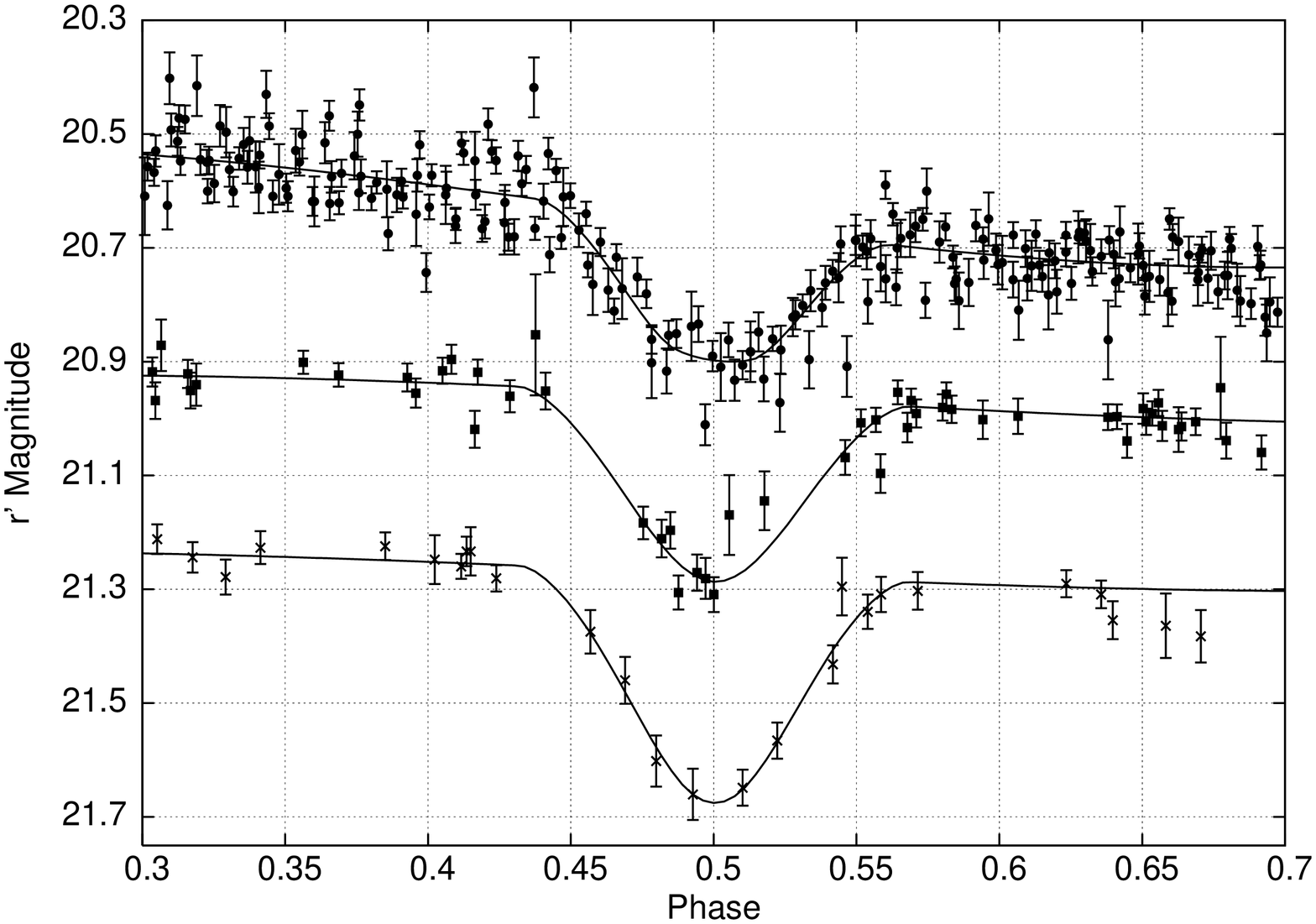,angle=270.0,width=0.5\linewidth}} \\
\end{tabular}
\caption{Possible long period cataclysmic variable.}
\end{figure*}

\subsection{Eclipsing Binaries Exhibiting Ellipsoidal Variations And Heating Effects}

In Fig. 8 we present 3 transit candidates that exhibit ellipsoidal variations and heating effects in their 
lightcurves which immediately implies that the companion is stellar. Fig.~8 shows the folded lightcurve for each 
transit candidate along with the best fit transiting planet model using Model 3. The lightcurves 
from different runs are offset vertically (in magnitude) from each other in order to highlight any changes in the 
amplitude of the out-of-eclipse variations. 

\begin{figure*}
\def\subfigtopskip{4pt}
\def\subfigbottomskip{8pt}
\def\subfigcapskip{4pt}
\centering
\begin{tabular}{cc}
\subfigure[{\bf EB-16} - PL, BL \& FTFIT - 1999-06, 1999-07 \& 2000-09]
{\epsfig{file=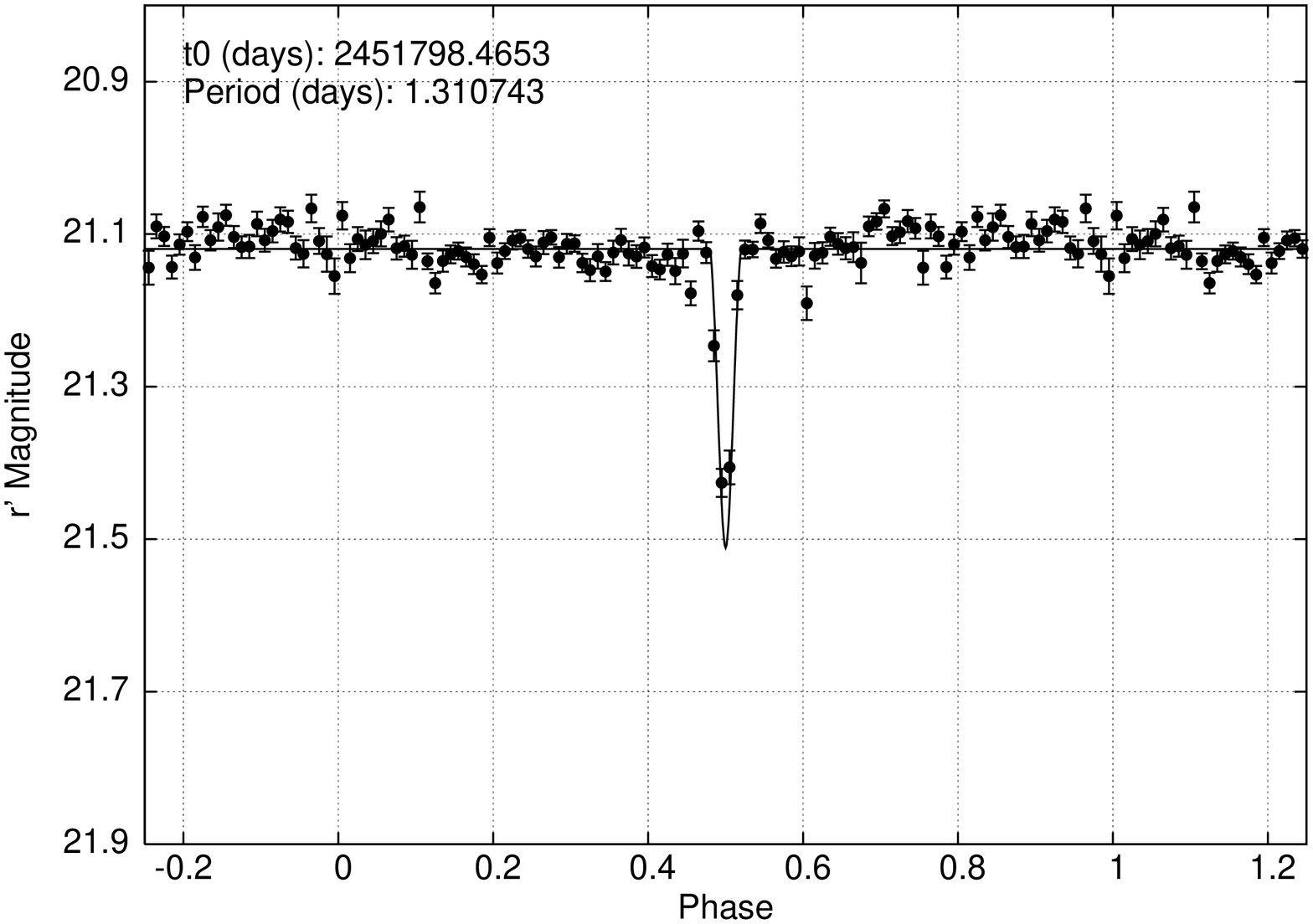,angle=270.0,width=0.5\linewidth}} &
\subfigure[{\bf EB-16} - CPL \& FTFIT - 1999-06, 1999-07 \& 2000-09]
{\epsfig{file=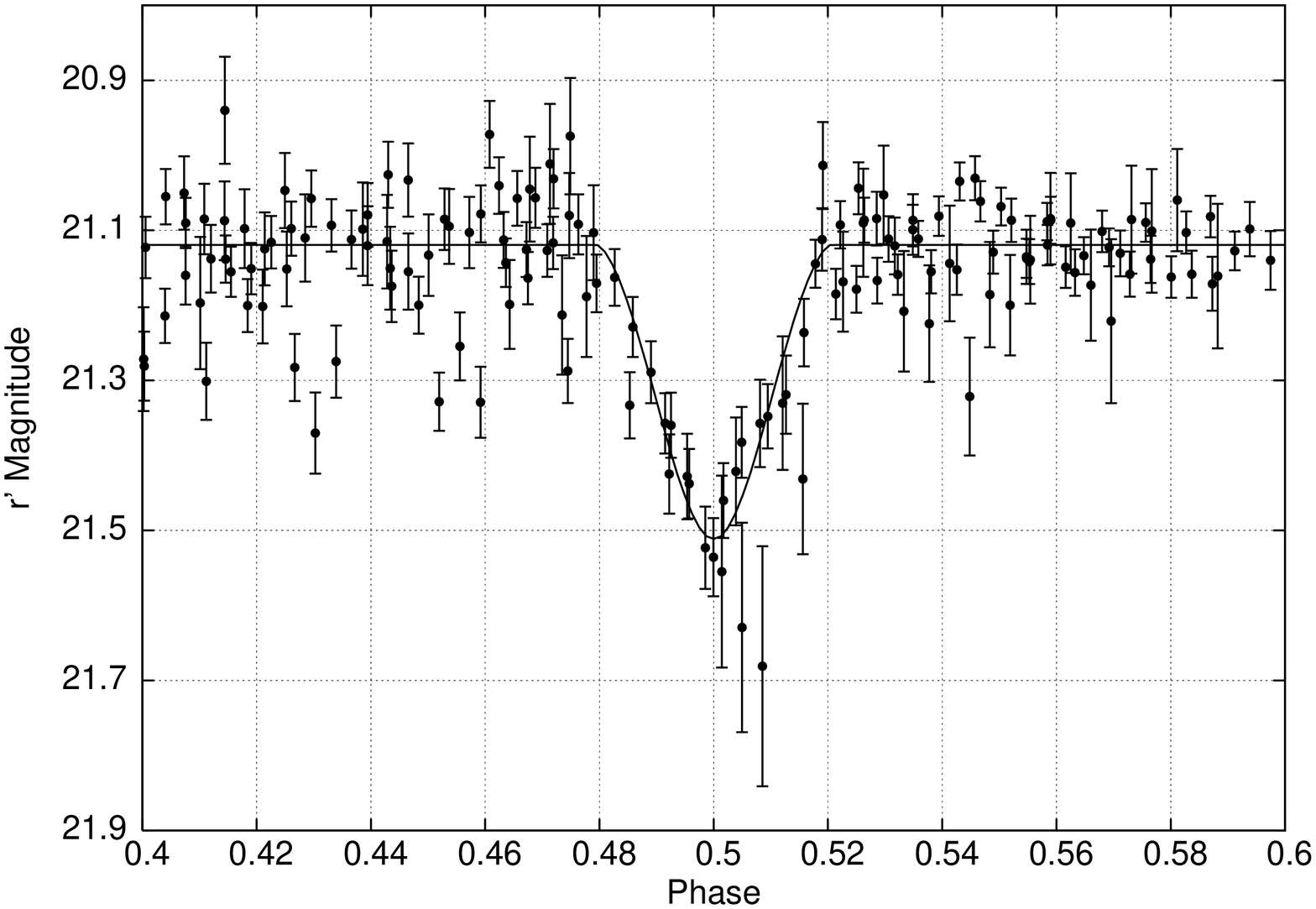,angle=270.0,width=0.5\linewidth}} \\
\subfigure[{\bf EB-17} - PL, BL \& FTFIT - 2000-09 (Top), 1999-07 (Middle with +0.2~mag offset) \& 
           1999-06 (Bottom with +0.4~mag offset)]
{\epsfig{file=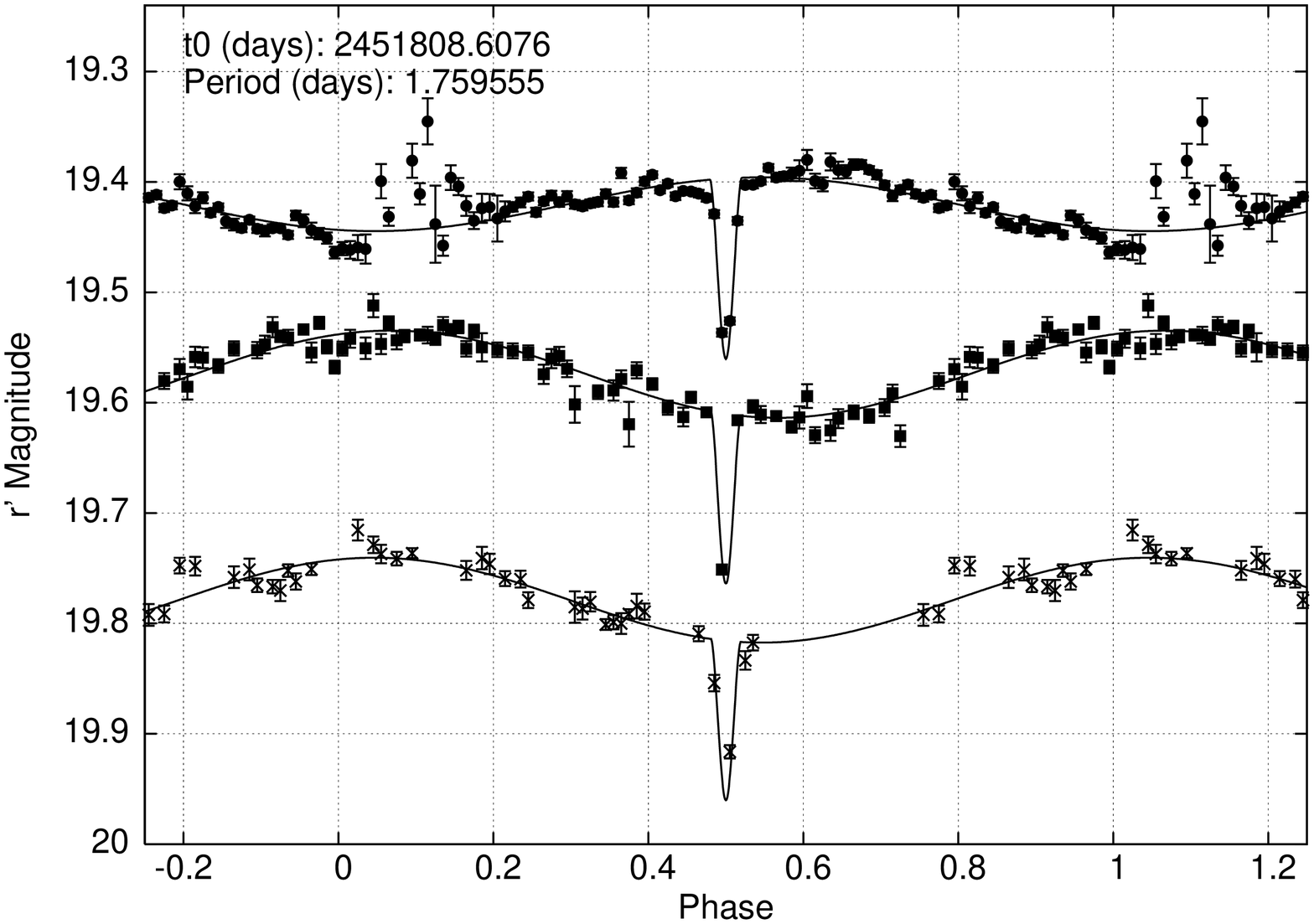,angle=270.0,width=0.5\linewidth}} &
\subfigure[{\bf EB-17} - CPL \& FTFIT - 2000-09 (Top), 1999-07 (Middle with +0.2~mag offset) \&  
           1999-06 (Bottom with +0.4~mag offset)]
{\epsfig{file=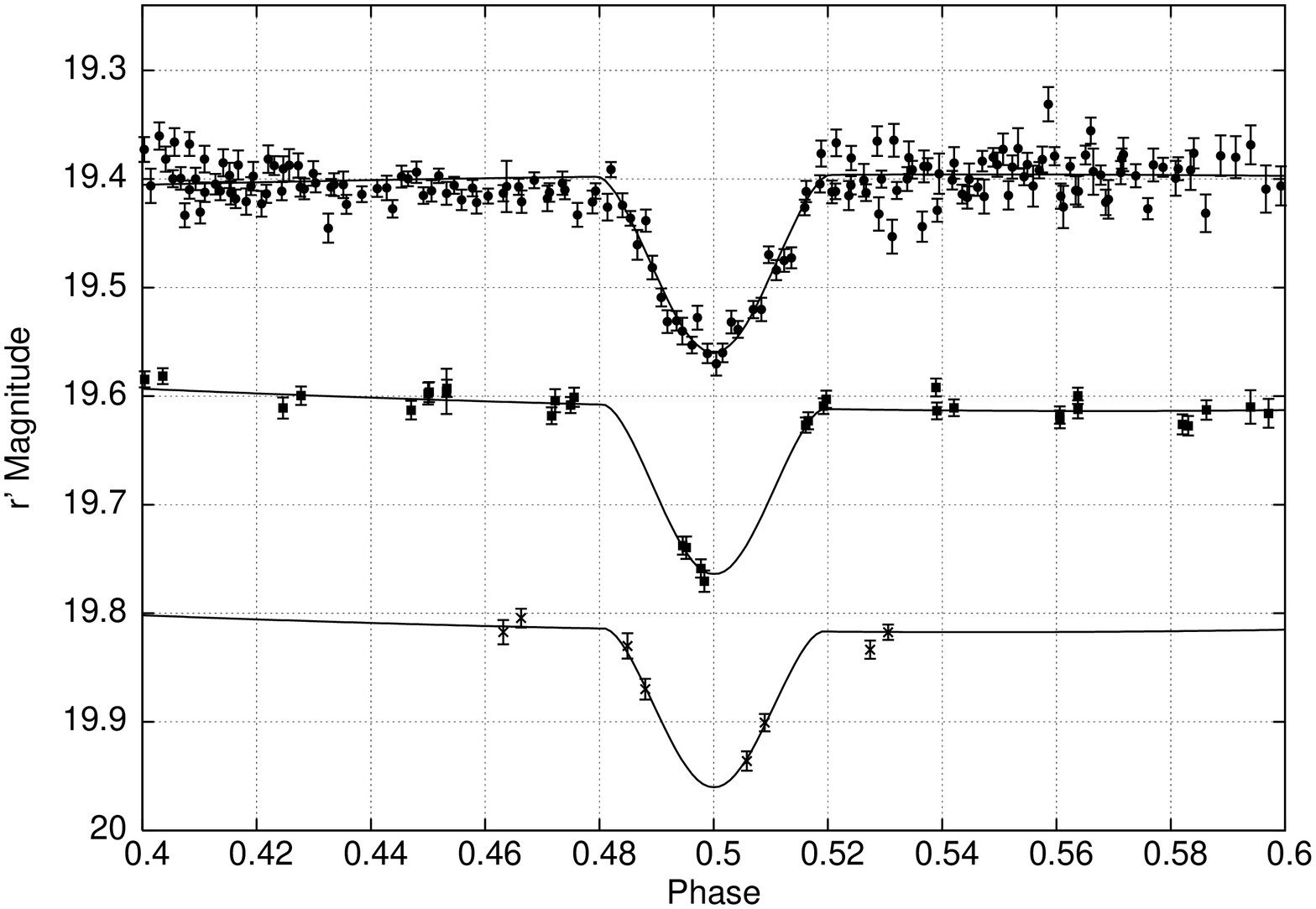,angle=270.0,width=0.5\linewidth}} \\
\subfigure[{\bf EB-18} - PL, BL \& FTFIT - 2000-09]
{\epsfig{file=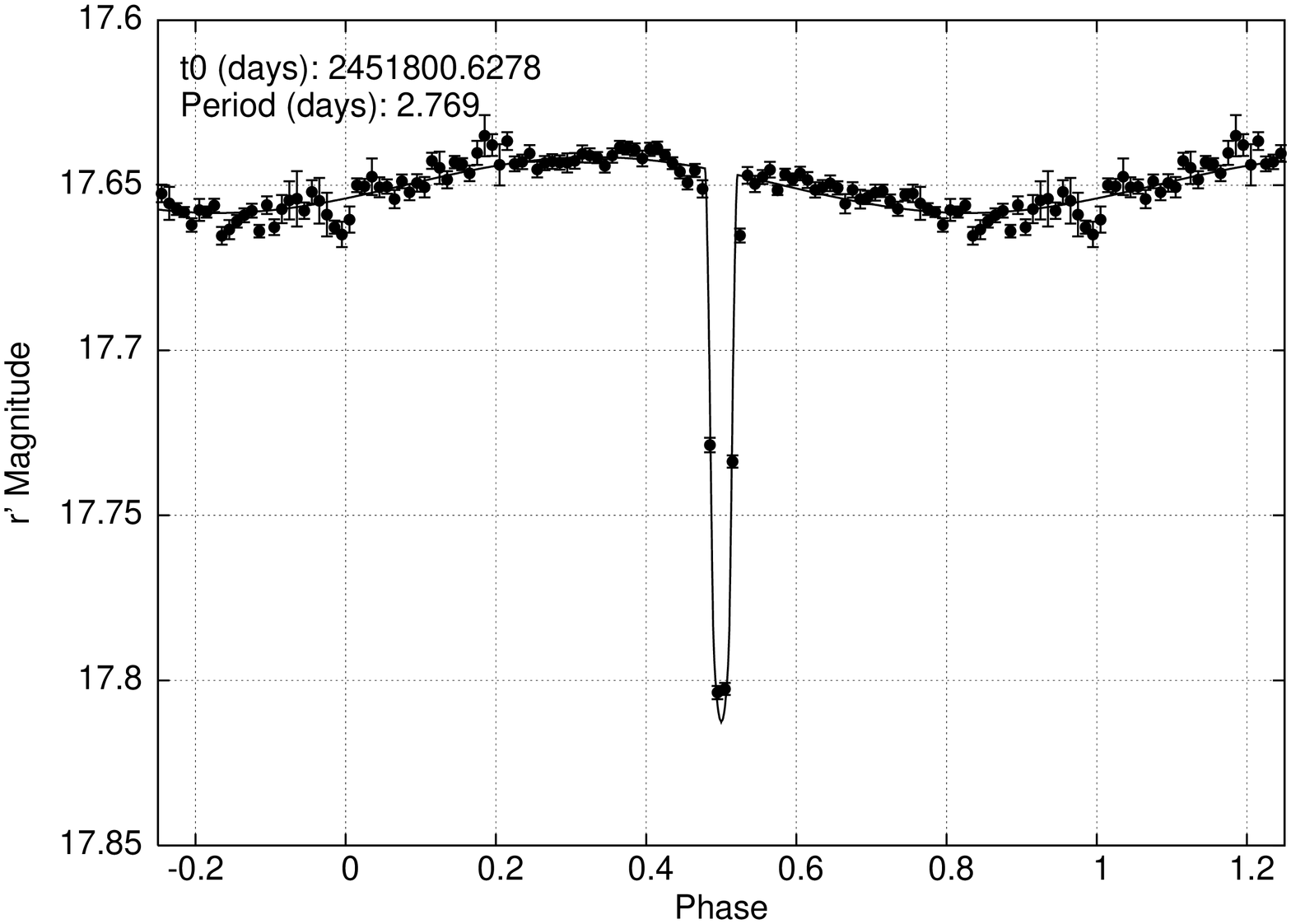,angle=270.0,width=0.5\linewidth}} &
\subfigure[{\bf EB-18} - CPL \& FTFIT - 2000-09]
{\epsfig{file=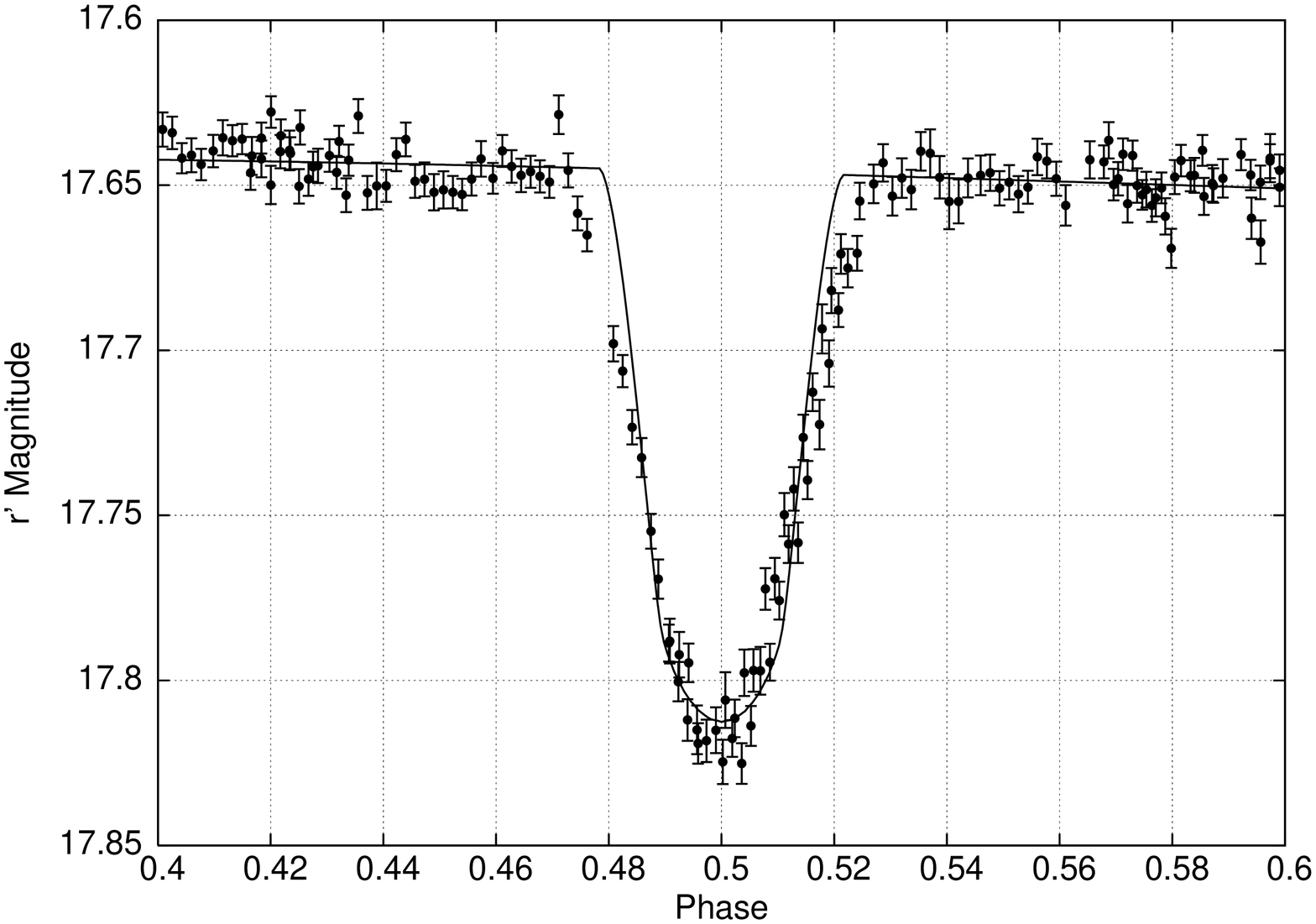,angle=270.0,width=0.5\linewidth}} \\
\end{tabular}
\caption{EB-16, EB-17 and EB-18.}
\end{figure*}

\begin{figure*}
\def\subfigtopskip{4pt}
\def\subfigbottomskip{8pt}
\def\subfigcapskip{4pt}
\centering
\begin{tabular}{cc}
\subfigure[{\bf EB-19} - PL, BL \& FTFIT - 2000-09]
{\epsfig{file=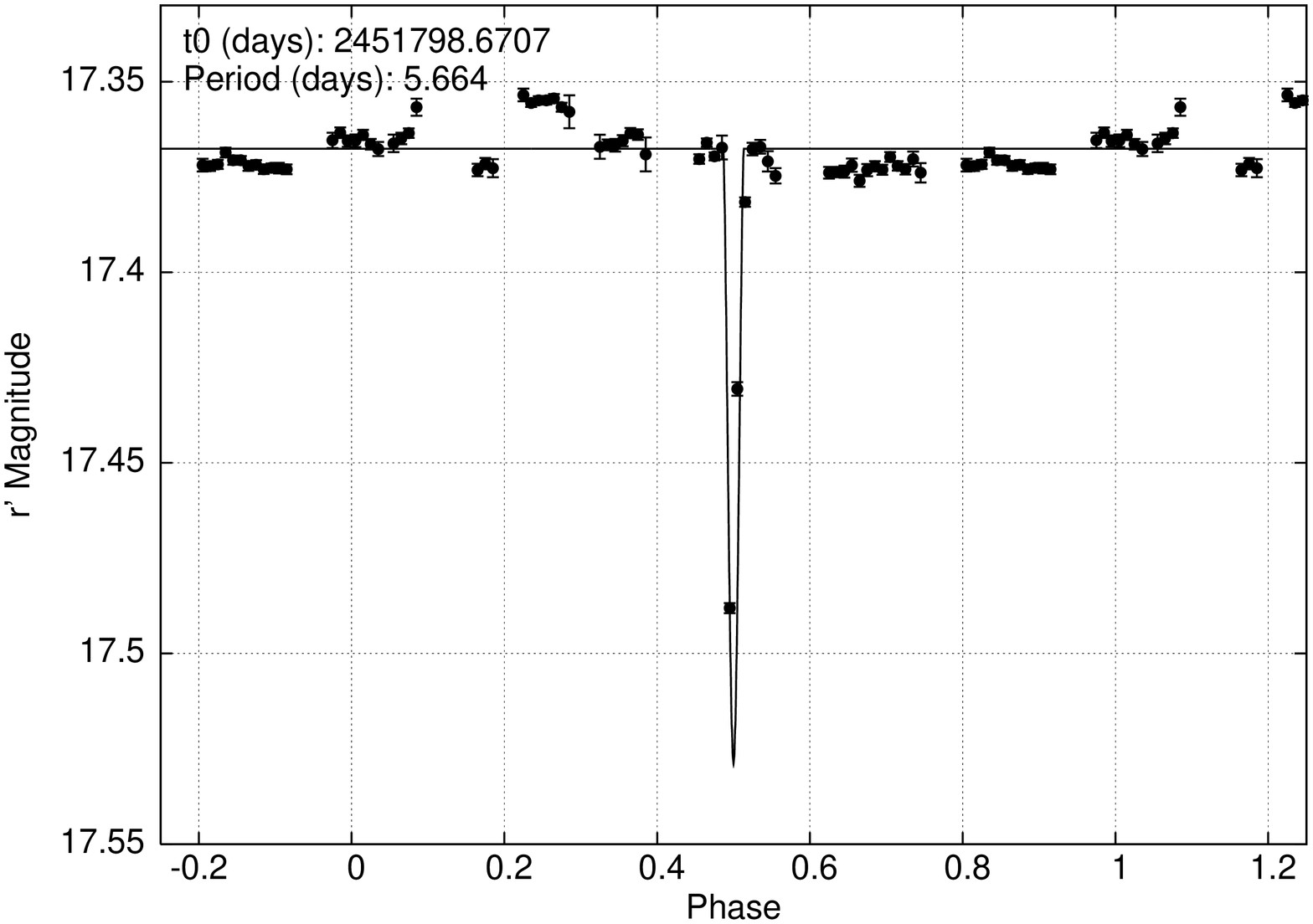,angle=270.0,width=0.5\linewidth}} &
\subfigure[{\bf EB-19} - CPL \& FTFIT - 2000-09]
{\epsfig{file=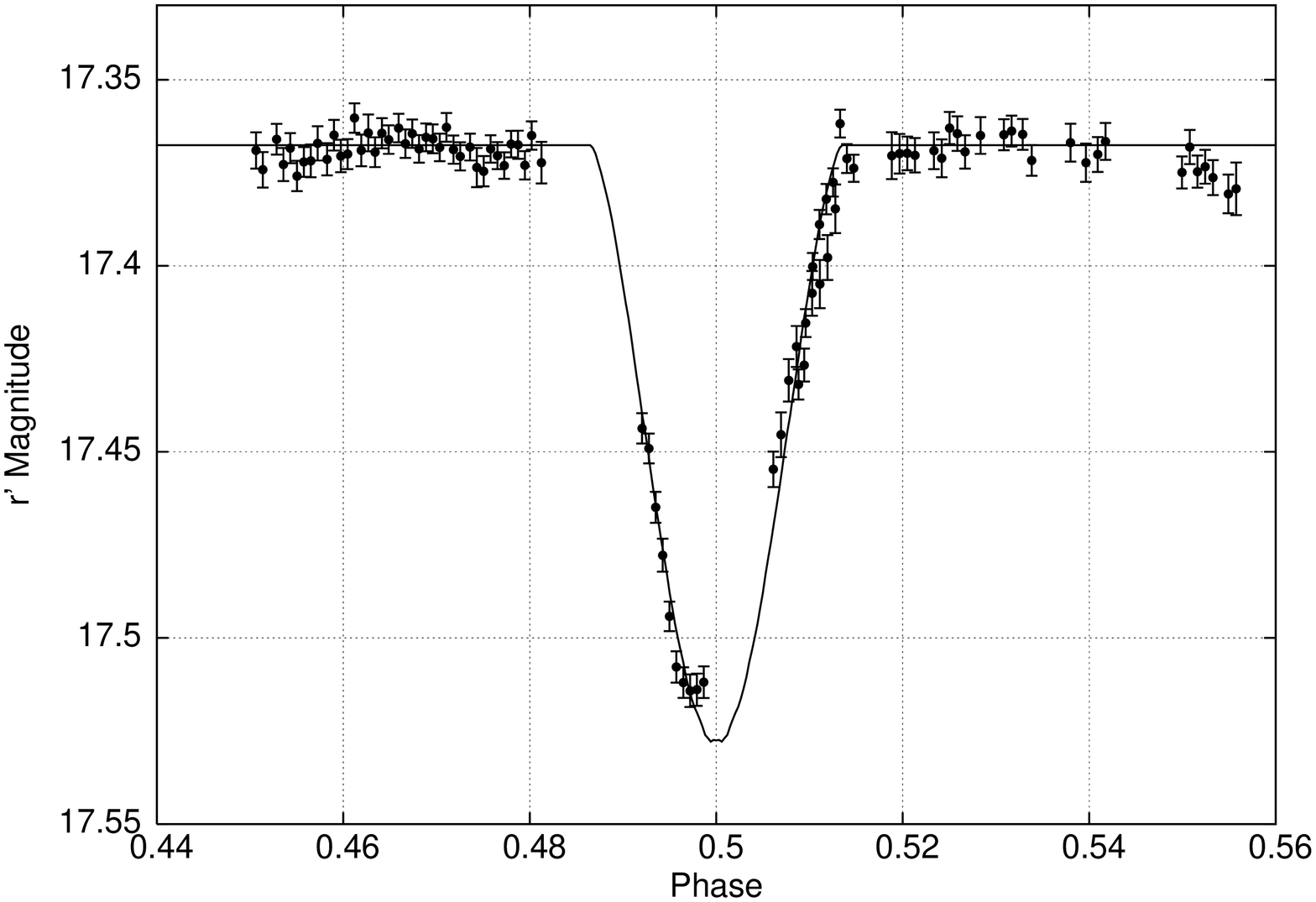,angle=270.0,width=0.5\linewidth}} \\
\subfigure[{\bf EB-20} - PL, BL \& FTFIT - 2000-09]
{\epsfig{file=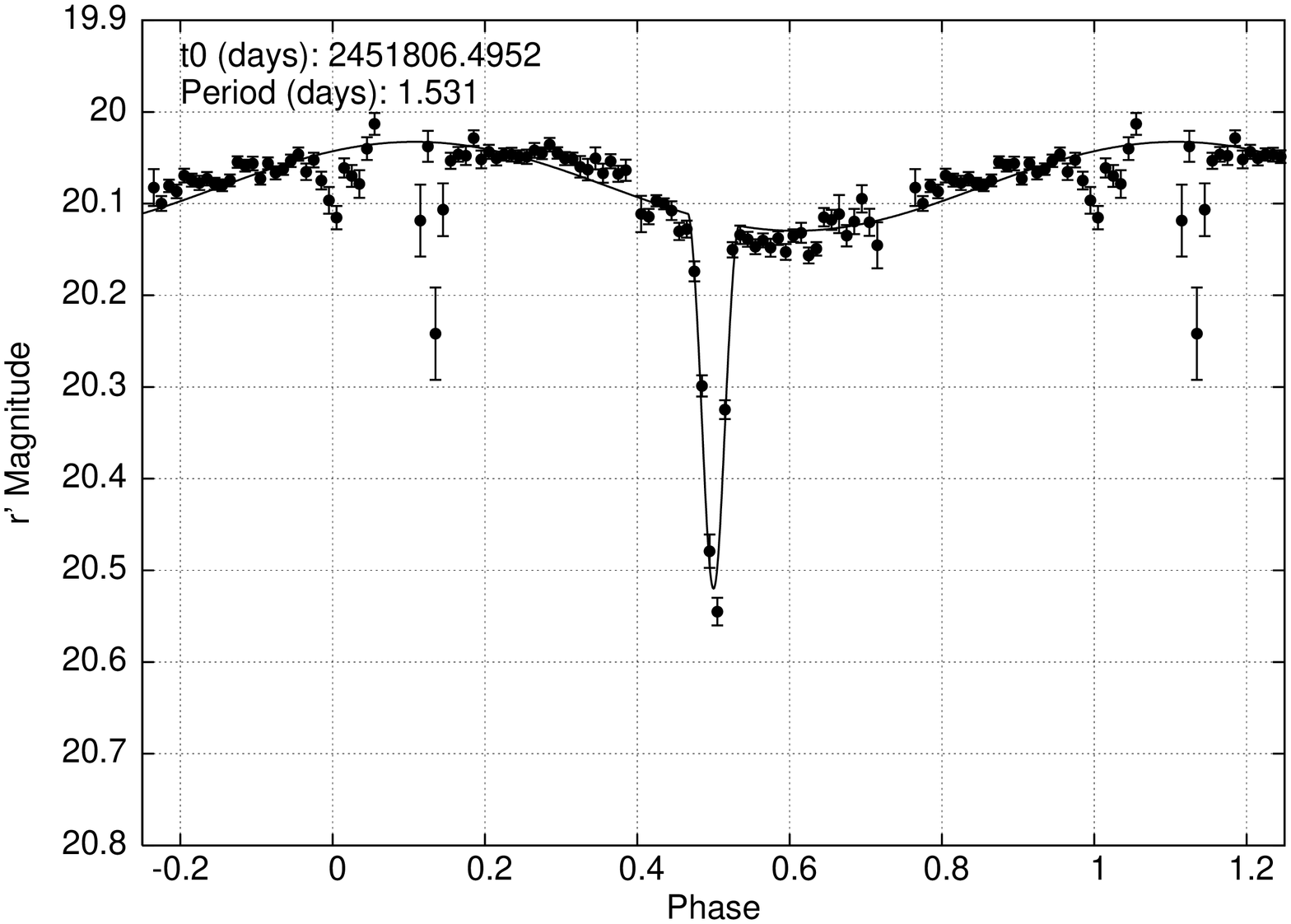,angle=270.0,width=0.5\linewidth}} &
\subfigure[{\bf EB-20} - CPL \& FTFIT - 2000-09]
{\epsfig{file=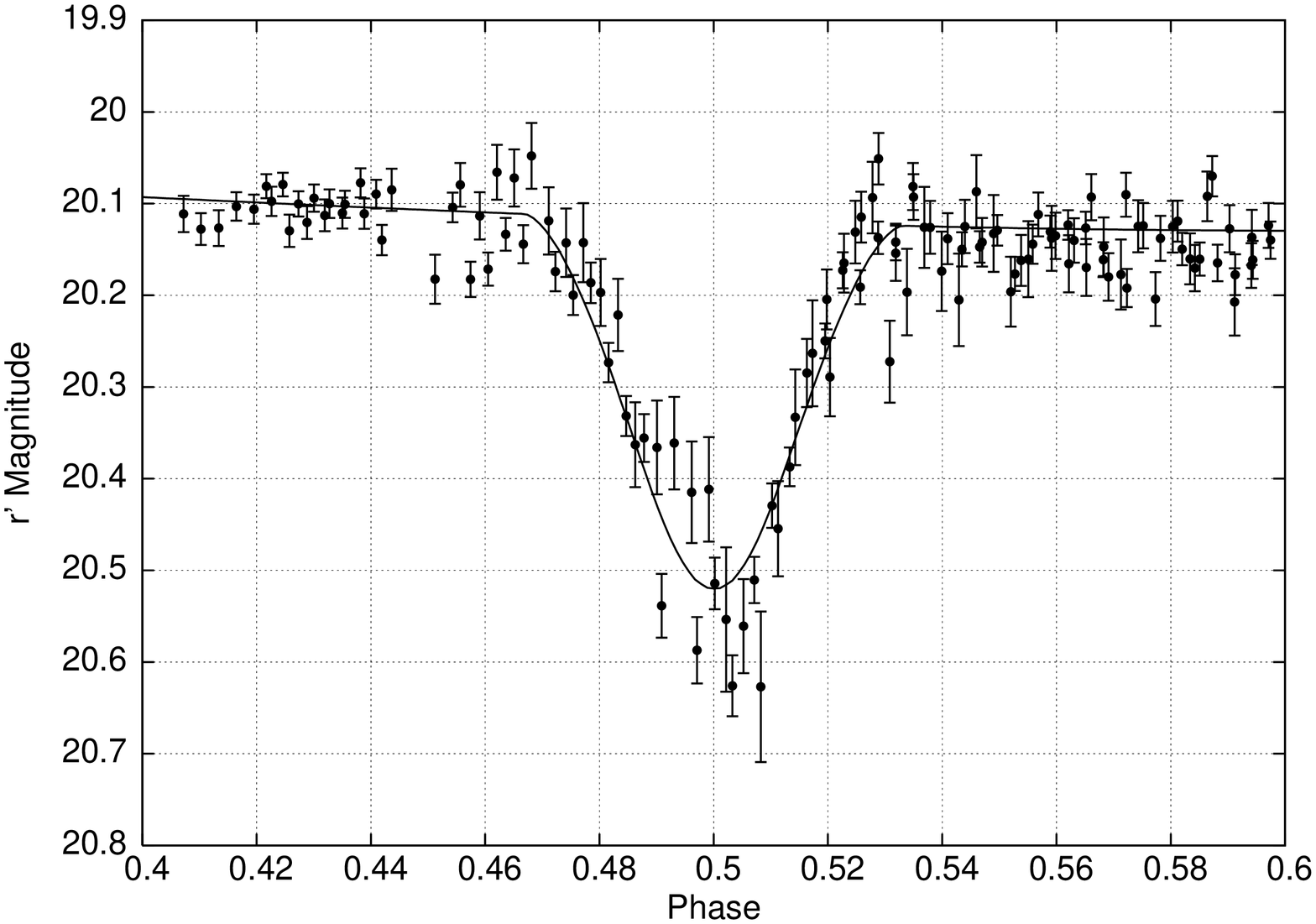,angle=270.0,width=0.5\linewidth}} \\
\subfigure[{\bf EB-21} - $\chi^{2}_{\mbox{\scriptsize ecl}}$ (dashed line) \& $\chi^{2}_{\mbox{\scriptsize fold}}$
           (continuous line) versus $R_{\mbox{\scriptsize c}} / R_{*}$ - 2000-09]
{\epsfig{file=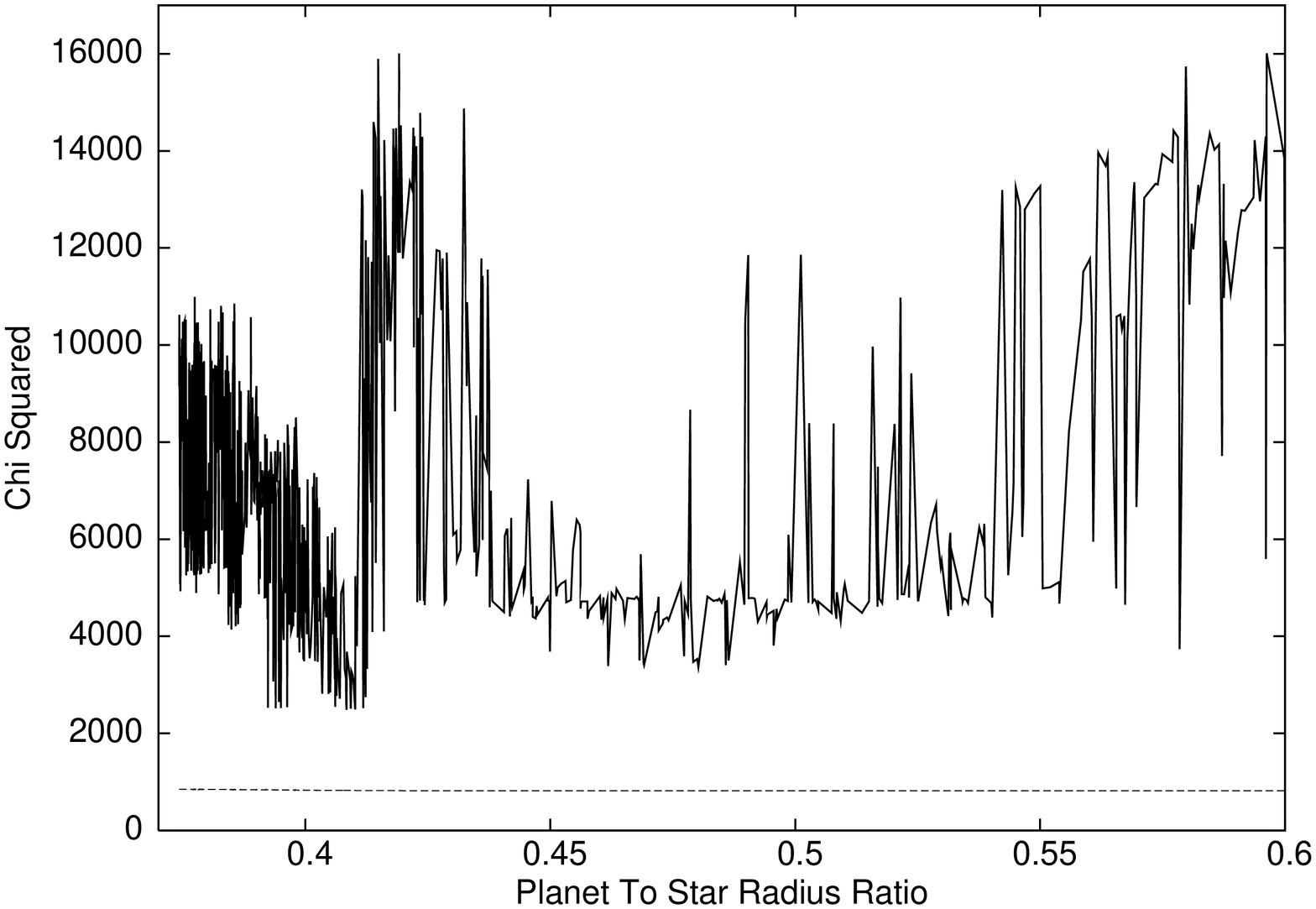,angle=270.0,width=0.5\linewidth}} &
\subfigure[{\bf EB-21} - L \& CTFIT - 2000-09 Night 4]
{\epsfig{file=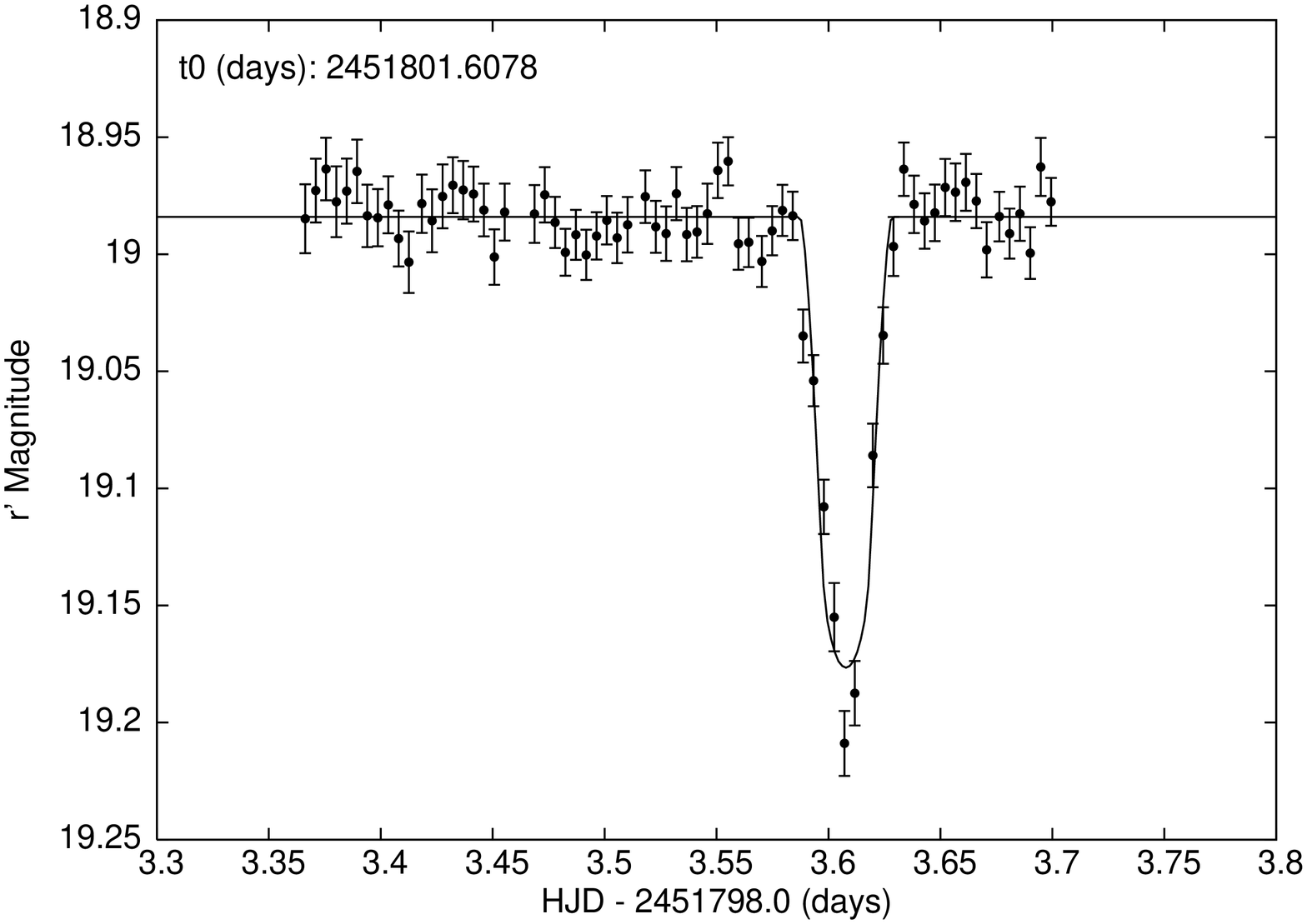,angle=270.0,width=0.5\linewidth}} \\
\end{tabular}
\caption{EB-19, EB-20 and EB-21.}
\end{figure*}

\subsection{A Possible Long Period Cataclysmic Variable}

{\bf EB-15} is a 10.8~h eclipsing binary (Fig.~9) that has round-bottomed eclipses lasting 0.1 in phase and orbital 
modulations that peak near phase 0.2. With $r^{\prime} \approx 20.62$~mag and $r^{\prime}-i^{\prime} \approx 1.13$~mag, 
the star falls close to the cluster main sequence (Fig.~2a). The colour index is consistent with a 0.62$M_{\sun}$ 
K7V star at $d = 2.4$kpc, a possible cluster member. Over the three runs the orbital modulations increase in amplitude 
from 0.1~mag to 0.2~mag, while the eclipse depth decreases from 0.40~mag to 0.24~mag. 

The orbital modulation could arise from spots on one or both stars, though this 
would require a preferred longitude that remains stable over 15 months. The orbital phasing is consistent with that
of an ``orbital hump'' that is often seen in quiescent dwarf novae, arising from the anisotropic emission of a
``hotspot'' on the rim of an accretion disk where the mass transfer stream from the companion star feeds material into 
the disk. The relatively shallow eclipse would then imply a moderate inclination so that the donor star eclipses only 
the near rim of the disk and possibly the hotspot. The eclipse shape is more symmetric than would be expected for 
eclipses of a hotspot, however, and a hotspot eclipse would become deeper rather than shallower as the orbital 
modulation increased. We are unable to decide which interpretation may be correct and recommend follow up observations 
to resolve this ambiguity. In any case the eclipse is too deep to be attributed to a planetary transit.

\subsection{More Eclipsing Binaries}

In this section we present 6 transit candidates exhibiting neither easily discernible secondary eclipses
nor orbital modulations consistent with ellipsoidal variations and heating effects.
Figs. 10 and 11 show for each star
a folded and binned lightcurve with the best fit transiting planet 
model, and an unbinned close-up of the folded lightcurve around the primary eclipse along with the best fit model.
For EB-17 the lightcurves from different runs are offset vertically to highlight
changes in the amplitude and phase of the out-of-eclipse variations.
We rule out the transiting planet model for stars EB-16 thru EB-20
because the full transit fit yields a companion radius greater than 0.2$R_{\sun}$. 
For EB-21 the full transit fit admits $R_{\mbox{\small c}}<0.2R_{\sun}$ but only for periods $P<1.1$~d that are ruled 
out and hence the transiting planet model is ruled out also for EB-21.

{\bf EB-16} is an interesting case in its own right. The lightcurve shows 4 V-shaped eclipses of depth 0.39~mag and duration 1.3~h over 
the three runs. Figs.~10a and 10b
show the lightcurve folded on the 1.31~d period. With $r^{\prime} \approx 21.12$~mag and $r^{\prime}-i^{\prime} 
\approx 1.85$~mag we derive a primary 0.24$M_{\sun}$ M5V
star at $d\sim$750pc, in front of the cluster. The full transit fit reveals that the companion is the same 
size as the primary at 0.245$R_{\sun}$ and that the eclipses are grazing. Hence the period is actually 2.62~d and we 
classify the system as
a grazing eclipsing binary consisting of a pair of M5V stars, an interesting discovery in that few such 
systems are known.

{\bf EB-21} is a difficult case in that the lightcurve data show one eclipse in the 1999-06 run with only 2 data 
points 
during the eclipse, and one well sampled eclipse in the 2000-09 run (Fig.~11f). The eclipse is V-shaped of depth 
0.2~mag and duration 1.0~h suggesting that it is likely to be a grazing eclipse. With $r^{\prime} \approx 18.98$~mag 
and $r^{\prime}-i^{\prime} \approx 1.52$~mag we find that
the primary is a 0.40$M_{\sun}$ M2V star that lies at only $\sim$530pc. A central transit fit to the 2000-09 data
(Fig.~11f) yields a minimum companion radius of 0.141$R_{\sun}$ due to the small size of the 
primary star (0.38$R_{\sun}$) from which we cannot rule out the transiting planet model. Analysis of the shape of the 
single eclipse in the 2000-09 data is possible 
due to the good time sampling of the observations and such an analysis may reveal whether the eclipse is the 
result of an annular occultation by a smaller companion or a grazing occultation by a larger companion.
Also, we may attempt to predict the orbital period $P$ of the planetary companion as a function of the impact parameter 
$b = a \cos i / R_{*}$ of the eclipse and subsequently use the lightcurve data from the whole run to determine which 
periods, and hence which values of $b$, may be ruled out. 

For EB-21 we created a grid for the impact parameter $b$ from 0.0 to 1.0. For each value of $b$ we fitted Model 1 to 
the single eclipse in the 2000-09 run, using the lightcurve data from the whole run, in order to determine a time of
mid-transit $t_{0}$, a constant magnitude $m_{0}$, a planetary radius $R_{\mbox{\small c}}$ and a transit duration
$\Delta t$. The chi squared $\chi^{2}_{\mbox{\small ecl}}$ of the fit was also calculated. The duration of a transit
event is given by:
\begin{equation}
\Delta t = \frac{P R_{*}}{\pi a} 
           \sqrt{ \left( 1 + \frac{R_{\mbox{\small c}}}{R_{*}} \right)^{2} - b^{2} }
\end{equation}
We also have Kepler's third law:
\begin{equation}
P^{2} = \frac{4 \pi^{2} a^{3}}{G M_{*}}
\end{equation}
We already know $M_{*}$ and $R_{*}$, and since we have $R_{\mbox{\small c}}$ and $\Delta t$ as functions of $b$
from our fits of the single eclipse, we may use Equations 17 and 18 to estimate $P$ (or $a$) as a function of $b$ for 
the transiting planet model. For each value of $b$ we folded the 2000-09 lightcurve of EB-21 on the predicted 
period $P$ using the fitted $t_{0}$, and calculated a new chi squared $\chi^{2}_{\mbox{\small fold}}$
using the fit to the single eclipse with period $P$. In general, if the predicted period is such that none of the folded
lightcurve data falls during the eclipse, then
$\chi^{2}_{\mbox{\small fold}} = \chi^{2}_{\mbox{\small ecl}}$. However, if the predicted period is such that some of 
the folded lightcurve data does fall during the eclipse, then
$\chi^{2}_{\mbox{\small fold}} \gg \chi^{2}_{\mbox{\small ecl}}$, ruling out that particular period, impact
parameter and corresponding eclipse solution.

Fig.~11e shows a plot of $\chi^{2}_{\mbox{\small ecl}}$ versus $R_{\mbox{\small c}} / R_{*}$ (dashed line), which 
appears constant due to the scale on the y-axis. The continuous line is a plot of $\chi^{2}_{\mbox{\small fold}}$
versus $R_{\mbox{\small c}} / R_{*}$, which clearly shows that
$\chi^{2}_{\mbox{\small fold}} \gg \chi^{2}_{\mbox{\small ecl}}$ for all values of $R_{\mbox{\small c}}$ from
0.141$R_{\sun}$ (the minimum companion radius with $b = 0$ and 
$R_{\mbox{\small c}} / R_{*} = 0.373$) to 0.227$R_{\sun}$ ($b = 0.92$ and $R_{\mbox{\small c}} / R_{*} = 0.600$).
This is due to the fact that the predicted period is less than 1.10~d for these values of $b$. This demonstrates that 
the transiting planet model is inconsistent with our observational data for this star, and hence a stellar companion is 
favoured. We have classified this system as an eclipsing binary.

\subsection{INT-7789-TR-1}

\begin{figure*}
\def\subfigtopskip{4pt}
\def\subfigbottomskip{8pt}
\def\subfigcapskip{4pt}
\centering   
\begin{tabular}{cc}
\subfigure[{\bf TR-1} - $\chi^{2}_{\mbox{\scriptsize ecl}}$ (dashed line) versus
           $R_{\mbox{\scriptsize c}} / R_{*}$ - 2000-09]
{\epsfig{file=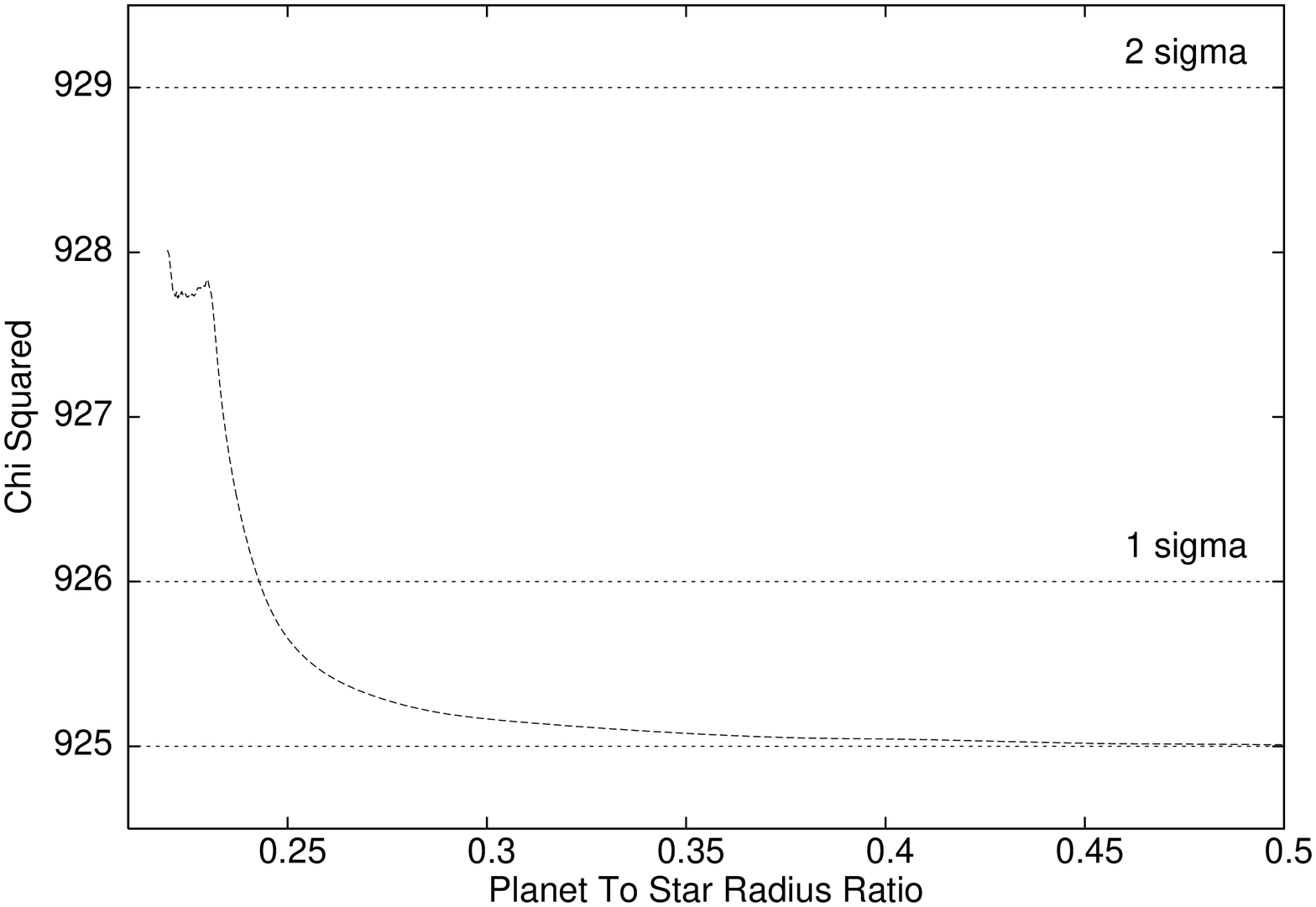,angle=270.0,width=0.4\linewidth}} &
\subfigure[{\bf TR-1} - L \& Single Eclipse Fit - 2000-09 Night 9]
{\epsfig{file=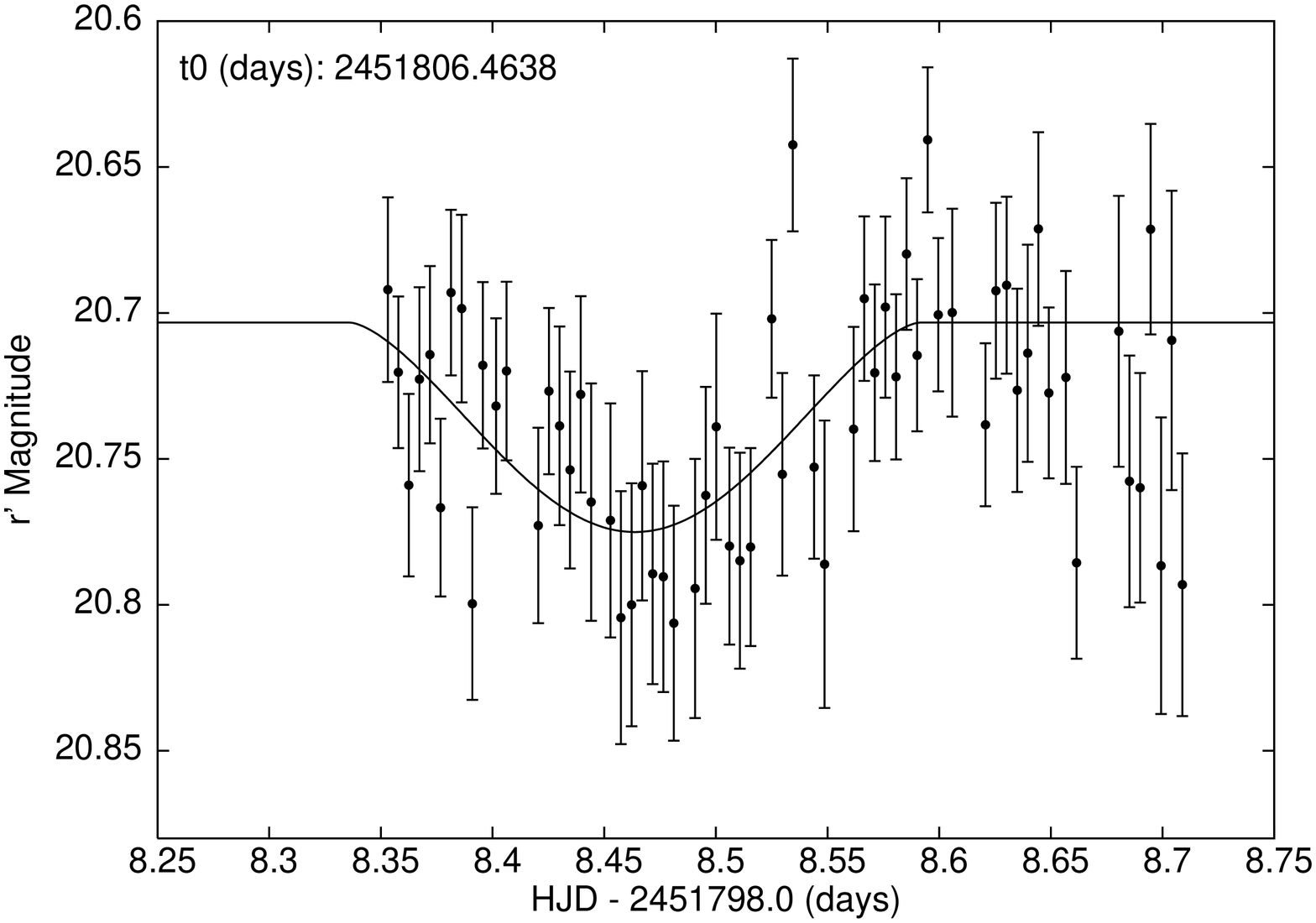,angle=270.0,width=0.4\linewidth}} \\
\subfigure[{\bf TR-2} - $\chi^{2}_{\mbox{\scriptsize ecl}}$ (dashed line) versus
           $R_{\mbox{\scriptsize c}} / R_{*}$ \& $\chi^{2'}_{\mbox{\scriptsize fold}}$ (continuous  
           line) versus $R_{\mbox{\scriptsize c}} / R_{*}$ - 2000-09]
{\epsfig{file=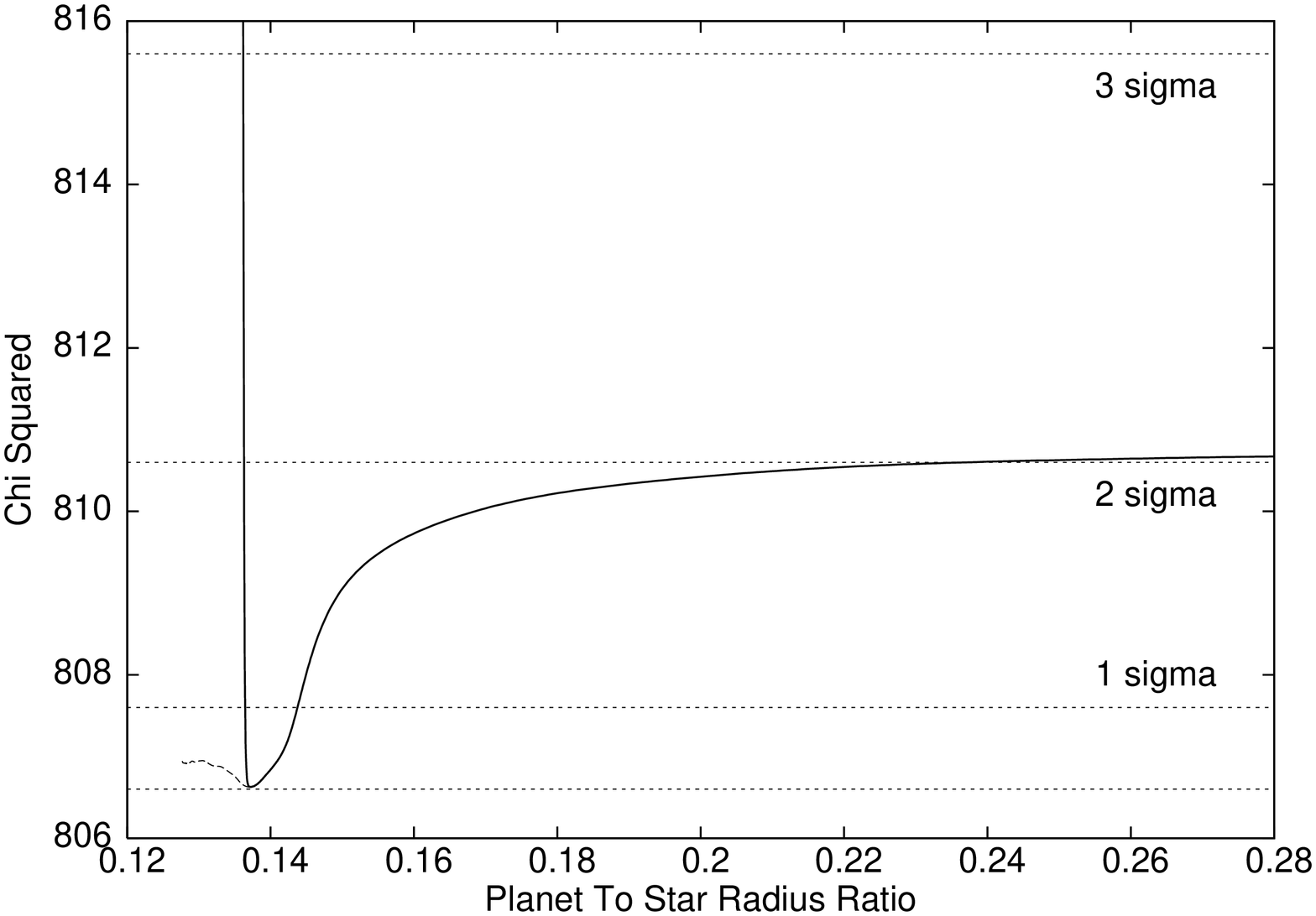,angle=270.0,width=0.4\linewidth}} &
\subfigure[{\bf TR-2} - L \& Single Eclipse Fit - 2000-09 Night 8]
{\epsfig{file=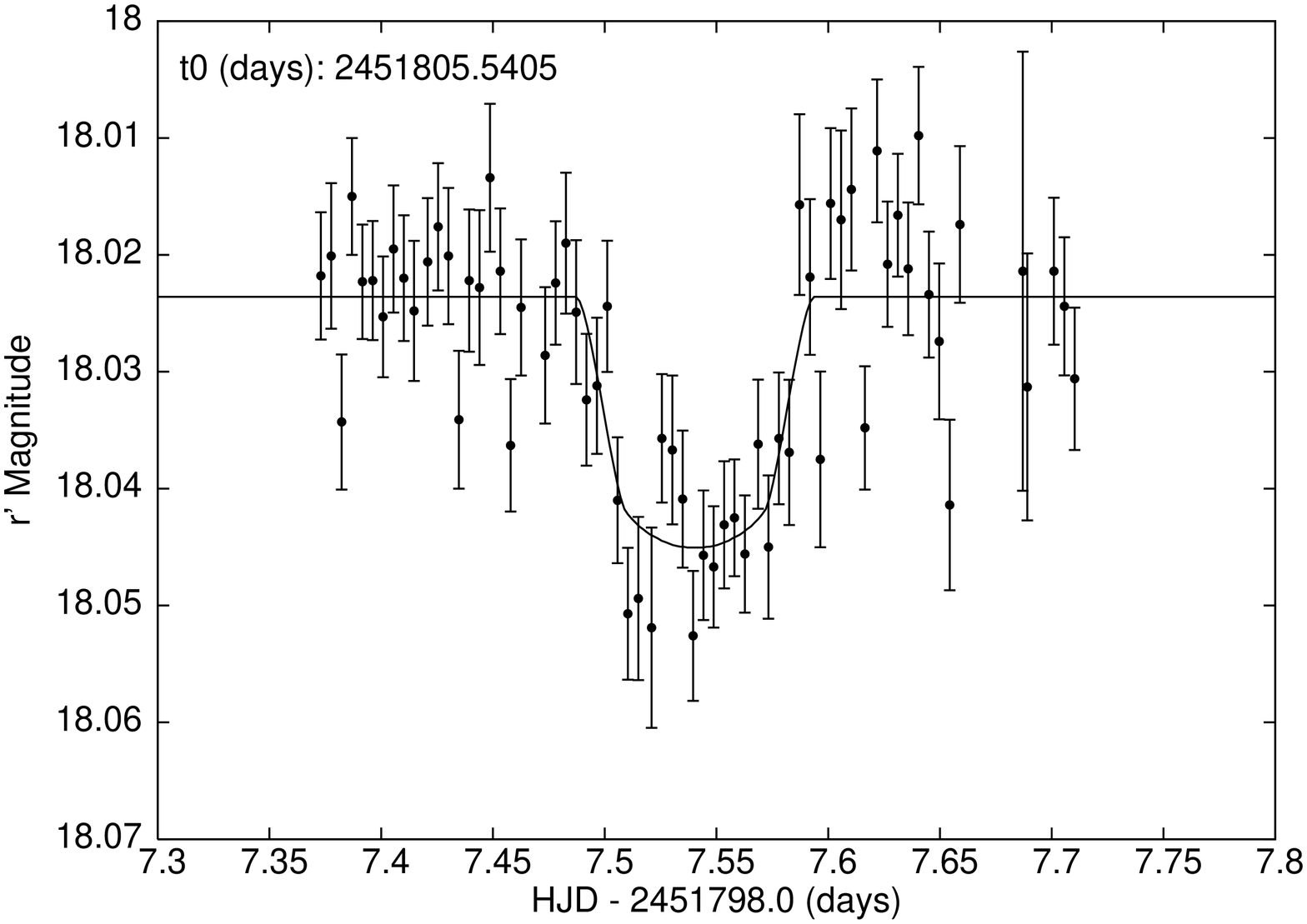,angle=270.0,width=0.4\linewidth}} \\
\subfigure[{\bf TR-3} - Eclipse Periodogram - 2000-09]
{\epsfig{file=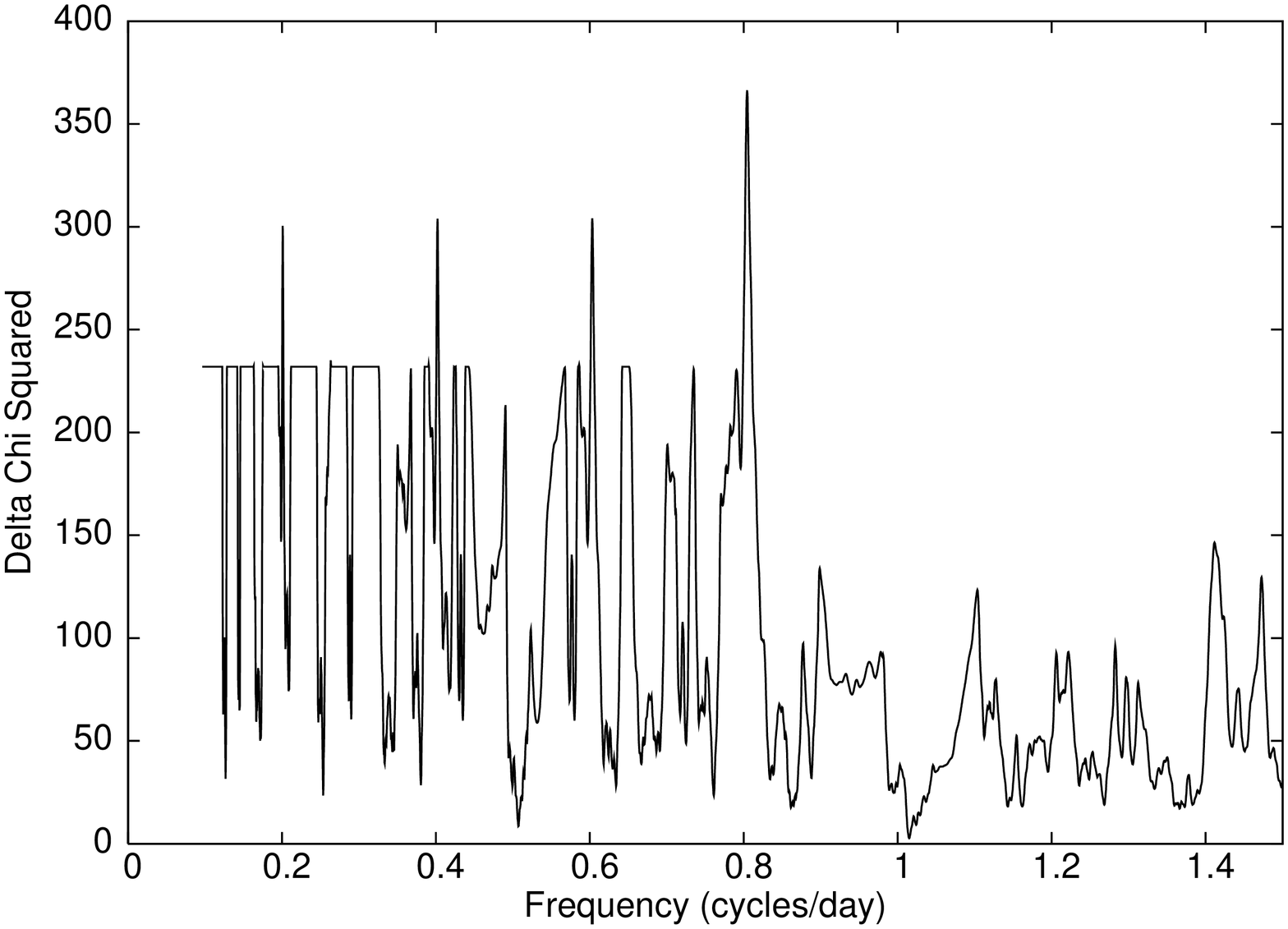,angle=270.0,width=0.4\linewidth}} &
\subfigure[{\bf TR-3} - CM - 2000-09]
{\epsfig{file=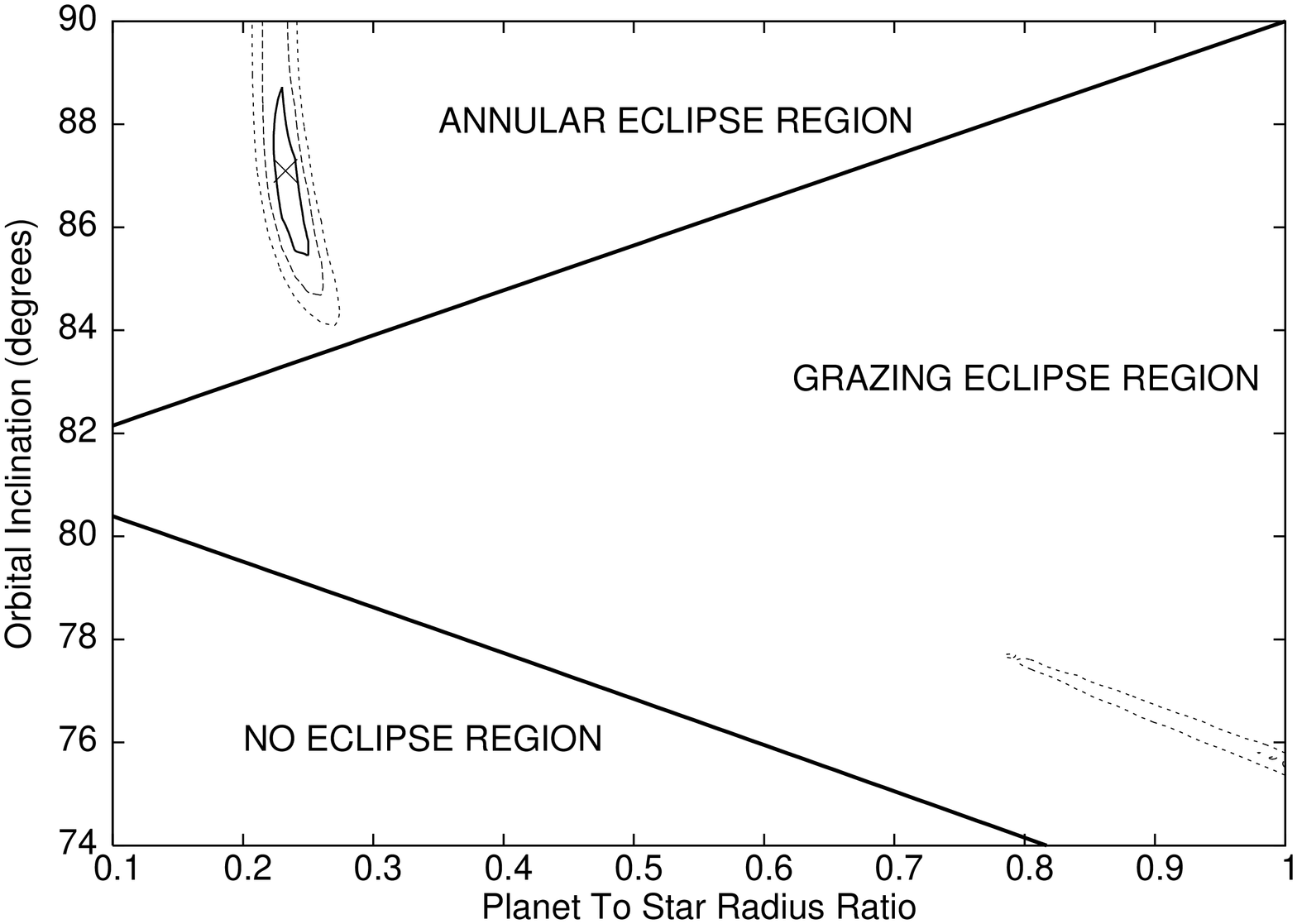,angle=270.0,width=0.4\linewidth}} \\
\subfigure[{\bf TR-3} - PL, BL \& FTFIT - 2000-09]
{\epsfig{file=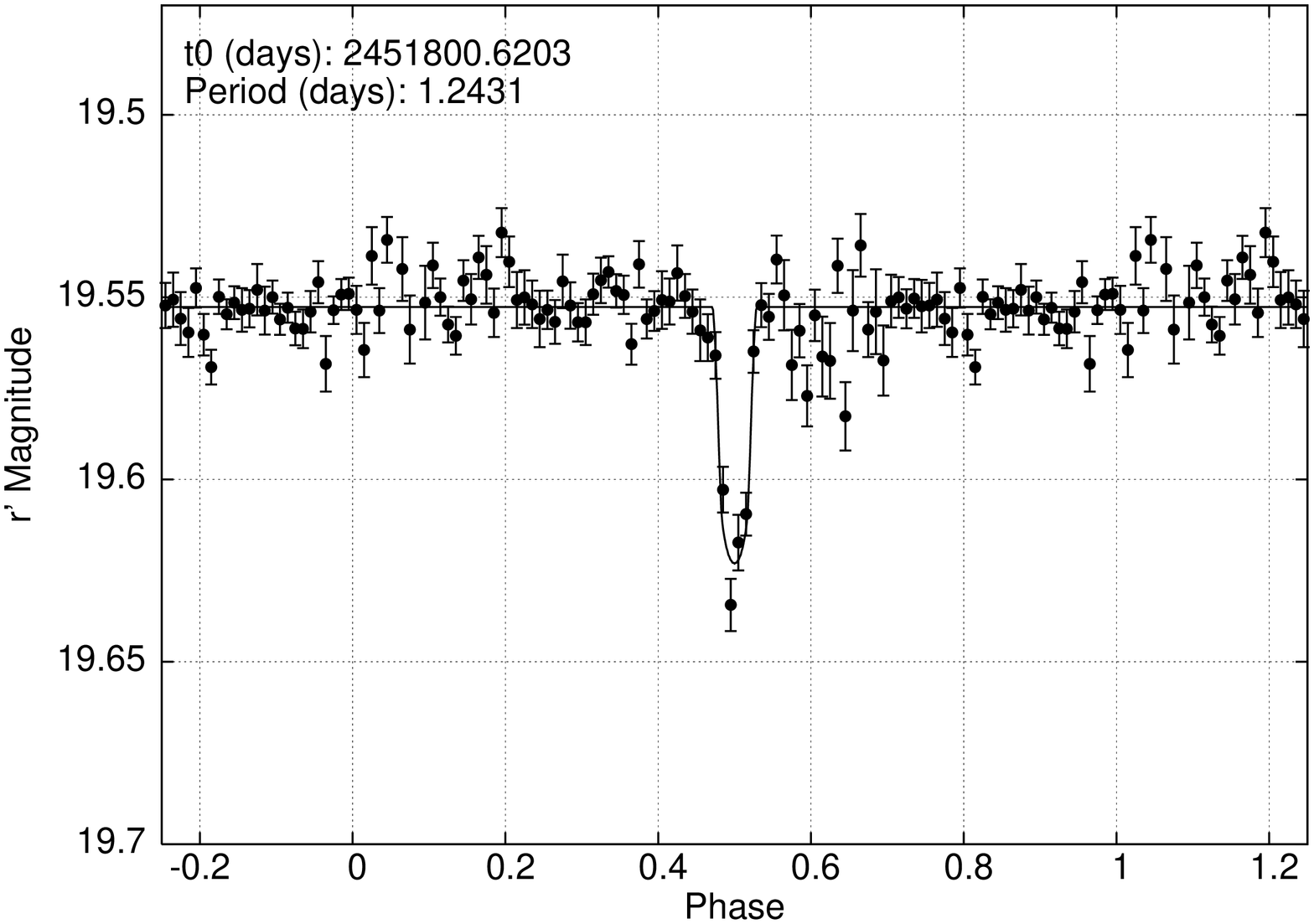,angle=270.0,width=0.4\linewidth}} &
\subfigure[{\bf TR-3} - CPL \& FTFIT - 2000-09]
{\epsfig{file=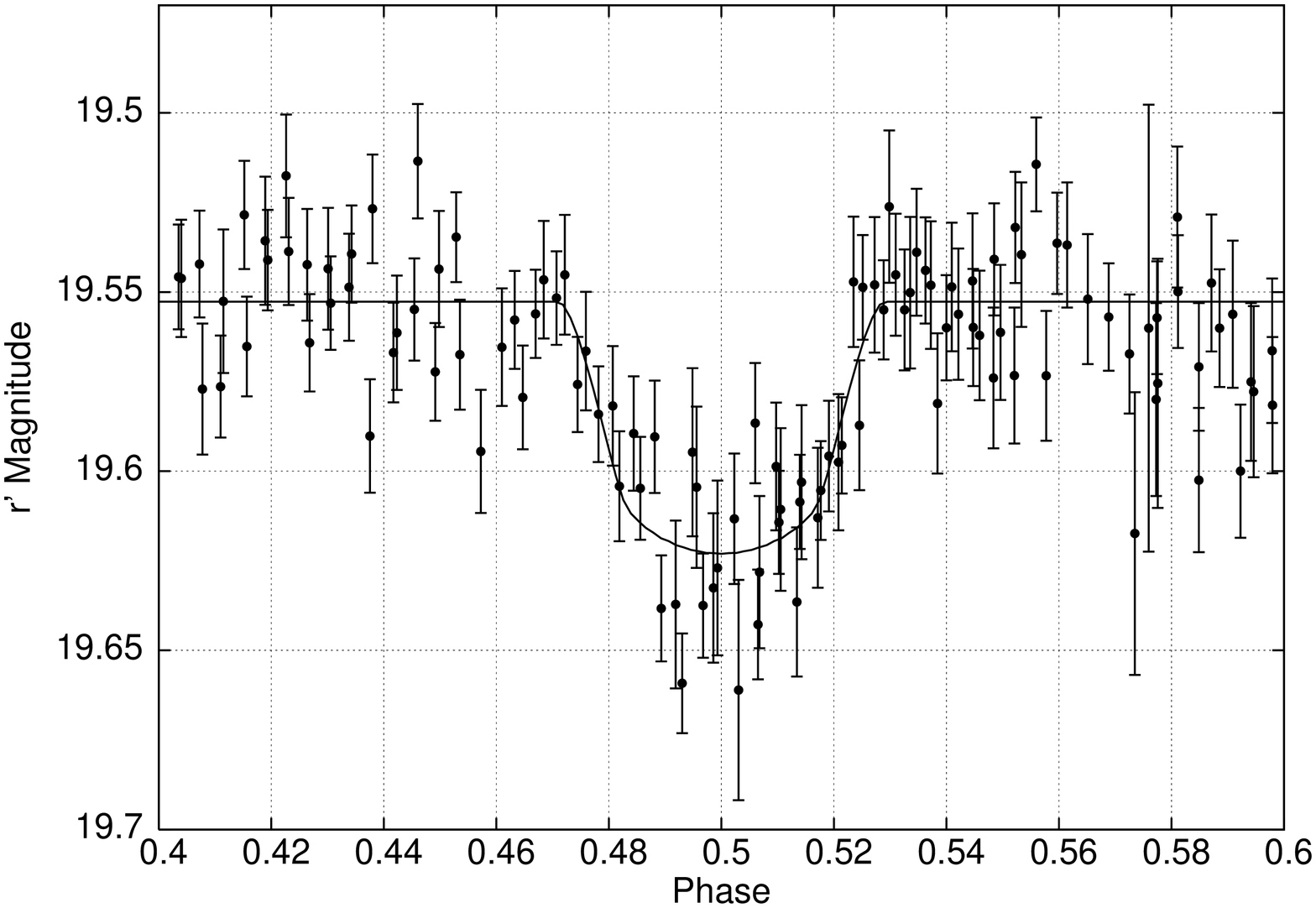,angle=270.0,width=0.4\linewidth}} \\
\end{tabular}
\caption{Planetary transit candidates TR-1, TR-2 and TR-3.}
\end{figure*}

\begin{figure*}
\def\subfigtopskip{4pt}
\def\subfigbottomskip{8pt}
\def\subfigcapskip{4pt}
\centering
\begin{tabular}{cc}
\subfigure[{\bf TR-3} - CM - 2000-09]
{\epsfig{file=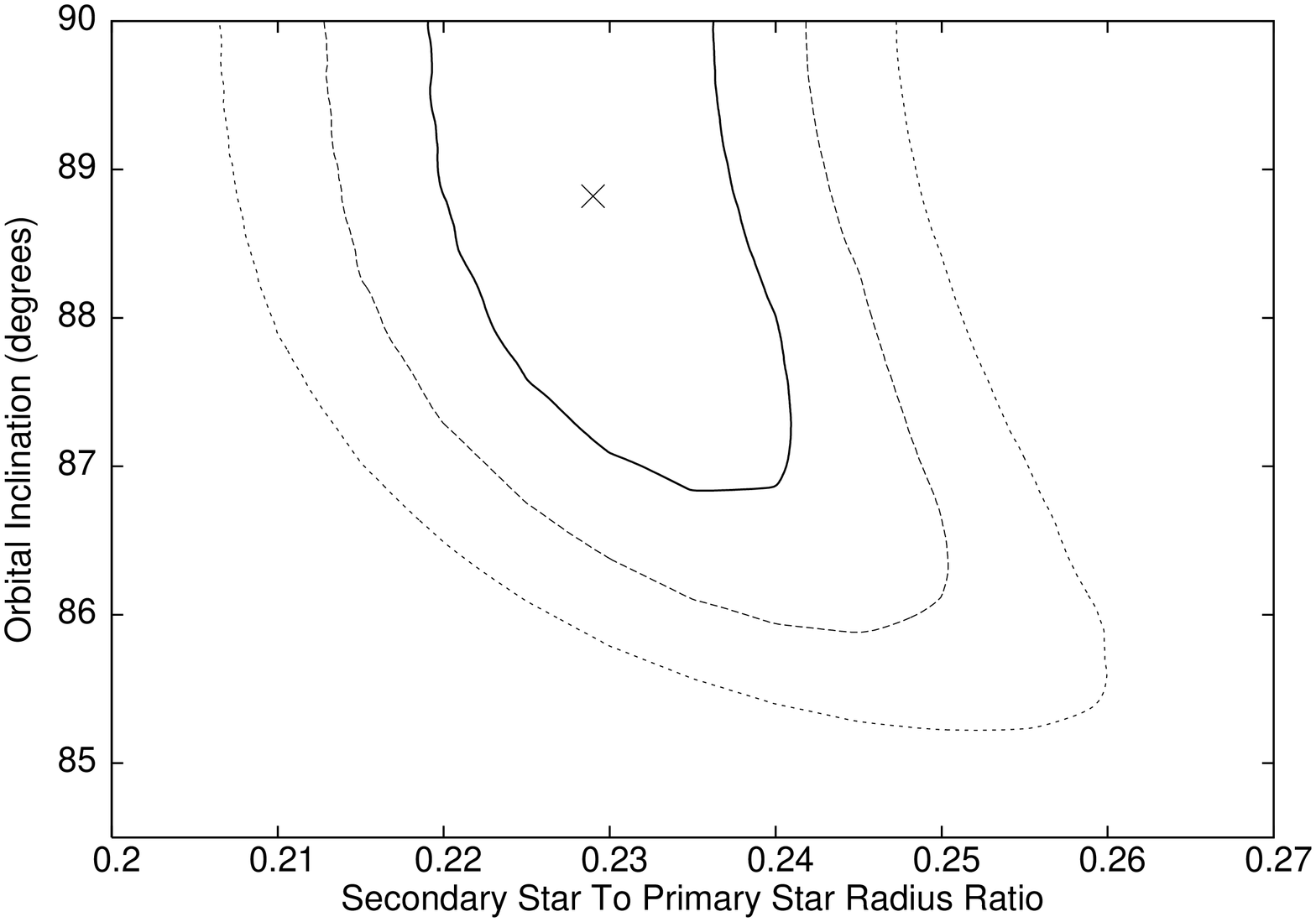,angle=270.0,width=0.4\linewidth}} &
\subfigure[{\bf TR-3} - CM - 2000-09]
{\epsfig{file=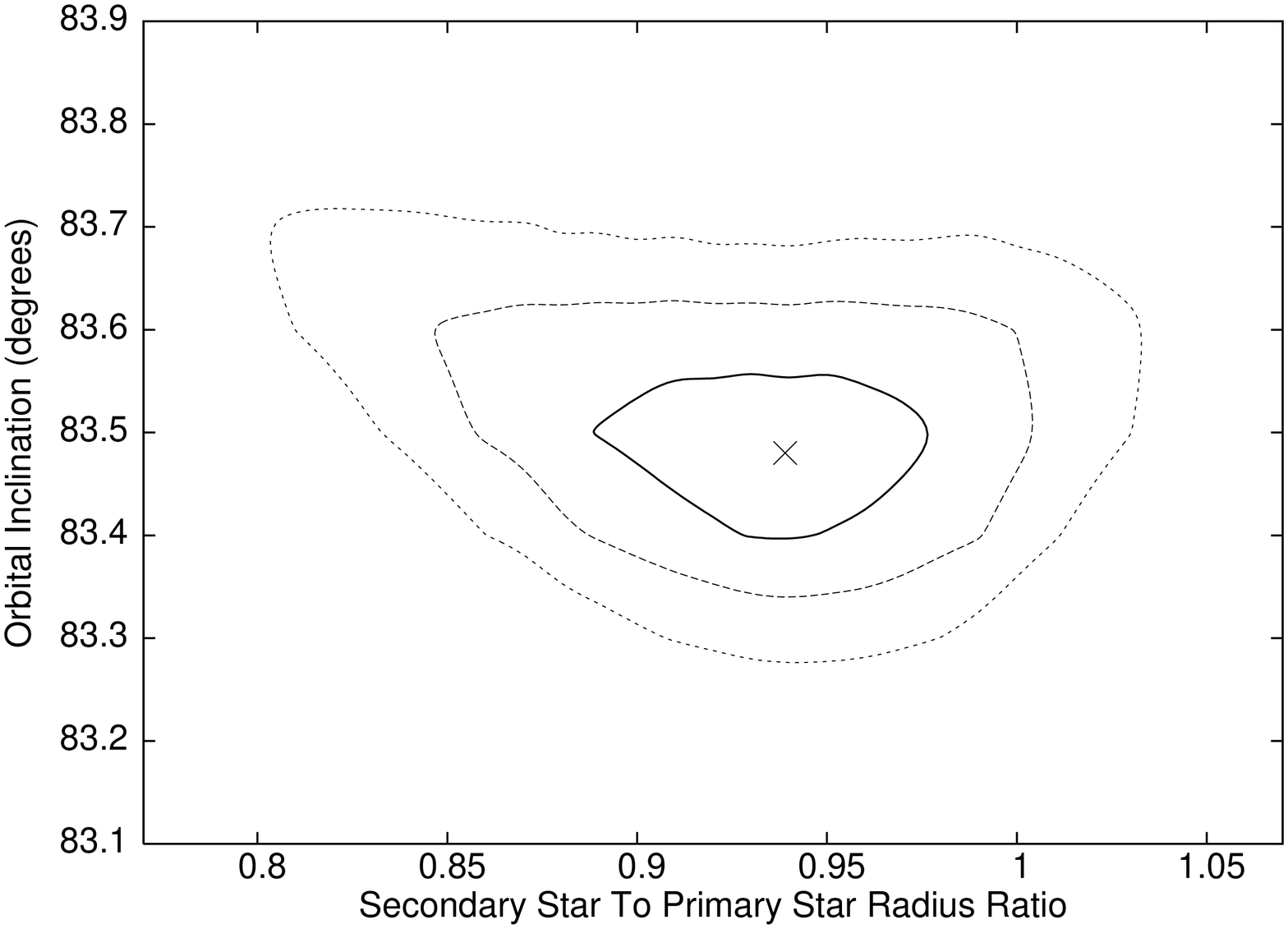,angle=270.0,width=0.4\linewidth}} \\
\subfigure[{\bf TR-3} - PL, BL \& Eclipsing Binary Fit - 2000-09]
{\epsfig{file=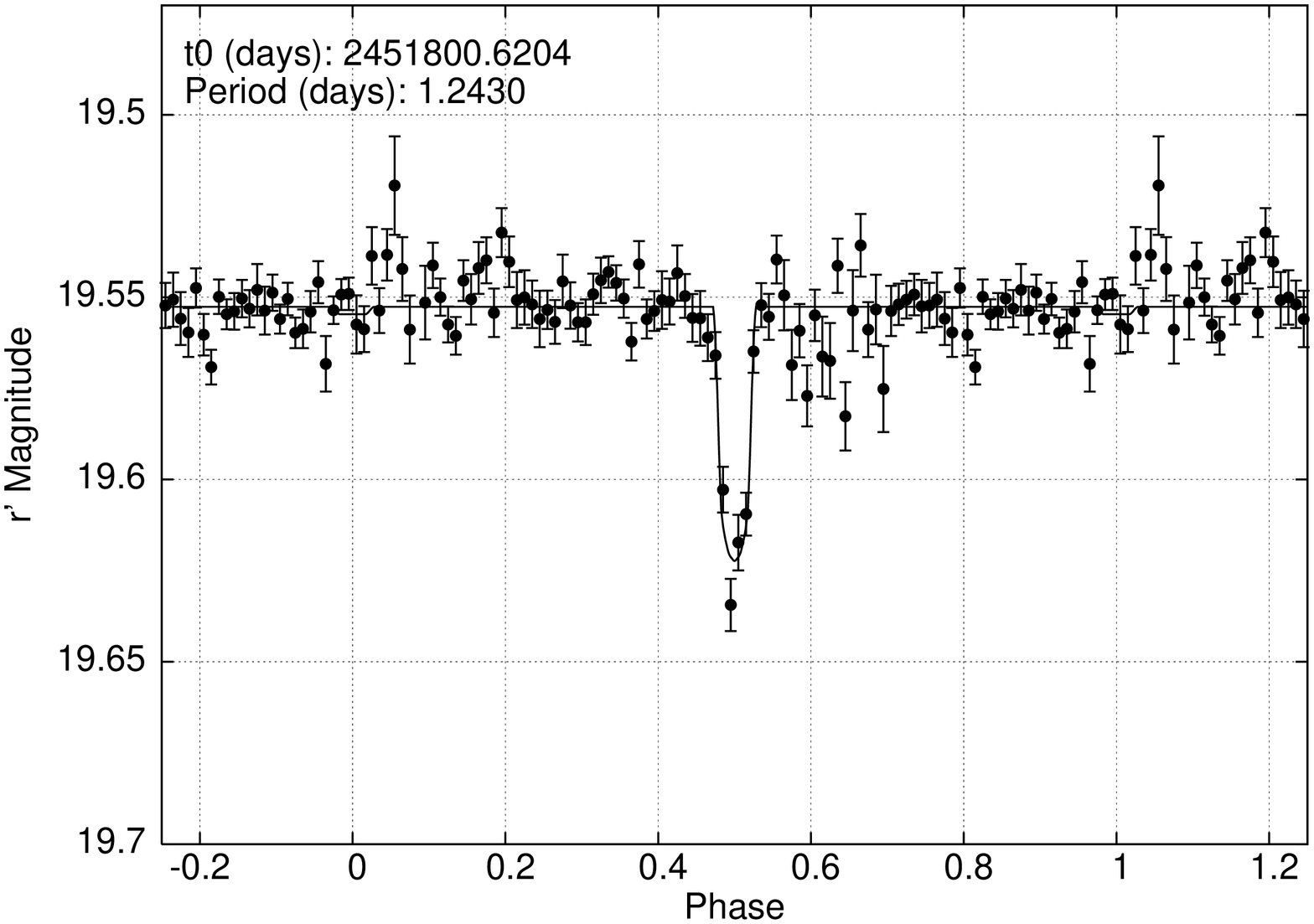,angle=270.0,width=0.4\linewidth}} &
\subfigure[{\bf TR-3} - PL, BL \& Eclipsing Binary Fit - 2000-09]
{\epsfig{file=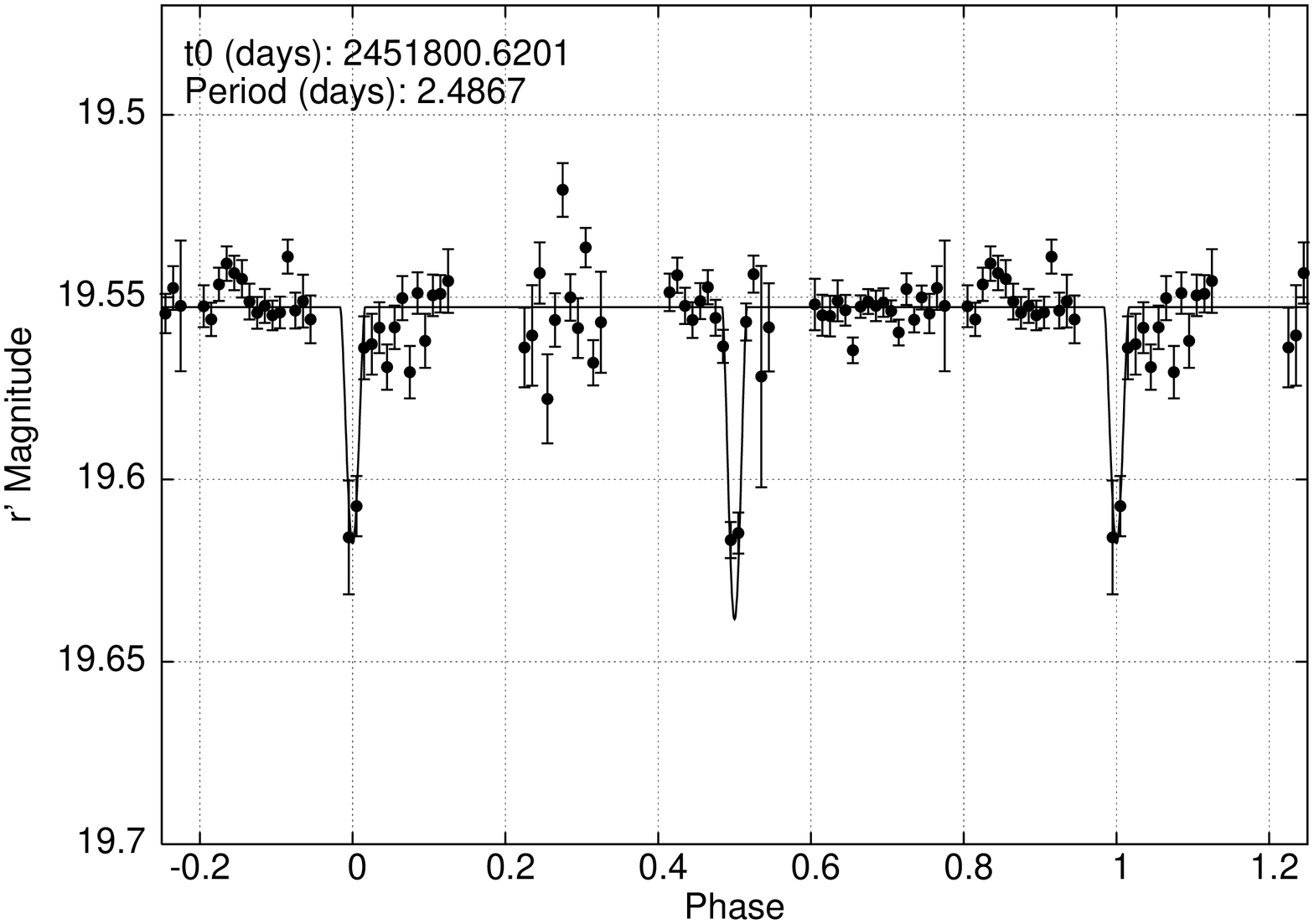,angle=270.0,width=0.4\linewidth}} \\
\subfigure[{\bf TR-3} - CPL \& Eclipsing Binary Fit - 2000-09]
{\epsfig{file=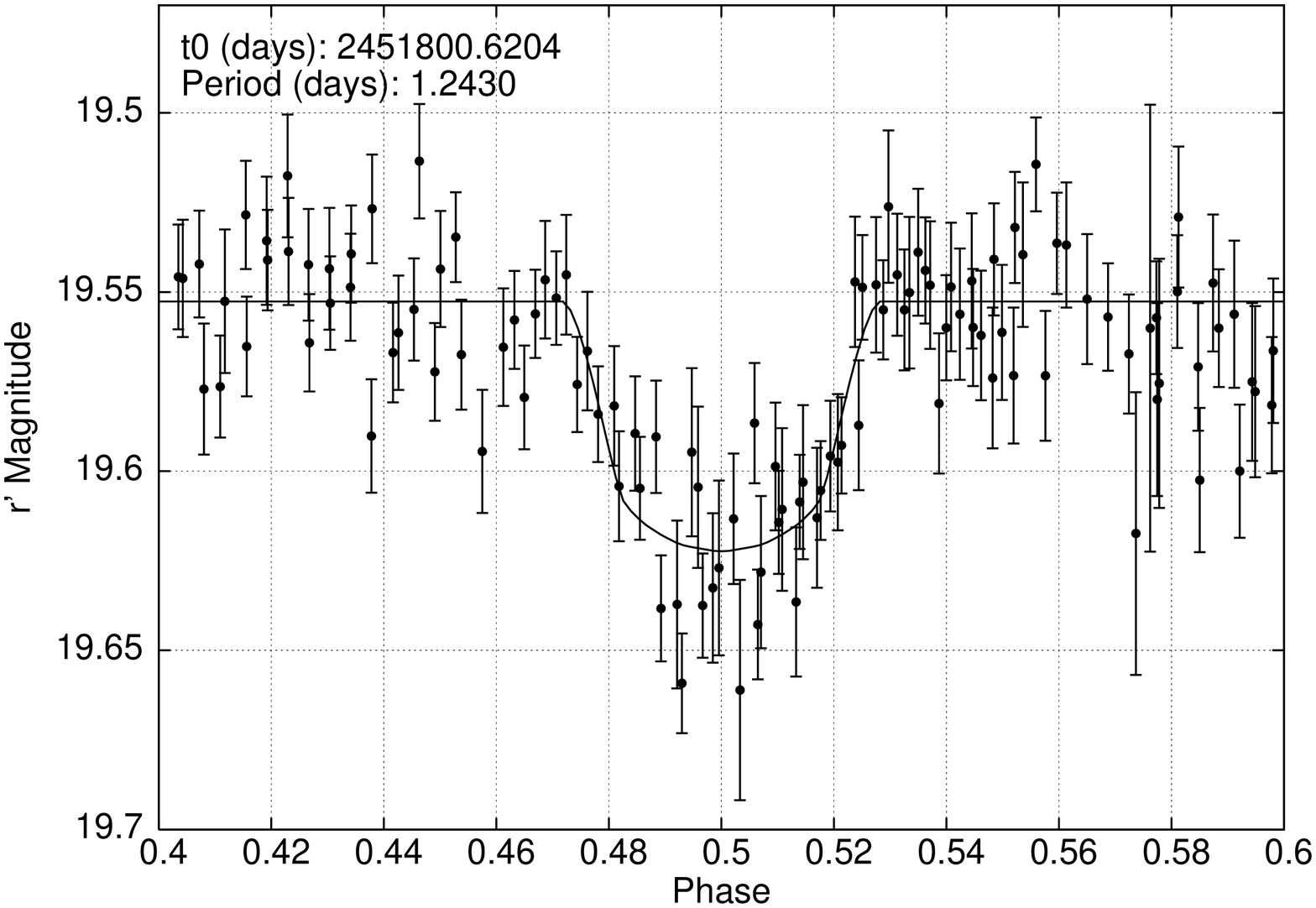,angle=270.0,width=0.4\linewidth}} &
\subfigure[{\bf TR-3} - CPL \& Eclipsing Binary Fit - 2000-09]
{\epsfig{file=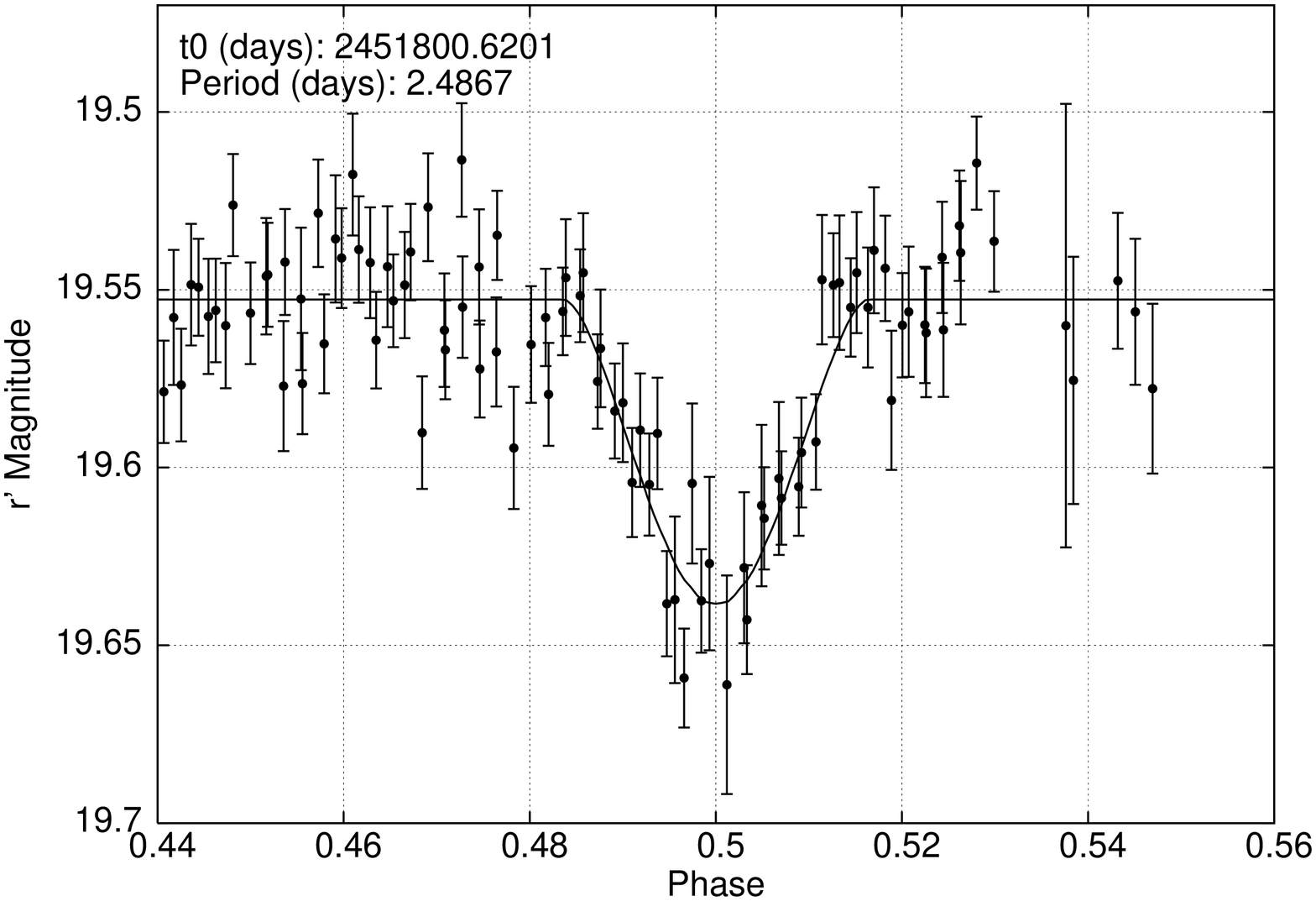,angle=270.0,width=0.4\linewidth}} \\
\subfigure[{\bf TR-3} - CPL \& Eclipsing Binary Fit - 2000-09]
{\epsfig{file=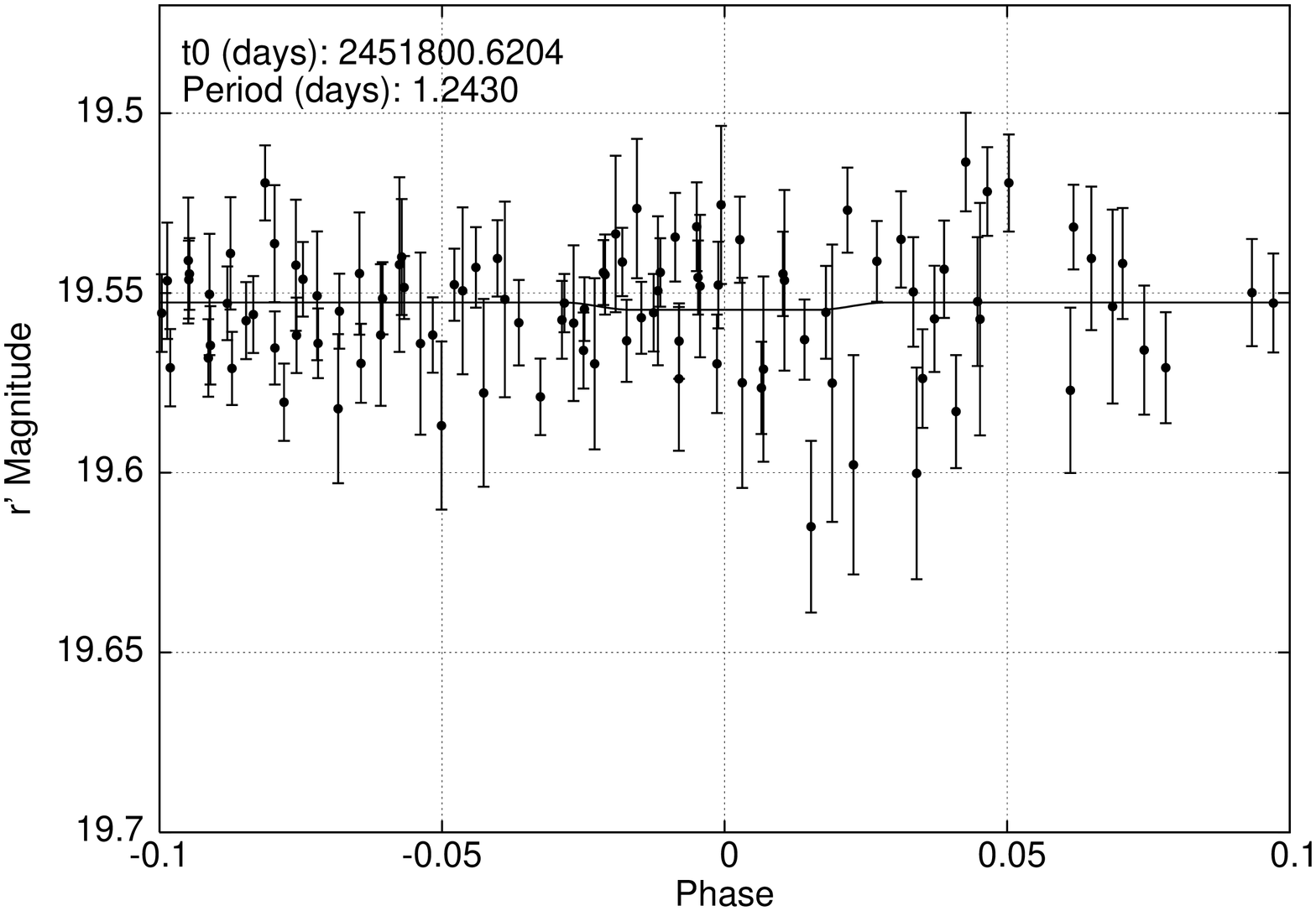,angle=270.0,width=0.4\linewidth}} &
\subfigure[{\bf TR-3} - CPL \& Eclipsing Binary Fit - 2000-09]
{\epsfig{file=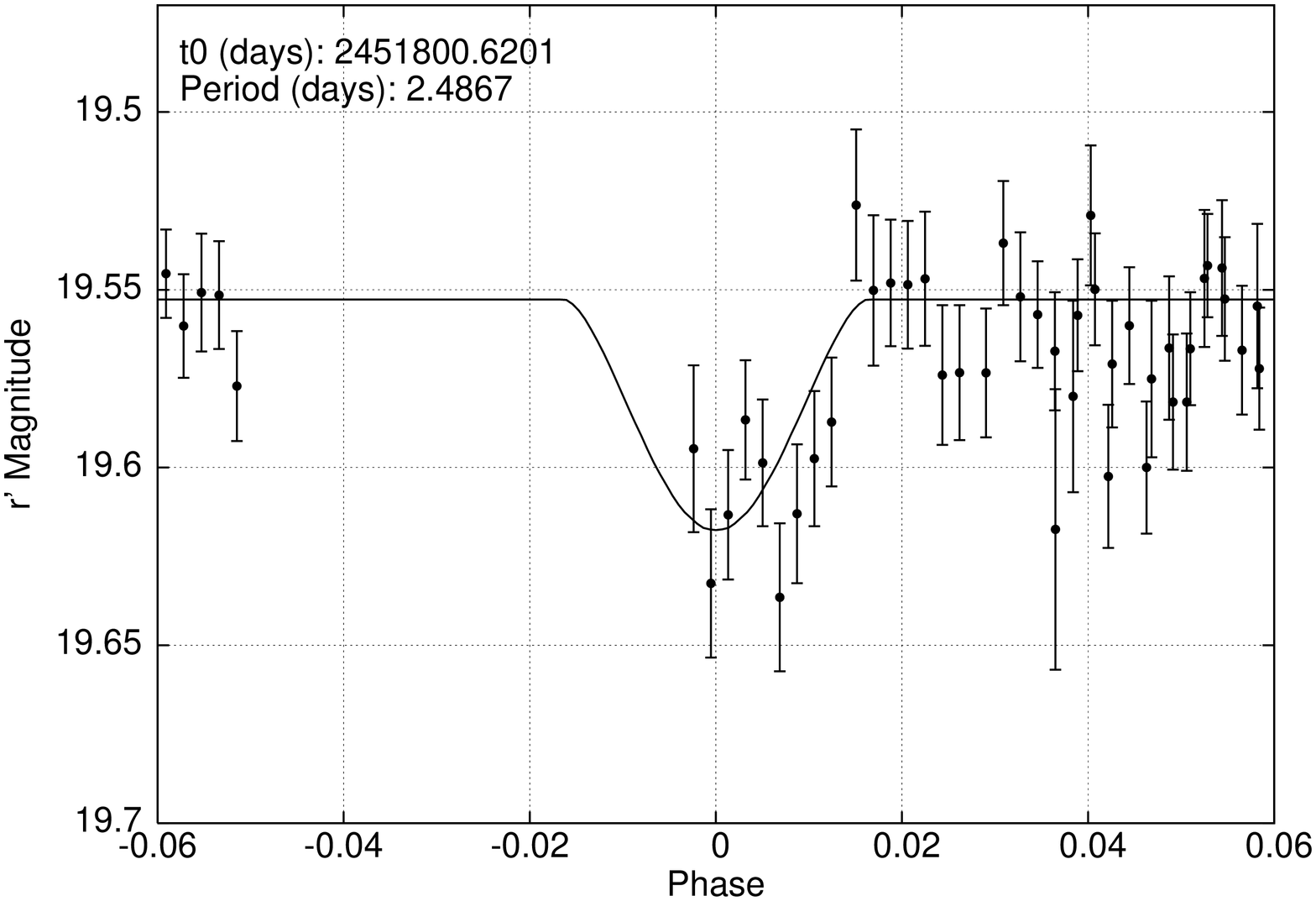,angle=270.0,width=0.4\linewidth}} \\
\end{tabular}
\caption{Eclipsing binary fits for TR-3.}
\end{figure*}

\begin{table*}
\centering
\caption{Star, companion and lightcurve properties for TR-3 as obtained from the various fits detailed in 
         Section 6.9. Column 4 is calibrated $r^{\prime}$ magnitude. $M_{\mbox{\scriptsize c}}$ and 
         $R_{\mbox{\scriptsize c}}$ are the companion mass and radius respectively. The ratio 
         $f_{\mbox{\scriptsize c}} / f_{*}$ is the flux ratio of the companion to the primary star in the Sloan
         $r^{\prime}$ waveband. The quantity $d$ 
         is the distance to the system.}
\begin{tabular}{@{}lc@{}c@{}c@{}ccccccc@{}c@{}cc}   
\hline
Model & $t_{0} - 2451000.0$ & $P$ & $m_{0}$ & $\Delta m$ & $\Delta t$ & $i$ & $M_{*}$ & $R_{*}$ & 
$M_{\mbox{\scriptsize c}}$ &
$R_{\mbox{\scriptsize c}}$ & $f_{\mbox{\scriptsize c}} / f_{*}$ & $d$ & $\chi^{2}$ \\
  & (d) & (d) & $\;\;$($r^{\prime}$ mag)$\;\;$ & (mag) & (h) & ($\degr$) & ($M_{\sun}$) & ($R_{\sun}$) & ($M_{\sun}$) & 
($R_{\sun}$) & $\;$(Sloan $r^{\prime}$)$\;$ & (pc) & \\  
\hline
A &
800.6201(7) & 2.4867(116) & 19.553 & 0.086 & 1.93 & 83.48(6) & 0.701(5) & 0.670(5) & 0.661(30) & 0.629(29) &
0.669  & 2788(6) & 725.88 \\
B &
800.6203(7) & 1.2431(58)  & 19.553 & 0.070 & 1.72 & 87.1(1.2) & 0.679(4) & 0.649(4) & 0.000     & 0.151(7)  &
0.000  & 1995(5) & 729.99 \\
C &
800.6204(7) & 1.2430(58)  & 19.553 & 0.070 & 1.68 & 88.8(1.5) & 0.682(4) & 0.651(4) & 0.126(6)  & 0.149(6)  &
0.0019 & 2021(5) & 731.73 \\
\hline
\end{tabular}
\raggedright
A = Similar Size Stellar Companion. B = Planetary Companion. C = Small Stellar Companion.
\end{table*}

{\bf TR-1} exhibits one poorly sampled partially observed eclipse during the 1999-07 run and one well sampled fully 
observed eclipse of 
depth 0.07~mag and duration 6.0~h in the 2000-09 run (Fig.~12b), and as a result we were unable to determine a 
period for the system. The star has
$r^{\prime}~\approx~20.70$~mag and $r^{\prime}-i^{\prime} \approx 0.63$~mag from which we derive a 0.87$M_{\sun}$ 
late G 
star primary that lies far behind the cluster ($d = 7.3$kpc). A central transit fit to the 2000-09 data
yields a minimum companion radius of 0.188$R_{\sun}$ from which we cannot rule out the transiting planet model. 

Analysing the single eclipse in the 2000-09 run using the same method as for EB-21 in Section 6.6 yields a 
predicted period of 7.0~d for $b = 0$ that increases 
rapidly with increasing values of $b$. In fact $\chi^{2}_{\mbox{\small fold}} = \chi^{2}_{\mbox{\small ecl}}$ for 
all $b > 0.09$. Fig.~12a shows a plot of $\chi^{2}_{\mbox{\small ecl}}$ versus $R_{\mbox{\small c}} / R_{*}$ 
(dashed line) for $b = 0$ ($R_{\mbox{\small c}} / R_{*} = 0.223$) to $b = 1.13$ ($R_{\mbox{\small c}} / R_{*} = 0.500$).
The minimum value of $\chi^{2}_{\mbox{\small ecl}}$
obtained is $\chi^{2}_{\mbox{\small ecl}} = 925.0$ corresponding to $R_{\mbox{\small c}} / R_{*} = 0.500$, and this is 
marked on Fig.~12a as a horizontal shorter dashed line, along with the chi squared values 
$\chi^{2}_{\mbox{\small ecl}} = 925.0 + 1.0$ and $\chi^{2}_{\mbox{\small ecl}} = 925.0 + 4.0$ corresponding to the 1 
and 2$\sigma$ confidence levels. 

One can see from Fig.~12a that $R_{\mbox{\small c}} / R_{*} \geq 0.243$ with a 
1$\sigma$ confidence. This is equivalent to stating that $R_{\mbox{\small c}} \geq 0.205 R_{\sun}$ with a 1$\sigma$  
confidence. As a result, we can only rule out the transiting planet model for this transit candidate at the 1$\sigma$
level and therefore further observations are required to confirm the conclusion that this system is an eclipsing binary. 
Fig.~12b shows a plot of the 
solution corresponding to the minimum value of $\chi^{2}_{\mbox{\small ecl}}$ along with the lightcurve data for the 
night on which the eclipse occurs.
This solution predicts a period of 58$\pm$12 d and an inclination of 89.1$\pm$0.3$\degr$.

\subsection{INT-7789-TR-2}

{\bf TR-2} shows a single 0.02~mag eclipse of duration 2.5~h during the 2000-09 run (Fig.~12d). With
$r^{\prime}~\approx~18.02$~mag and $r^{\prime}-i^{\prime}~\approx~0.47$~mag we find that the primary is a 
1.20$M_{\sun}$
F star at $d~=~5.3$kpc, behind the cluster. A central transit fit to the 
2000-09 data yields a minimum companion radius of 0.174$R_{\sun}$ from which we cannot rule out the transiting planet 
model.

Applying the same analysis as for EB-21 in Section 6.6 to the single eclipse yields a predicted period of 0.62~d 
for $b = 0$ that increases slowly with increasing values of $b$. For $b < 0.66$, 
$\chi^{2}_{\mbox{\small fold}}~\gg~\chi^{2}_{\mbox{\small ecl}}$ and for $b \geq 0.66$, $\chi^{2}_{\mbox{\small fold}}$ 
oscillates between the states 
$\chi^{2}_{\mbox{\small fold}}~\gg~\chi^{2}_{\mbox{\small ecl}}$ and 
$\chi^{2}_{\mbox{\small fold}} = \chi^{2}_{\mbox{\small ecl}}$. Hence we can be sure that $b \geq 0.66$, which
corresponds to $R_{\mbox{\small c}} / R_{*} \geq 0.137$. In Fig.~12c we plot $\chi^{2}_{\mbox{\small ecl}}$
versus $R_{\mbox{\small c}} / R_{*}$ (dashed line) and $\chi^{2'}_{\mbox{\small fold}}$
versus $R_{\mbox{\small c}} / R_{*}$ (continuous line) where:
\begin{equation}
\chi^{2'}_{\mbox{\small fold}} =
  \begin{cases}
    \chi^{2}_{\mbox{\small fold}} & \text{if $b < 0.66$} \\
    \chi^{2}_{\mbox{\small ecl}} & \text{if $b \geq 0.66$}   
  \end{cases}
\end{equation}
The minimum value of $\chi^{2}_{\mbox{\small ecl}}$ obtained is $\chi^{2}_{\mbox{\small ecl}} = 806.6$ corresponding to 
$R_{\mbox{\small c}} / R_{*} = 0.137$, and this is 
marked on Fig.~12c as a horizontal shorter dashed line, along with the chi squared values  
$\chi^{2}_{\mbox{\small ecl}} = 806.6 + 1.0$, $\chi^{2}_{\mbox{\small ecl}} = 806.6 + 4.0$ and
$\chi^{2}_{\mbox{\small ecl}} = 806.6 + 9.0$ corresponding to the 1, 2 and 3$\sigma$ confidence levels.

One can see from Fig.~12c that $0.137~\leq~R_{\mbox{\small c}}/R_{*}~\leq~0.144$ with a 
1$\sigma$ confidence. This is equivalent to stating that $R_{\mbox{\small c}}~=~0.185^{+0.009}_{-0.000}~R_{\sun}$.
Hence the conclusion at the 1$\sigma$ level is that this is a possible transiting planet in orbit around a 
1.20$M_{\sun}$ F star that merits follow up observations. 
Fig.~12d shows a plot of the 
solution corresponding to the minimum value of $\chi^{2}_{\mbox{\small ecl}}$ along with the lightcurve data for the 
night on which the eclipse occurs. This solution predicts a period of 1.8$\pm$1.3 d and an inclination of
81.9$\pm$1.6$\degr$.

\subsection{INT-7789-TR-3}

{\bf TR-3} exhibits two fully observed eclipses and one partially observed eclipse during the 2000-09 run. The 
eclipses have a depth of 0.07~mag and duration 1.7~h with a period of 1.24~d (Figs.~12e, g \& h). The star has
$r^{\prime}~\approx~19.55$~mag and $r^{\prime}-i^{\prime} \approx 0.97$~mag from which we derive a 0.68$M_{\sun}$ 
primary 
star of spectral type K5V that lies slightly in front of the cluster ($d = 2.0$kpc). A full 
transit fit to the 2000-09 lightcurve data yields a best fit companion radius of 0.151$\pm$0.007$R_{\sun}$ consistent 
with the radius of a transiting planet (Fig.~12f). This solution is reported in Table 7 under Model B, and the 
$\chi^{2}$ of the fit is 729.99.

However, there are two other models for TR-3 that should be considered to see if they produce a better 
$\chi^{2}$ value for the fit to the lightcurve data. It is possible that the companion is a smaller and less luminous  
star than the primary star that produces a secondary eclipse which is not visible in the lightcurve folded at the 
$\sim$1.24 day period, and it is also possible that the companion is a star of similar size and luminosity to the 
primary star and that the system actually has an orbital period of $\sim$2.48 days. In order to test these models for 
TR-3 we have developed an eclipsing binary model based on the same assumptions as for the star and planet system 
presented in Section 6.1 except that we assume that the companion is now luminous and massive, and that our 
theoretical main sequence relationships adopted in Section 4.3 apply to the companion. 

The eclipsing binary model has five parameters to optimise: orbital period $P$, time of mid-eclipse $t_{0}$, orbital
inclination $i$, companion to primary star radius ratio $R_{\mbox{\small c}} / R_{*}$ and a constant magnitude $m_{0}$.
We fitted this model to the lightcurve of TR-3 by calculating the $\chi^{2}$ for a grid in $i$ and 
$R_{\mbox{\small c}} / R_{*}$. For each
value of $R_{\mbox{\small c}} / R_{*}$, we had to recalculate the distance $d$ to the system, the values of $M_{*}$ and 
$R_{*}$ for the primary star, and the mass of the companion $M_{\mbox{\small c}}$. This was done by constructing a theoretical binary 
main sequence for the current value of $R_{\mbox{\small c}} / R_{*}$ and then finding the distance $d$ such that this 
model passes through the 
position of TR-3 in the colour-magnitude domain. The initial 
value of $P$ was either 1.24~d or 2.48~d corresponding to the smaller or similar size stellar companion models 
respectively. Table 7 reports the results of 
these fits. Fig.~13 shows a chi 
squared contour map, a folded and binned lightcurve with the best fit eclipsing binary model, an unbinned close-up 
of the folded lightcurve around the primary eclipse along with the best fit model and another unbinned close-up 
around the secondary 
eclipse along with the best fit model. The left hand column of diagrams in Fig.~13 applies to the case of the smaller 
stellar companion and the right hand column of diagrams in Fig.~13 applies to the case of the similar size stellar 
companion.

Table~7 shows that the best model for TR-3 is the eclipsing binary model with a similar size stellar companion 
since this model attains the smallest $\chi^{2}$ of 725.88. All three models require exactly 5 parameters to be 
optimised and hence we calculate a likelihood ratio of $\sim$7.8 for the eclipsing binary model with a similar 
size stellar companion compared to the transiting planet model, and we calculate a likelihood ratio of $\sim$18.6 for 
the eclipsing binary model with a similar size stellar companion compared to the eclipsing binary model with a smaller 
stellar companion. Finally, we calculate a likelihood ratio of $\sim$2.4 for the transiting planet model compared to 
the eclipsing binary model with a smaller stellar companion. Hence our conclusion is that this system is most likely to be  
a grazing eclipsing binary with period 2.49~d consisting of a K4V star primary and a K5V star secondary 
that lies at $d = 2.8$kpc, slightly behind the cluster. However, further observations will be 
required confirm this conclusion and categorically rule out the transiting planet model.

\section{Conclusions}

In the search for our transit candidates we have developed an accurate, efficient and fast photometry 
pipeline employing the technique of difference image analysis. Raw data from the telescope are processed by the 
pipeline with minimal user input in order to directly produce lightcurves and colour magnitude diagrams.
This is especially important considering the high quantity of data that may arise from a transit survey.

Our analysis of the colour magnitude diagrams by including the treatment of extinction for the open cluster 
NGC~7789 has allowed us to assign a model-dependent mass, radius and distance to each star. Such information is vital 
in 
the subsequent analysis of the transit candidates since it 
allows a direct estimate of the companion radius. We detected 24 transit candidates which warranted a detailed 
analysis of their lightcurves and we were able to determine periods for 14 of these candidates. Of the 10 candidates 
without periods, we could rule out the transiting planet model for 7 of them by determining the minimum companion radius 
and for another one by predicting the orbital period. For INT-7789-TR-1, it was only at the 1$\sigma$ level 
that we could rule out the transiting planet model based on the shape of the best eclipse. For INT-7789-TR-2
we found a companion radius of $0.185^{+0.009}_{-0.000} R_{\sun}$ ($1.81^{+0.09}_{-0.00} 
R_{\mbox{\small J}}$) based on the analysis of the best eclipse. Follow-up observations (see below) will be required 
for both of these candidates in order confirm that INT-7789-TR-1 is an eclipsing binary and in order to determine the 
nature of INT-7789-TR-2.

For the 14 transit candidates with well determined periods, we could rule out the transiting planet model for 4 of 
them from the detection of previously disguised secondary eclipses, and for 3 of them from the observation that the 
out-of-eclipse lightcurve data exhibit ellipsoidal variations and heating effects. One of the candidates is possibly a 
new cataclysmic variable with a long period (10.8 h) which could be a cluster member, worthy of follow up observations in 
its own right. All of the 8 above mentioned candidates plus another 5 may be ruled out as having planetary companions 
by considering that the companion radius obtained from the full transit fit is greater than 0.2$R_{\sun}$. 
For INT-7789-TR-3, none of the above arguments may be used to rule 
out the transiting planet model. However, on application of an eclipsing binary model to the lightcurve we find that 
the model consisting a pair of grazing K dwarf stars is $\sim$7.8 times more likely than the transiting planet model. 
This is by no means a definitive conclusion that INT-7789-TR-3 is an eclipsing binary since there is a 
non-negligible probability that the transiting planet model is still valid. Follow-up observations will 
be required to confirm that INT-7789-TR-3 is the type of eclipsing binary that we predict in this paper.

Future photometric observations of the three transit candidates for which we could not rule out the transiting planet 
model with confidence should consist of time series observations in two different filters. Eclipsing binary status may 
be confirmed by the observation of different eclipse depths in different filters since a planetary transit is an 
achromatic event. INT-7789-TR-1 and INT-7789-TR-2 also require the observations of multiple eclipses in 
order to determine their period and whether they exhibit secondary eclipses or not. If these follow up photometric 
observations still allow the possibility that the transiting planet model is valid, then radial velocity observations
may be used to place an upper limit on the mass of the orbiting companion, hopefully low enough to rule out a stellar 
or brown dwarf companion. The fact that these candidates are so faint ($r^{\prime} \approx 20.7$~mag for 
INT-7789-TR-1, 
$r^{\prime} \approx 18.0$~mag for INT-7789-TR-2 and $r^{\prime} \approx 19.6$~mag for INT-7789-TR-3) makes it very 
unlikely that radial velocity observations with 10-m class telescopes will achieve the accuracy 
required to determine the actual mass of the companion \citep{cha2003}.

From a simple signal-to-noise argument presented in Section 4.5, we expected to detect $\sim$2
transiting hot Jupiters. At most we have detected 3 transiting hot Jupiters, 
but our analysis of these candidates shows that this is very unlikely. Follow up observations will most likely show 
that our candidates are eclipsing binaries, which means that our transit survey will have produced a null 
result. We are currently modelling in more detail the number of hot Jupiters that we expected to detect as a function of 
planetary radius, orbital period, star mass and detection threshold via Monte     
Carlo simulations. The results of these simulations are consistent with our simple estimate of the expected number of detections, 
and by combining this information with the results of further observations on the three remaining transit candidates, we 
will be able to estimate the hot Jupiter fraction of the cluster field, a very important result for the testing of star 
and planetary formation theories.

\section*{Acknowledgements}

This research made use of the SIMBAD data base operated at CDS, Strasbourg, France and the WEBDA data base operated at 
the University Of Lausanne, Switzerland. This paper was based on observations made with the Isaac Newton 
Telescope operated by the Isaac Newton Group on the island of La Palma in the Spanish Observatorio del Roque 
de los Muchachos. D.M.Bramich was funded by a PPARC research studentship during the course of this work, and would like 
to thank both St. Andrews University, Scotland, and the Instituto de Astrofisica de Canarias, Spain, for their 
hospitality during this time. A special thanks goes to Rocio Mu\~niz Santacoloma who provided the inspiration and support 
necessary in order to complete this work.

\label{lastpage}

\end{document}